\documentclass[conference]{IEEEtran}

\usepackage[utf8]{inputenc}
\usepackage[T1]{fontenc}
\usepackage[letterpaper,margin=1in]{geometry}
\usepackage{amsmath,amssymb,amsfonts}
\usepackage{amsthm}
\usepackage{booktabs}
\usepackage{array}
\usepackage{tabularx}
\usepackage{makecell}
\usepackage{adjustbox}
\usepackage{multirow}
\usepackage{graphicx}
\usepackage[table]{xcolor}
\usepackage{url}
\usepackage{cite}
\usepackage{microtype}
\usepackage{enumitem}
\usepackage{mathtools}
\usepackage{pifont}
\usepackage{longtable}
\usepackage[most]{tcolorbox}
\usepackage{placeins}
\usepackage{float}
\usepackage{algorithm}
\usepackage{algpseudocode}
\usepackage{tikz}
\usepackage[hidelinks]{hyperref}
\hypersetup{
  pdftitle={Targeted Counterfactual Fingerprinting for Black-Box LLM Ownership Verification},
  pdfauthor={Yutong Wu, Xiaofan Bai, Shixin Li, Pingyi Hu, Ziqi Zhou, Zilong Wang, Xiaojing Ma, Songfeng Lu, Yuhong Li, Jin Xuan, Yi Wang, Dongmei Zhang, Bin Benjamin Zhu}
}

\graphicspath{{./}}
\makeatletter
\def\input@path{{./}}
\makeatother

\definecolor{HeaderGray}{gray}{0.92}
\definecolor{familybg}{RGB}{252,235,235}
\definecolor{indepbg}{RGB}{235,248,235}
\definecolor{BestGreen}{RGB}{224,244,232}

\definecolor{TCFBlue}{RGB}{35,91,128}
\definecolor{TCFBlueLight}{RGB}{241,248,252}
\definecolor{TCFGreen}{RGB}{43,126,91}
\definecolor{TCFGreenLight}{RGB}{242,250,246}
\definecolor{TCFOrange}{RGB}{190,99,48}
\definecolor{TCFOrangeLight}{RGB}{255,248,239}
\definecolor{TCFCream}{RGB}{255,251,239}
\definecolor{TCFDark}{RGB}{67,69,73}

\newtcolorbox{tcfabstractbox}{%
  enhanced,colback=TCFBlueLight,colframe=TCFBlue,
  boxrule=0.55pt,borderline west={2.2pt}{0pt}{TCFBlue},
  arc=1.4mm,boxsep=2pt,left=4pt,right=4pt,top=4pt,bottom=4pt,
  before skip=2pt,after skip=5pt}
\newtcolorbox{tcfcodebox}{%
  enhanced,colback=TCFGreenLight,colframe=TCFGreen,
  boxrule=0.55pt,borderline west={2.2pt}{0pt}{TCFGreen},
  arc=1.4mm,boxsep=2pt,left=4pt,right=4pt,top=3.5pt,bottom=3.5pt,
  before skip=2pt,after skip=6pt}
\newtcolorbox{tcfhighlightbox}[1]{%
  enhanced,breakable,title={#1},
  colback=TCFOrangeLight,colframe=TCFOrange,
  colbacktitle=TCFOrange,coltitle=white,
  fonttitle=\bfseries\footnotesize,boxrule=0.55pt,arc=1.4mm,
  boxsep=2pt,left=4pt,right=4pt,top=3.5pt,bottom=3.5pt,
  before skip=5pt,after skip=6pt}
\newtcolorbox{tcfresultbox}[1]{%
  enhanced,breakable,title={#1},
  colback=TCFGreenLight,colframe=TCFGreen,
  colbacktitle=TCFGreen,coltitle=white,
  fonttitle=\bfseries\footnotesize,boxrule=0.55pt,arc=1.4mm,
  boxsep=2pt,left=4pt,right=4pt,top=3.5pt,bottom=3.5pt,
  before skip=5pt,after skip=6pt}
\newtcolorbox{tcfpromptbox}[1]{%
  enhanced,breakable,title={#1},
  colback=TCFCream,colframe=TCFDark,
  colbacktitle=TCFDark,coltitle=white,
  fonttitle=\bfseries\small,boxrule=0.55pt,arc=1.7mm,
  boxsep=2pt,left=6pt,right=6pt,top=5pt,bottom=5pt,
  before skip=6pt,after skip=7pt}
\newtcolorbox{tcfappendixbox}[1]{%
  enhanced,breakable,title={#1},
  colback=TCFBlueLight,colframe=TCFBlue,
  colbacktitle=TCFBlue,coltitle=white,
  fonttitle=\bfseries\small,boxrule=0.55pt,arc=1.7mm,
  boxsep=2pt,left=6pt,right=6pt,top=5pt,bottom=5pt,
  before skip=6pt,after skip=8pt}
\newcommand{\hf}[1]{{\ttfamily\scriptsize #1}}

\newcolumntype{C}[1]{>{\centering\arraybackslash}p{#1}}

\newcommand{\method}{TCF}
\newcommand{\best}[1]{\cellcolor{BestGreen}\textbf{#1}}

\newtheorem{lemma}{Lemma}
\newtheorem{theorem}{Theorem}[section]
\newtheorem{assumption}[theorem]{Assumption}

\theoremstyle{definition}
\newtheorem{definition}[theorem]{Definition}

\newcommand{\AffiliationLogoStrip}{%
  \begin{minipage}{\textwidth}
    \raggedright
    \raisebox{-0.010in}[0.34in][0pt]{%
      \includegraphics[height=0.325in,keepaspectratio,trim=0 0 722.95 0,clip]{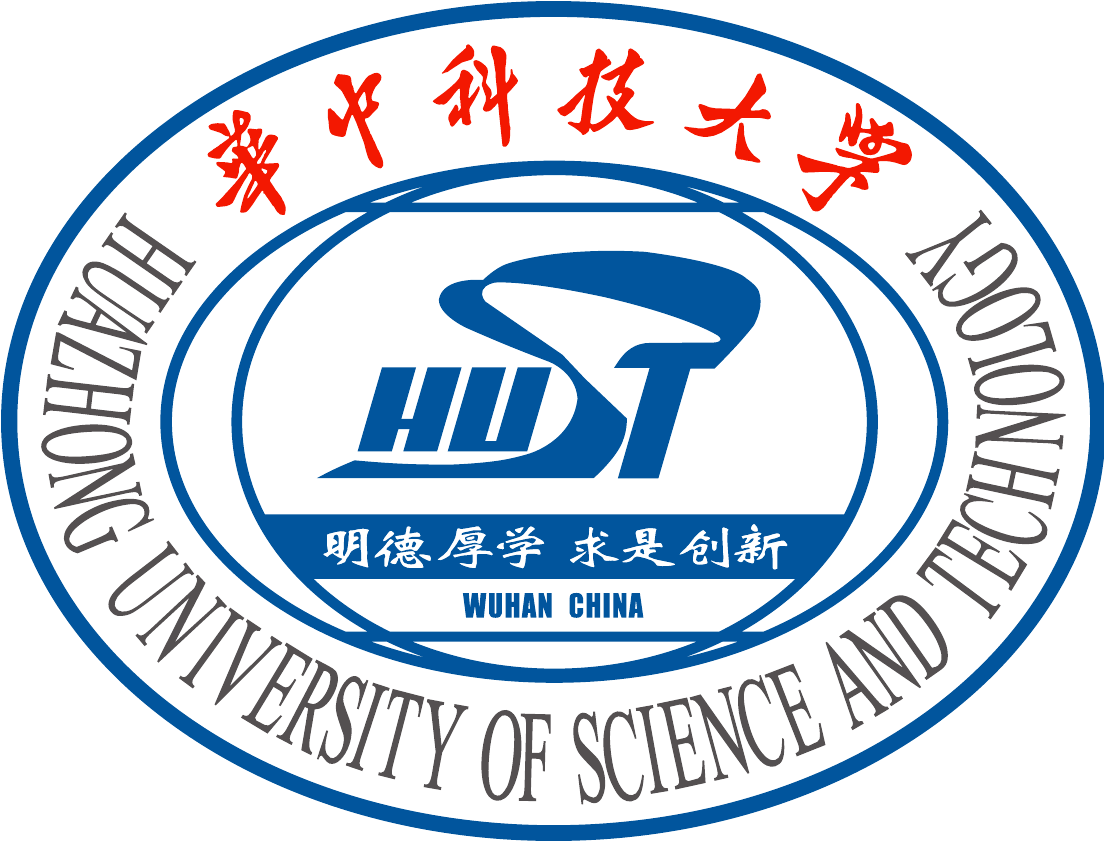}}%
    \hspace{0.30cm}%
    \raisebox{0.025in}[0.27in][0pt]{%
      \includegraphics[height=0.235in,keepaspectratio]{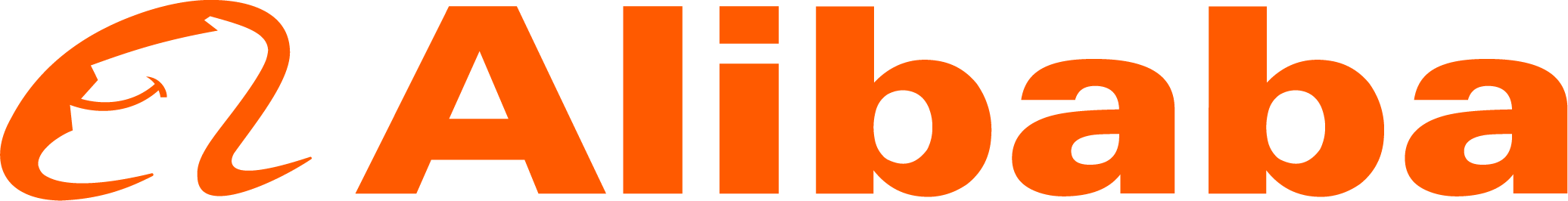}}%
    \hspace{0.30cm}%
    \raisebox{0.025in}[0.27in][0pt]{%
      \includegraphics[height=0.235in,keepaspectratio]{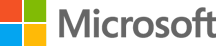}}%
    \par\vspace{0.10in}%
    \noindent\rule{\textwidth}{0.55pt}%
  \end{minipage}%
}

\title{\vspace{0.82in}Targeted Counterfactual Fingerprinting for Black-Box LLM Ownership Verification}
\author{%
\IEEEauthorblockN{%
\begin{minipage}{0.98\textwidth}
\centering
Yutong Wu$^{1,*}$, Xiaofan Bai$^{2,*}$, Shixin Li$^{1}$, Pingyi Hu$^{1}$,
Ziqi Zhou$^{1}$, Zilong Wang$^{1}$, Xiaojing Ma$^{1}$\\
Songfeng Lu$^{1}$, Yuhong Li$^{2}$, Jin Xuan$^{2}$,
Yi Wang$^{3}$, Dongmei Zhang$^{3}$, Bin Benjamin Zhu$^{3}$
\end{minipage}}
\IEEEauthorblockA{%
\begin{minipage}{0.98\textwidth}
\centering\normalsize
\mbox{$^{1}$Huazhong University of Science and Technology \enspace
$^{2}$Alibaba Group \enspace
$^{3}$Microsoft Corporation}\\[1pt]
\mbox{\texttt{wyt@hust.edu.cn, baixiaofan.bxf@alibaba-inc.com}\enspace $^{*}$Equal contribution.}
\end{minipage}}
}

\begin{document}
\maketitle

\begin{tikzpicture}[remember picture,overlay]
\node[anchor=north west,xshift=1.00in,yshift=-0.55in]
  at (current page.north west) {\AffiliationLogoStrip};
\end{tikzpicture}

\begin{tcfabstractbox}
\footnotesize\noindent\textbf{Abstract---}
Large language models (LLMs) are high-value assets that can be derived through redeployment, fine-tuning, quantization, or further alignment. Because deployed LLMs are commonly exposed only through query APIs, ownership verification must often rely on black-box text responses. This setting is difficult: generations are open-ended and can vary across repeated queries, while existing black-box fingerprints rely on signals that are fragile under a final-response interface, including full-text matching, soft behavioral features, or model-specific prompts designed not to transfer. The central challenge is to obtain auditable evidence that separates derivatives of a protected model from independently trained models using only black-box queries.

We propose \method{} (\emph{Targeted Counterfactual Fingerprinting}), a black-box LLM fingerprinting framework that converts open-ended generation comparison into constrained-answer targeted counterfactual transfer. \method{} restricts each verification query to a finite answer space, reducing the surface-form ambiguity that enters the verification score, and optimizes a prompt perturbation toward a counterfactual target different from the protected model's clean answer on the original prompt. The resulting perturbed prompt is a fingerprint: verification reduces to checking whether the suspect model's parsed final answer matches the recorded target. To construct discriminative fingerprints offline, we introduce the \emph{source-model counterfactual margin} (SCM), a protected-model-only quantity that certifies the target is unlikely before the perturbation and likely after it; SCM controls target selection, perturbation stopping, and fingerprint filtering. Under explicit derived-preservation and independent-transfer budgets motivated by local behavioral closeness, we derive a target-accuracy gap between derived and independent models. The bounds explain when targeted counterfactual transfer becomes ownership evidence, while empirical SCM diagnostics motivate an interior feasible operating region rather than arbitrarily large margins. Across four LLM families, \method{} achieves an average AUC of \textbf{0.9861}, improving over state-of-the-art black-box fingerprints TRAP, ProFLingo, and ZeroPrint by $0.07$--$0.19$.
\end{tcfabstractbox}

\begin{tcfcodebox}
\footnotesize\raggedright
\textbf{Project available at: }
\href{https://github.com/Underflow-1/TCF-LLM-Fingerprinting}{\nolinkurl{https://github.com/Underflow-1/TCF-LLM-Fingerprinting}}
\end{tcfcodebox}

\section{Introduction}
\label{sec:introduction}

Large language models (LLMs) are high-value intellectual property assets, but verifying whether a deployed model derives from a protected source remains a critical unsolved challenge. Training state-of-the-art LLMs demands massive data, exascale computation, and costly post-training alignment, yet third parties can reuse these models with minimal friction—wrapping checkpoints in APIs, fine-tuning for downstream tasks, quantizing for efficiency, or adapting via preference tuning. All such reuse is typically delivered as a black-box cloud service: auditors can only submit prompts and read text responses, with no access to weights, hidden states, logits, or decoding configurations \cite{gubri2024trap,jin2024proflingo,shao2026zeroprint,carlini2024stealing}. This creates a fundamental challenge: enforcement of licenses and usage policies requires reliable, auditable evidence, but the only available signal is ambiguous, variable text output.

Existing ownership verification methods fall into two categories: watermarking and fingerprinting. \emph{Watermarking} embeds ownership signals by modifying training, fine-tuning, decoding, or model parameters \cite{kirchenbauer2023watermark,abdelnabi2021adversarial,hu2023unbiased}. While effective when owners control the full pipeline, it is invasive and cannot be retroactively applied to released, leaked, or third-party-modified models. \emph{Fingerprinting}, by contrast, constructs verification queries from pre-trained models without modification, making it a natural non-invasive approach for auditing deployed derivatives \cite{jin2024proflingo,gubri2024trap,shao2026zeroprint}.

The core challenge for LLM fingerprinting is open-ended generation. Unlike classifiers with fixed label spaces, LLMs produce variable, discrete responses under stochastic decoding: semantically identical answers differ in wording, decoding settings alter formats, and semantic similarity scoring is brittle and hard to justify as audit evidence. Existing methods lean on signals that weaken under strict final-text interface:
ProFLingo relies on full-text response matching \cite{jin2024proflingo}, which breaks under paraphrasing or formatting changes.
TRAP uses GCG-optimized suffixes for exact identity verification \cite{gubri2024trap,zou2023universal}, but its non-transferable design misses fine-tuned, quantized, or modified derivatives.
ZeroPrint estimates soft behavioral features via perturbations \cite{shao2026zeroprint,chen2017zoo}, requiring embeddings and similarity scoring rather than parsed text responses.

We propose Targeted Counterfactual Fingerprinting (\method{}), a framework that reduces generation ambiguity at the scoring stage by reformulating verification as a constrained-answer problem. Each prompt restricts outputs to a finite set $\mathcal{A}=\{a_1,\ldots,a_K\}$ (implemented as $\{A,B,C,D\}$ multiple-choice labels), and a deterministic parser maps responses to valid labels or an invalid symbol. This converts noisy text generation into reproducible hard-label events, eliminating semantic scoring and reducing sensitivity to surface-form variation, decoding randomness, and format differences.

Constrained answers alone do not provide ownership evidence—unrelated models may share public knowledge and answer natural questions identically. \method{} therefore constructs fingerprints as \emph{targeted counterfactual transfers}: for each base question, we record the protected model's clean answer, then use GCG optimization to synthesize a text perturbation that flips the model's answer to a preselected unlikely target. The resulting prompt encodes a source-specific transition: a target that was improbable under the clean prompt becomes probable after perturbation.

All stages of fingerprint construction are governed by the \emph{source-model counterfactual margin} (SCM), a single quantity computed exclusively from the protected model. SCM quantifies both the unlikelihood of the target under the clean prompt and its likelihood after perturbation, unifying target selection, perturbation stopping, and fingerprint filtering. For the analysis, local behavioral closeness motivates explicit budgets for how much a derived model may weaken the source target preference and how much an independent model may acquire through generic transfer. Conditioned on these budgets, SCM yields quantitative lower and upper bounds on target transfer and a sufficient condition for a positive target-accuracy gap. Empirically, very strong perturbations can also increase generic transfer, motivating a feasible SCM operating region rather than maximal optimization.

\begin{tcfhighlightbox}{TCF at a glance}
\small
\textbf{Construct offline:} choose a clean-unlikely answer, optimize the source model toward that target, and retain only SCM-qualified prompts.
\textbf{Verify online:} send each private prompt to the suspect API, parse one final A/B/C/D label, and aggregate target hits.
\textbf{Ownership signal:} derivatives preserve more source-induced counterfactual transitions than independently trained models.
\end{tcfhighlightbox}

This framing fundamentally differs from prior adversarial-prompt work. TRAP optimizes for exact model identity \cite{gubri2024trap}, while ownership verification requires detecting family membership—derivatives may differ from the original but still be unauthorized copies. \method{} therefore treats \textbf{preservation of source-specific counterfactual transitions itself as ownership evidence}. Verification is strictly black-box and hard-label: we compute target accuracy (TA) across the private fingerprint set, using no suspect-model logits, embeddings, gradients, or hidden states.
Across four LLM families, \method{} achieves an average AUC of \textbf{0.9861} (perfect 1.0000 on three families), outperforming TRAP (0.7931), ProFLingo (0.8880), and ZeroPrint (0.9128) by 0.07–0.19. These results validate our central thesis: constrained-answer restriction reduces generation ambiguity at the verification interface, while SCM-guided counterfactual perturbation produces a highly discriminative signal for source-family ownership verification.

Our key contributions are:
\begin{itemize}[leftmargin=*]
    \item \textbf{Problem reformulation.} We recast black-box LLM ownership verification from fragile open-ended comparison into constrained-answer targeted counterfactual transfer, producing a robust, auditable hard-label signal.
    \item \textbf{Unified source-only design.} We introduce SCM, a single quantity computed exclusively from the protected model that unifies all stages of fingerprint construction, eliminating the need for surrogates or reference models.
    \item \textbf{Formal theoretical foundation.} Under explicit derived-preservation and independent-transfer budgets motivated by local behavioral closeness, we derive quantitative target-transfer bounds and a sufficient condition for a positive target-accuracy gap; empirical diagnostics identify a useful feasible SCM operating region.
    \item \textbf{State-of-the-art performance.} Under strict hard-label constraints, \method{} matches or outperforms the strongest evaluated baseline across the four source families and maintains strong performance across diverse derivative subgroups.
\end{itemize}

\section{Related Work}
\label{sec:related_work}

Model ownership verification divides into two paradigms: \emph{invasive watermarking} and \emph{non-invasive fingerprinting}.

\subsection{Watermarking and Invasive Ownership Signals}
Watermarking embeds verifiable ownership signals by intervening in the model lifecycle. Classical DNN watermarking uses backdoor triggers, parameter-level signatures, adversarial frontiers, or output modulations \cite{adi2018turning, uchida2017embedding,zhang2018protecting,rouhani2019deepsigns,lemerrer2019frontier}; LLM watermarking extends this to generation statistics, fine-tuning data, or trigger-response pairs \cite{kirchenbauer2023watermark,abdelnabi2021adversarial, hu2023unbiased,gu2022watermarking,li2023plmmark,li2024double,xu2025markllm}. Recent variants include instructional backdoors \cite{xu2024instructional}, cross-turn contextual backdoors \cite{xu2025ctcc}, and probabilistic provenance injection \cite{xu2025evertracer}.

These methods share two critical limitations for our threat model: (1) they cannot be retroactively applied to released, leaked, or third-party-modified models; (2) they require altering the protected model or its training pipeline. Additionally, watermarking faces unresolved tradeoffs in utility preservation, stealthiness, and attack resistance \cite{jovanovic2024watermark,chakraborty2023possibilities,sadasivan2023can}. Notably, text watermarking targets content provenance (identifying generated text) rather than model ownership (verifying a suspect service derives from a protected source), making it orthogonal to our problem.

\subsection{Non-Invasive Fingerprinting}
Fingerprinting constructs verification queries from pre-trained models without modification, addressing watermarking's core limitations. For DNNs, prior work leverages decision boundaries, conferrable adversarial examples, universal perturbations, adversarial trajectories, and robustness signatures \cite{cao2021ipguard,chen2022copy,lukas2021deep,peng2022fingerprinting,xu2024united,yang2023naturalfinger,yan2025enhancing}. These methods motivate our use of transferable behavioral signals, but they are designed for fixed-label classifiers with continuous inputs—LLMs present unique challenges due to discrete prompts, autoregressive generation, and surface-form variability from paraphrasing, formatting, and stochastic decoding.

LLM fingerprinting further splits by access level:
- \textbf{White-box/representation-level methods} require internal model information: HuRef extracts human-readable fingerprints from invariant parameters \cite{zeng2023huref}; REEF compares representation similarity across models \cite{zhang2025reef}. While these reveal stable internal structures in derivatives, they are incompatible with strict black-box APIs that expose only final text outputs. Our method is complementary: we use source-side access exclusively for offline fingerprint construction and verify suspects using only parsed text labels.

\subsection{Black-Box Behavioral Fingerprinting}
Our work focuses on the most restrictive and practically relevant setting: non-invasive, hard-label black-box verification using only final text responses. ProFLingo \cite{jin2024proflingo} constructs queries that elicit target responses from the source model, then checks for exact reproduction. It relies on full-text matching, which is fragile to paraphrasing, decoding variation, and formatting changes. TRAP \cite{gubri2024trap,zou2023universal} optimizes GCG adversarial suffixes to force the source model to output random strings. Designed for exact identity verification, its suffixes are intentionally non-transferable, failing to detect fine-tuned, quantized, or modified derivatives. ZeroPrint \cite{shao2026zeroprint,chen2017zoo} estimates local Jacobian-like fingerprints via semantic-preserving perturbations and output embeddings. While it captures richer behavioral features, it requires embedding comparisons and repeated perturbation queries, violating strict hard-label constraints.

\method{} addresses these limitations by reformulating verification as constrained-answer targeted counterfactual transfer. Unlike prior work, we restrict outputs to a finite label space (reducing surface-form ambiguity at scoring time) and use only parsed hard labels for verification. This produces a simple, reproducible signal compatible with text-only APIs that permit ordinary prompt queries, while SCM-guided perturbation ensures discriminative power for family-level ownership verification.

Table~\ref{tab:rw_position} summarizes the taxonomy of LLM ownership verification methods, highlighting our unique position as a non-invasive, strict black-box, hard-label approach.

\begin{table}[htp]
\centering
\caption{LLM ownership-verification methods. ``Modified?'' indicates whether the protected model or its training/decoding process is modified. ``Strict black box'' means that verification can be performed from the text responses only.}
\label{tab:rw_position}
\resizebox{\columnwidth}{!}{%
\begin{tabular}{lcccc}
\toprule
Method class & Examples & Modified? & Suspect access & Strict black box \\
\midrule
Text watermarking & \cite{abdelnabi2021adversarial,kirchenbauer2023watermark,hu2023unbiased} & Yes & Text/API & Partial \\
Backdoor / active watermark & \cite{gu2022watermarking,xu2024instructional,li2024double,xu2025ctcc} & Yes & Text/API & Often yes \\
White-box / representation fingerprint & \cite{zeng2023huref,zhang2025reef} & No & Weights/rep./circuits & No/partial \\
Black-box behavioral fingerprint & \cite{jin2024proflingo,gubri2024trap,shao2026zeroprint} & No & Text/API & Yes \\
\method{} & Ours & No & Text/API & Yes \\
\bottomrule
\end{tabular}}
\end{table}

\section{Threat Model and Problem Formulation}
\label{sec:threat_model}

\subsection{Threat Model}

We consider an LLM owner who protects a source model $M_0$ and audits a suspect service $M_s$. The goal is to determine whether $M_s$ is either the same model as $M_0$ or a derivative of $M_0$. The owner can access $M_0$ during offline fingerprint construction, but can only query $M_s$ through its public text interface during verification.

\textbf{Owner capability.}
The owner has full access to the protected source model $M_0$ before deployment or auditing. Therefore, the owner may use source-model probabilities, gradients, and decoding behavior to construct fingerprints. This access is only assumed for $M_0$ and is never assumed for the suspect model.

\textbf{Suspect-model access.}
The suspect model $M_s$ is a black-box text-generation service. During verification, the owner submits prompts and observes only the final generated text. The owner cannot access the suspect model's parameters, logits, token probabilities, hidden states, tokenizer internals, system prompt, decoding configuration, training data, or post-training procedure.

\begin{tcfhighlightbox}{Strict online evidence contract}
\small
The suspect-side audit uses only ordinary text queries and the final parsed label. No suspect-model probability, embedding, hidden state, gradient, tokenizer access, or semantic-similarity model enters the ownership score. Invalid or ambiguous responses are counted as failures rather than being repaired by a soft scorer.
\end{tcfhighlightbox}
The positive class contains $M_0$ and models derived from $M_0$. Such derived models may be produced by supervised fine-tuning, instruction tuning, RLHF, DPO-style preference tuning, LoRA adaptation, pruning, quantization, or lightweight post-processing. The negative class contains independently trained or unrelated LLMs. These models may still share public data, similar architectures, comparable sizes, or similar benchmark performance with $M_0$. Therefore, the verification task is not to test general capability similarity, but to identify whether the suspect service preserves source-family behavior of $M_0$.

\textbf{Adversary.} The suspect provider may deploy a modified version of the source model, adjust decoding parameters, alter system prompts, change response formatting, apply output post-processing, or use a fine-tuned, compressed, quantized, merged, or otherwise transformed model. The provider is not assumed to reveal any implementation details. We assume the private fingerprint set is not known to the suspect provider before verification. The suspect service is expected to answer ordinary user prompts, including the verifier's prompts, through its text-generation interface.

\subsection{Verification Objective}
\label{sec:Verification_Objective}
The owner constructs a private fingerprint set
\begin{equation}
\mathcal{F}=\{(p_i,y_i,t_i)\}_{i=1}^{n},
\label{eq:fingerprint_set_threat}
\end{equation}
where $p_i$ is the verification prompt, $y_i\in\mathcal{A}$ is the source model's clean answer associated with the underlying question, and $t_i\in\mathcal{A}\setminus\{y_i\}$ is the target answer recorded by the owner. During verification, the owner queries the suspect model $M_s$ with each $p_i$ and counts a hit only when the parsed response equals the recorded target $t_i$.
The verification score is the fingerprint target accuracy:
\begin{equation}
\operatorname{TA}(M_s;\mathcal{F})
=
\frac{1}{n}
\sum_{i=1}^{n}
\mathbb{I}
\left[
\operatorname{Parse}(r_s(p_i))=t_i
\right].
\label{eq:fta_threat}
\end{equation}
The owner declares $M_s$ as a derived model from $M_0$ if its target accuracy reaches a pre-chosen verification threshold $\zeta$.

\textbf{Verification goal.}
The goal is not to prove byte-level equality between $M_s$ and $M_0$. Instead, the goal is to decide whether $M_s$ exhibits sufficiently strong source-family fingerprint behavior under the strict black-box interface.

\section{Theoretical Analysis}
\label{sec:theory}

In this section, we develop an LLM-specific theory for our fingerprinting
method. The analysis is deliberately separated from the practical
construction of Section~\ref{sec:method}. We are aiming to distinguish two
model classes---models derived from the protected source $M_{0}$ and
independently trained models---and we ask: under what structural property
does a \emph{perturbed verification input}, constructed only from $M_{0}$,
induce a target-answer accuracy gap between these two classes? Our analysis
centers on the \emph{source-model counterfactual margin} (SCM), the
source-side target preference created by the optimized text perturbation.
A large SCM certifies a strong source-side counterfactual transition. The
locality premise motivates explicit derived-preservation and independent
transfer budgets, and the formal separation results below are conditioned on
those budgets. Under fixed budgets, larger SCM strengthens the separation
bounds. Empirically, however, stronger perturbations can also increase generic
transfer to independent models, which motivates a feasible operating region
and early stopping rather than maximal optimization.
The proof details, modeling qualifications, and cross-model validation are provided in Appendix~\ref{app:theory_details} and Appendix~\ref{app:cross_model_theory_validation}.

\subsection{Finite Constrained-Answer Space}
\label{subsec:legal_answer_channel}

Let $M_{0}$ denote the protected source LLM and let $M$ denote a suspect
LLM. For a prompt $p$, the raw output of $M$ is a text response, denoted
by $R_{M}(p)$. Our prompt template restricts valid answers to a finite
constrained answer space
\begin{equation}
\mathcal{A}=\{a_{1},a_{2},\ldots,a_{K}\},
\qquad K<\infty .
\label{eq:legal_answer_space}
\end{equation}
In our implementation, $\mathcal{A}$ is instantiated by multiple-choice
labels such as $\{A,B,C,D\}$.

Let $\operatorname{Parse}(\cdot)$ be a deterministic parser that maps a raw
response to one valid label in $\mathcal{A}$, or to $\bot$ if the response
is invalid, ambiguous, or unparsable. We define the \emph{constrained-answer
distribution} of model $M$ on prompt $p$ as
\begin{equation}
q_{M}(a\mid p)
=
\Pr\left[\operatorname{Parse}(R_{M}(p))=a\right],
\qquad a\in\mathcal{A}.
\label{eq:legal_answer_channel}
\end{equation}
Invalid responses are not assigned to any valid label, so
$\sum_{a\in\mathcal{A}}q_{M}(a\mid p)$ may be smaller than one. For a target
label $t\in\mathcal{A}$ and a prompt $p$, we define the target log-odds $L$
on model $M$ as a target preference signal
\begin{equation}
L_{M}(p,t)
=
\log
\frac{q_{M}(t\mid p)}{1-q_{M}(t\mid p)} .
\label{eq:target_log_odds}
\end{equation}
$L_{M}(p,t)>0$ means that the
target answer has a probability larger than $1/2$, while a larger value
means a stronger target preference. Details on probability estimation and clipping are deferred to Appendix~\ref{app:finite_channel_details}.

\subsection{Source-model Counterfactual Margin}
\label{subsec:scm_definition}

For a question $x$, let $p^{0}(x)$ be the clean prompt without any
perturbation. Let $u$ be a discrete input perturbation, constructed only
from $M_{0}$, and let
\begin{equation}
p^{u}(x)=\operatorname{Template}(u,x)
\label{eq:prefixed_prompt}
\end{equation}
be the perturbed fingerprint prompt. The clean answer of the source model is
\begin{equation}
y(x)=\arg\max_{a\in\mathcal{A}} q_{M_{0}}(a\mid p^{0}(x)).
\label{eq:source_clean_answer}
\end{equation}
A target answer $t(x)$ is drawn from $\mathcal{A}\setminus\{y(x)\}$, but
differing from the clean answer is not by itself enough: the fingerprint
should make this target answer unlikely before the perturbation and likely
after it. The counterfactual requirement is therefore certified by the
quantity below rather than by the choice $t(x)\neq y(x)$ alone.

\begin{definition}[Source-model counterfactual margin]
\label{def:scm}
For a candidate fingerprint $(x,u,t)$, its \textbf{S}ource-model
\textbf{C}ounterfactual \textbf{M}argin (SCM) is
\begin{equation}
\Gamma(x,u,t)
=
\min
\left\{
L_{M_{0}}(p^{u}(x),t), \thinspace
-L_{M_{0}}(p^{0}(x),t)
\right\}.
\label{eq:scm_definition}
\end{equation}
\end{definition}

The two terms in Eq.~\ref{eq:scm_definition} correspond to two sides of
the same counterfactual transition. The first term requires the target
answer to be likely after the perturbation is applied. The second term
requires the same target answer to be unlikely for the clean question.
Therefore, a large SCM certifies
\begin{equation}
q_{M_{0}}(t\mid p^{0}(x))\ \text{is small},
\qquad
q_{M_{0}}(t\mid p^{u}(x))\ \text{is large}.
\label{eq:scm_intuition}
\end{equation}

The SCM also provides an explicit probability interpretation. Let
$\sigma(z)=1/(1+e^{-z})$ be the sigmoid function. Since
Eq.~\ref{eq:target_log_odds} is the logit of $q_{M}(t\mid p)$, we have
\begin{equation}
q_{M}(t\mid p)=\sigma\left(L_{M}(p,t)\right).
\label{eq:probability_logit_relation}
\end{equation}
Thus, if $\Gamma(x,u,t)\ge \Gamma_{\min}$, Eq.~\ref{eq:scm_definition}
and Eq.~\ref{eq:probability_logit_relation} imply
\begin{equation}
q_{M_{0}}(t\mid p^{u}(x))
\ge
\sigma(\Gamma_{\min}),
\label{eq:prefixed_source_probability_control}
\end{equation}
and
\begin{equation}
q_{M_{0}}(t\mid p^{0}(x))
\le
\sigma(-\Gamma_{\min}).
\label{eq:clean_source_probability_control}
\end{equation}
Here, $\Gamma_{\min}$ denotes the minimum required SCM for accepting a
fingerprint. A candidate is called SCM-qualified if
\begin{equation}
\Gamma(x,u,t)\ge \Gamma_{\min}.
\label{eq:scm_qualified_threshold}
\end{equation}
Increasing $\Gamma_{\min}$ strengthens individual fingerprints but can
reduce the number and diversity of retained fingerprints. The detailed
 feasibility discussion is in Appendix~\ref{app:scm_feasibility_details}.

\subsection{How the Perturbation Controls SCM}
\label{subsec:gcg_controls_scm}

The SCM is useful because it can be controlled using only the protected
source model. The clean term $-L_{M_{0}}(p^{0}(x),t)$ is controlled by
selecting a target answer that is unlikely under the clean source prompt.
The perturbed term $L_{M_{0}}(p^{u}(x),t)$ is controlled by optimizing the
input perturbation $u$ against $M_{0}$. Let the source-model perturbation
loss be
\begin{equation}
\mathcal{L}_{\mathrm{src}}(u;x,t)
=
-\log q_{M_{0}}(t\mid p^{u}(x)).
\label{eq:gcg_loss}
\end{equation}

\begin{lemma}[Loss threshold for SCM control]
\label{lem:loss_threshold_for_scm}
Suppose the target label $t$ satisfies
\begin{equation}
-L_{M_{0}}(p^{0}(x),t)
\ge
\Gamma_{\min}.
\label{eq:clean_side_threshold}
\end{equation}
If the perturbation $u$ satisfies
\begin{equation}
\mathcal{L}_{\mathrm{src}}(u;x,t)
\le
\log\left(1+e^{-\Gamma_{\min}}\right),
\label{eq:gcg_loss_threshold}
\end{equation}
then the fingerprint satisfies
\begin{equation}
\Gamma(x,u,t)
\ge
\Gamma_{\min}.
\label{eq:scm_threshold_satisfied}
\end{equation}
\end{lemma}

\begin{figure}[htp]
    \centering
    \includegraphics[width=0.98\linewidth]{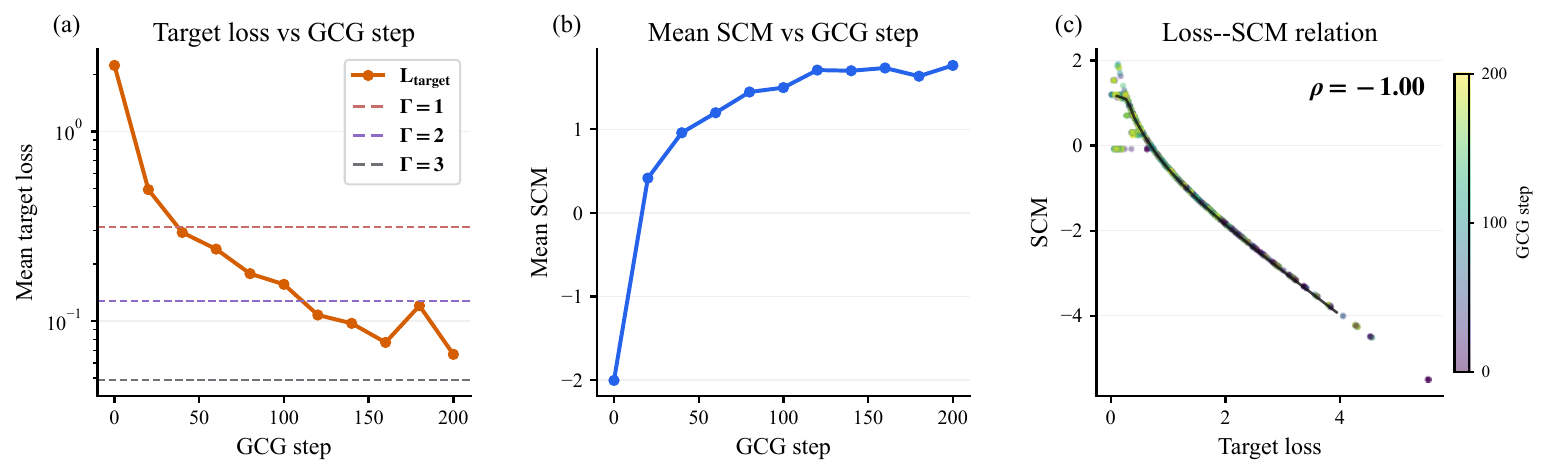}
    \caption{The relationship between target loss and SCM. Optimizing the source-side input perturbation reduces the source target loss, increases the mean SCM, and produces a strong negative relation.}
    \label{fig:scm_support_bundle}
\end{figure}

Lemma~\ref{lem:loss_threshold_for_scm} gives an explicit stopping rule for
the perturbation optimizer: after selecting a clean-unlikely target, the
perturbation should be optimized until the source target loss falls below
the threshold in Eq.~\ref{eq:gcg_loss_threshold}. Equivalently, the
optimization should make the source target probability exceed
$\sigma(\Gamma_{\min})$ and hence satisfy
Eq.~\ref{eq:scm_threshold_satisfied}. To verify Lemma~\ref{lem:loss_threshold_for_scm}, Fig.~\ref{fig:scm_support_bundle} records how the source target loss and SCM evolve during GCG optimization. Panel~(a) plots the target loss over steps, panel~(b) shows the mean SCM over the same trajectory, and panel~(c) directly compares target loss with SCM. The result is consistent across the three views: target loss decreases, SCM increases, and the two show a strong negative relation. This supports using target-loss minimization as a practical source-side proxy for increasing SCM, while the early stopping rule avoids unnecessary over-optimization. Additional details are in Appendix~\ref{app:empirical_scm_control}.

\begin{tcfresultbox}{Takeaway}
\small
Target-loss minimization is an effective operational proxy for increasing SCM.
\end{tcfresultbox}

\subsection{Local Behavioral Closeness}
\label{subsec:local_closeness}

Lemma~\ref{lem:loss_threshold_for_scm} focuses on the protected model
$M_{0}$. To reason about how the target transition induced by text input
perturbation $u$ behaves on other models, the analysis needs one structural
premise. Let $z_{M}(p)\in\mathbb{R}^{K}$ be the answer-logit vector that
$M$ assigns to the labels in $\mathcal{A}$ on prompt $p$, and define the
\emph{target margin}
\begin{equation}
m_{M}(p,t)
=
z_{M}(p)[t]-\max_{a\neq t} z_{M}(p)[a].
\label{eq:target_margin}
\end{equation}
A positive margin means $t$ is the model's parsed hard answer in
$\mathcal{A}$; a large margin means the hard answer is stable.

\begin{assumption}[Local behavioral closeness]
\label{ass:local_closeness}
There exist a closeness radius $\beta\ge0$ and a small slack
$\eta\in[0,1)$ such that, for a derived model $M_{D}$, a fraction at least
$1-\eta$ of fingerprint inputs $p\in\{p^{0}(x),p^{u}(x)\}$ obey
\begin{equation}
\left|m_{M_{D}}(p,t)-m_{M_{0}}(p,t)\right|\le \beta.
\label{eq:local_closeness_bound}
\end{equation}
No such uniform high-probability closeness to $M_{0}$ holds for an
independent model $M_{I}$.
\end{assumption}

Assumption~\ref{ass:local_closeness} states that a derived model's local
decision behavior stays near the protected model's at each evaluated input,
including both the clean and the perturbed prompt. The premise is
probabilistic: it need only hold for a large fraction of fingerprints, and
the verifier aggregates many fingerprints so that a per-fingerprint edge
concentrates into a reliable separation. This premise motivates the transfer diagnostics below but is not, by itself, a numerical bound on an unknown suspect; the formal results are conditioned on the explicit $\epsilon_D$ and $\epsilon_I$ budgets. Appendix~\ref{app:local_closeness_details} gives the detailed interpretation and targeted-transfer intuition.

\subsection{Targeted Transferability of Derived Models}
\label{subsec:derived_constraints}

We now instantiate the derived side of Assumption~\ref{ass:local_closeness}
on the parsed-log-odds scale used by the verifier. For a derived model
$M_D$ and a candidate fingerprint $(x,u,t)$, define
\begin{small}
\begin{equation}
\begin{aligned}
\Delta^{+}_{D}(M_D,x,u,t)
&=\max\Big\{0,\; L_{M_0}(p^{u}(x),t)\;\\
&\hspace{4.5em}-L_{M_D}(p^{u}(x),t)\Big\}.
\end{aligned}
\label{eq:derived_positive_drift}
\end{equation}
\end{small}
This quantity is the loss of perturbed target log-odds when moving from
the source model to the derivative. For any budget
$\epsilon_D(M_D,x,u,t)$ satisfying
\begin{equation}
\epsilon_D(M_D,x,u,t)\ge \Delta^{+}_{D}(M_D,x,u,t),
\label{eq:derived_valid_budget}
\end{equation}
we have the deterministic inequality
\begin{equation}
L_{M_D}(p^{u}(x),t)
\ge
L_{M_0}(p^{u}(x),t)-\epsilon_D(M_D,x,u,t).
\label{eq:derived_log_odds_preservation}
\end{equation}
Smaller budgets mean that the considered derivative class preserves the
source prefixed target preference more strongly.

\begin{figure}[htp]
    \centering
    \includegraphics[width=0.98\linewidth]{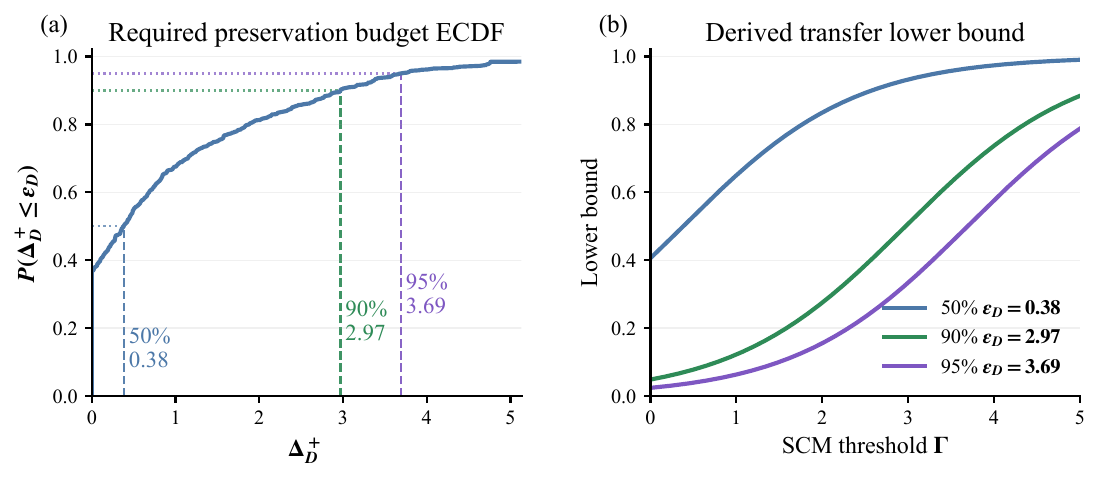}
    \caption{Validation of derived-model local closeness (Assumption~\ref{ass:local_closeness}). (a) The empirical weakening budget $\epsilon_D$ from $\Pr(\Delta_D^+\le\epsilon_D)$ over derived model--fingerprint pairs estimates the covered fraction on which the premise holds. (b) The resulting conditional transfer lower bound $\sigma(\Gamma-\epsilon_D)$ under the marked empirical budgets.}
    \label{fig:eq25_logodds_drift}
\end{figure}

\begin{lemma}[Conditional derived-model targeted-transfer lower bound]
\label{lem:derived_lower_bound}
Let $\Gamma=\Gamma(x,u,t)$. If
$\epsilon_D(M_D,x,u,t)\ge\Delta_D^+(M_D,x,u,t)$, then
\begin{equation}
q_{M_D}(t\mid p^{u}(x))
\ge
\sigma\!\bigl(\Gamma-\epsilon_D(M_D,x,u,t)\bigr).
\label{eq:derived_target_probability_lower_actual_gamma}
\end{equation}
In particular, if $\Gamma\ge\Gamma_{\min}$, then
\begin{equation}
q_{M_D}(t\mid p^{u}(x))
\ge
\sigma\!\bigl(\Gamma_{\min}-\epsilon_D(M_D,x,u,t)\bigr).
\label{eq:derived_target_probability_lower}
\end{equation}
\end{lemma}

Lemma~\ref{lem:derived_lower_bound} shows how SCM compensates for
post-training drift. To guarantee a derived-model target-hit probability
at least $1-\tau_D$ under budget $\epsilon_D$, it is sufficient that
\begin{equation}
\Gamma
\ge
\epsilon_D+\log\frac{1-\tau_D}{\tau_D}.
\label{eq:derived_threshold_guideline}
\end{equation}
To verify Lemma~\ref{lem:derived_lower_bound}, Fig.~\ref{fig:eq25_logodds_drift} empirically examines whether derived models preserve the source-induced target preference. We compute the weakening drift $\Delta_D^+$ over derived model--fingerprint pairs, summarize its ECDF in panel~(a), and plug representative budgets into the lower bound $\sigma(\Gamma-\epsilon_D)$ in panel~(b). The result shows that smaller budgets correspond to stronger preservation, and that increasing SCM gives a stronger lower bound on derived-model target-hit probability. Higher-coverage budgets are more conservative. Appendix~\ref{app:derived_validation_details} further summarizes these results, and the corresponding cross-model ECDF diagnostics are reported in Appendix~\ref{app:cross_model_theory_validation}.

\begin{tcfresultbox}{Takeaway}
\small
Derived models preserve enough of the source-induced target preference to support conditional transfer.
\end{tcfresultbox}

\subsection{Targeted Transferability of Independent Models}
\label{subsec:independent_constraints}

To operationalize the independent-model side of Assumption~\ref{ass:local_closeness}, we allow an explicit generic-transfer slack rather than assuming a fixed numerical distance. An independent model need not stay uniformly close
to $M_{0}$ at the perturbed input. The perturbation $u$ is optimized only
against $M_{0}$, so it follows $M_{0}$'s local gradient geometry; on a
model whose local geometry is misaligned with $M_{0}$'s, the same direction
produces only a weak, generic effect. For an independent model $M_I$ and a
candidate fingerprint $(x,u,t)$, define the generic-transfer slack
\begin{small}
\begin{equation}
\begin{aligned}
\Delta_I^+(M_I,x,u,t)
&=\max\Big\{0,\; L_{M_I}(p^{u}(x),t)\;\\
&\hspace{4.5em}-L_{M_0}(p^{0}(x),t)\Big\}.
\end{aligned}
\label{eq:independent_positive_slack}
\end{equation}
\end{small}
For any budget $\epsilon_I(M_I,x,u,t)$ satisfying
\begin{equation}
\epsilon_I(M_I,x,u,t)\ge\Delta_I^+(M_I,x,u,t),
\label{eq:independent_valid_budget}
\end{equation}
we have the deterministic inequality
\begin{equation}
L_{M_I}(p^{u}(x),t)
\le
L_{M_0}(p^{0}(x),t)+\epsilon_I(M_I,x,u,t).
\label{eq:independent_log_odds_upper}
\end{equation}
The slack explicitly permits accidental transfer and generic prompt effects.

\begin{figure}[ht]
    \centering
    \includegraphics[width=0.98\linewidth]{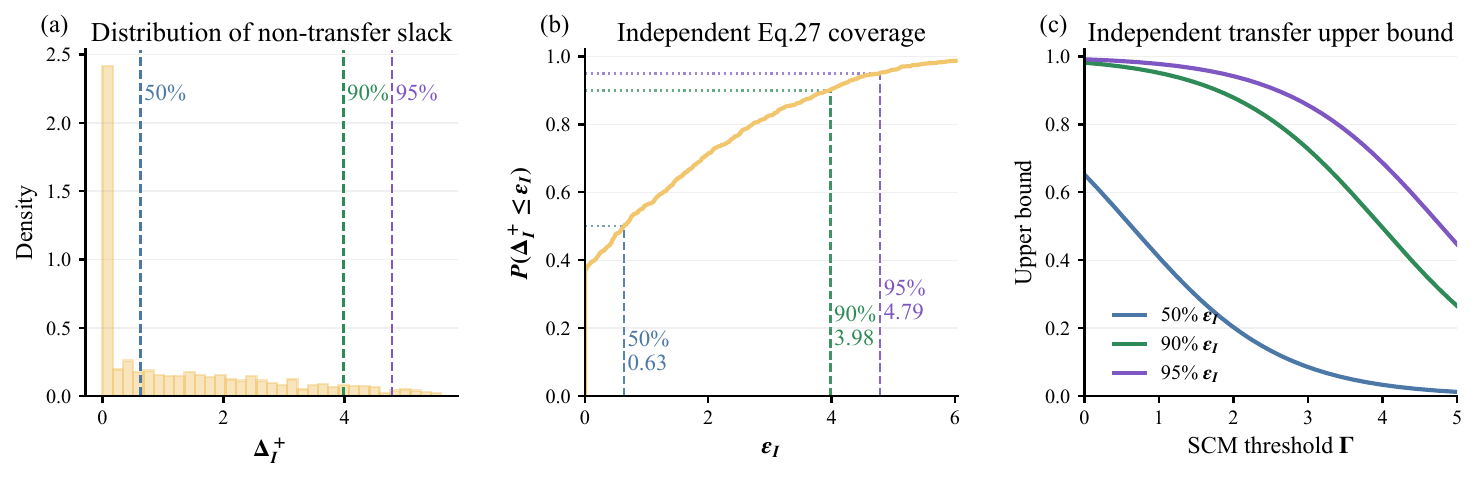}
    \caption{Validation of independent-model targeted transfer. (a) Distribution of the required generic-transfer slack $\Delta_I^+$. (b) Empirical coverage $\Pr(\Delta_I^+\le\epsilon_I)$ over independent model--fingerprint pairs. (c) Conditional independent-model transfer upper bound $\sigma(-\Gamma+\epsilon_I)$ under the marked empirical slack budgets.
    }
    \label{fig:eq31_nontransfer}
\end{figure}

\begin{lemma}[Conditional independent-model targeted-transfer upper bound]
\label{lem:independent_upper_bound}
Let $\Gamma=\Gamma(x,u,t)$. If
$\epsilon_I(M_I,x,u,t)\ge\Delta_I^+(M_I,x,u,t)$, then
\begin{equation}
q_{M_I}(t\mid p^{u}(x))
\le
\sigma\!\bigl(-\Gamma+\epsilon_I(M_I,x,u,t)\bigr).
\label{eq:independent_target_probability_upper_actual_gamma}
\end{equation}
In particular, if $\Gamma\ge\Gamma_{\min}$, then
\begin{equation}
q_{M_I}(t\mid p^{u}(x))
\le
\sigma\!\bigl(-\Gamma_{\min}+\epsilon_I(M_I,x,u,t)\bigr).
\label{eq:independent_target_probability_upper}
\end{equation}
\end{lemma}

For fixed $\epsilon_I$, increasing SCM tightens the independent upper
bound. If $\epsilon_I$ itself grows with SCM because stronger prefixes have
larger generic prompt effects, empirical independent accuracy may increase
even though the fixed-slack envelope decreases. This distinction is
important to understand SCM-bucket plots in Fig.~\ref{fig:scm_separability_bounds}. To empirically verify Lemma~\ref{lem:independent_upper_bound}, Fig.~\ref{fig:eq31_nontransfer} measures how much a source-optimized perturbation transfers to independent models. We compute the generic-transfer slack $\Delta_I^+$, show its distribution in panel~(a), convert it to an ECDF in panel~(b), and plug the marked budgets into the upper bound $\sigma(-\Gamma+\epsilon_I)$ in panel~(c). The result shows that larger slack budgets cover more independent model--fingerprint pairs, but also give looser non-transfer guarantees. For a fixed slack, increasing SCM tightens the independent-model upper envelope, although empirical independent accuracy can still rise when stronger prefixes create generic transfer. Appendix~\ref{app:independent_nontransfer_details} summarizes the diagnostic, and the corresponding cross-model non-transfer results are reported in Appendix~\ref{app:cross_model_theory_validation}.

\begin{tcfresultbox}{Takeaway}
\small
Independent-model transfer requires explicit generic-transfer slack, explaining residual false positives under over-optimization.
\end{tcfresultbox}

\subsection{SCM-Induced Separability of Target Accuracy}
\label{subsec:scm_separability}

Before aggregating fingerprints, consider the fixed-budget single-fingerprint
gap envelope
\begin{equation}
g(\Gamma;\epsilon_D,\epsilon_I)
=
\sigma(\Gamma-\epsilon_D)-\sigma(-\Gamma+\epsilon_I).
\label{eq:fixed_budget_gap_function}
\end{equation}
For fixed $\epsilon_D$ and $\epsilon_I$, larger SCM increases the fixed-budget gap envelope; Appendix~\ref{app:separability_details} gives the detailed fixed-budget interpretation and its empirical qualification.
For a fingerprint set
\begin{equation}
\mathcal{F}=\{(x_i,u_i,y_i,t_i)\}_{i=1}^{n},
\label{eq:fingerprint_set_theory}
\end{equation}
define the hard-label target-accuracy statistic of a suspect model $M$ as
\begin{equation}
\operatorname{TA}(M)
=
\frac{1}{n}\sum_{i=1}^{n}
\mathbb{I}\!
\left[
\operatorname{Parse}(R_M(p^{u_i}(x_i)))=t_i
\right].
\label{eq:fta_definition}
\end{equation}
This is exactly the statistic used by the online verifier.

\begin{theorem}[SCM yields a target-accuracy separation under explicit transfer budgets]
\label{thm:scm_gap}
Let $\Gamma_i=\Gamma(x_i,u_i,t_i)$. For a derived model $M_D$, choose
valid budgets $\epsilon_{D,i}\ge\Delta_D^+(M_D,x_i,u_i,t_i)$. For an
independent model $M_I$, choose valid budgets
$\epsilon_{I,i}\ge\Delta_I^+(M_I,x_i,u_i,t_i)$. Then
\begin{equation}
\mathbb{E}[\operatorname{TA}(M_D)]
\ge
\frac{1}{n}\sum_{i=1}^{n}\sigma(\Gamma_i-\epsilon_{D,i}),
\label{eq:derived_fta_lower}
\end{equation}
and
\begin{equation}
\mathbb{E}[\operatorname{TA}(M_I)]
\le
\frac{1}{n}\sum_{i=1}^{n}\sigma(-\Gamma_i+\epsilon_{I,i}).
\label{eq:independent_fta_upper}
\end{equation}
Consequently, the expected separability gap satisfies
\begin{equation}
\begin{aligned}
\mathcal{G}
&\triangleq
\mathbb{E}[\operatorname{TA}(M_D)]-
\mathbb{E}[\operatorname{TA}(M_I)] \\
&\ge
\frac{1}{n}\sum_{i=1}^{n}
\left[\sigma(\Gamma_i-\epsilon_{D,i})-
\sigma(-\Gamma_i+\epsilon_{I,i})\right].
\end{aligned}
\label{eq:fta_gap_lower}
\end{equation}
\end{theorem}

\begin{lemma}[Fixed-budget separability]
\label{lem:fixed_budget_separability}
If every fingerprint satisfies $\Gamma_i\ge\Gamma_{\min}$,
$\epsilon_{D,i}\le\epsilon_D$, and $\epsilon_{I,i}\le\epsilon_I$, then
\begin{small}
\begin{equation}
\begin{aligned}
\mathbb{E}[\operatorname{TA}(M_D)]&\ge \sigma(\Gamma_{\min}-\epsilon_D),\\
\mathbb{E}[\operatorname{TA}(M_I)]&\le \sigma(-\Gamma_{\min}+\epsilon_I).
\end{aligned}
\label{eq:fixed_budget_ta_bounds}
\end{equation}
\end{small}
The fixed-budget lower bound on the expected gap is positive whenever
\begin{equation}
2\Gamma_{\min}>\epsilon_D+\epsilon_I.
\label{eq:positive_gap_condition}
\end{equation}
\end{lemma}

By Lemma~\ref{lem:fixed_budget_separability}, the derived flip probability
is lower bounded while the independent flip probability is upper bounded,
so the operating point can be chosen inside the feasible band to obtain a
positive expected gap. Averaging $n$ diverse fingerprints then concentrates
$\operatorname{TA}$ around the two class-specific rates, converting a
per-fingerprint edge into a thresholdable separation.
\begin{figure}[htp]
    \centering
    \includegraphics[width=0.98\linewidth]{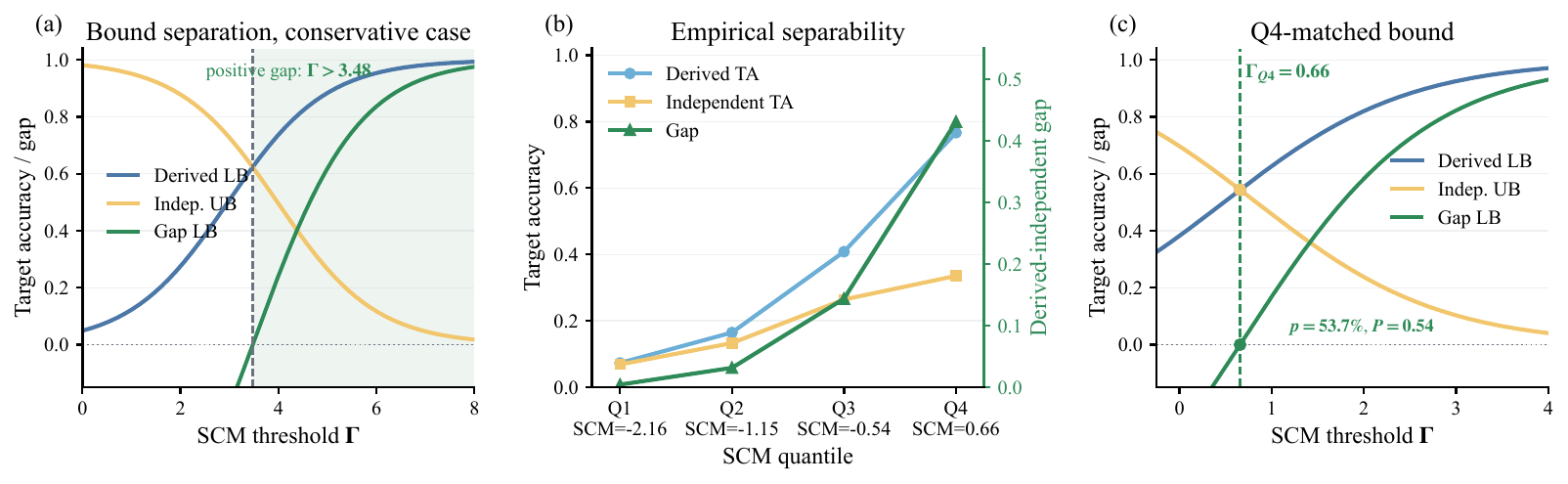}
    \caption{SCM-induced separability and the feasible margin band. (a) Conservative fixed-budget envelope using the $90\%$ empirical budgets $\epsilon_D\approx2.97$ and $\epsilon_I\approx3.98$; the resulting positive-gap condition $\Gamma>3.48$ is sufficient, not necessary. (b) Empirical SCM-quantile trend on Qwen3-1.7B Base: independent target accuracy may also rise in the highest-SCM buckets because a stronger perturbation manufactures generic transfer, so the derived--independent gap is maximized at an interior operating margin (the band sweet spot) rather than by pushing SCM arbitrarily high. (c) Q4-matched fixed-slack envelope.}
    \label{fig:scm_separability_bounds}
\end{figure}
To verify Lemma~\ref{lem:fixed_budget_separability}, Fig.~\ref{fig:scm_separability_bounds}  empirically validates whether the derived and independent bounds translate into an observable target-accuracy gap. Panel~(a) combines the derived lower bound and independent upper bound using the $90\%$ empirical budgets, panel~(b) compares this with the empirical SCM-bucket trend, and panel~(c) repeats the bound calculation with high-SCM matched budgets. The result supports the feasible-band interpretation: higher-SCM fingerprints generally increase the derived--independent gap, but the highest-SCM bucket can also increase independent target accuracy when the perturbation becomes too generic. Thus, SCM should be strong enough to survive derived-model weakening, but not pushed indefinitely. Appendix~\ref{app:empirical_separability_details} gives the detailed empirical interpretation, and model-wise SCM separability results are reported in Appendix~\ref{app:cross_model_theory_validation}.

\begin{tcfresultbox}{Takeaway}
\small
SCM is most useful inside a feasible margin band, supporting early stopping rather than maximal optimization.
\end{tcfresultbox}

\subsection{Threshold Selection and Verification Reliability}
\label{subsec:threshold_and_reliability}

The previous subsection gives expected target-accuracy bounds. Define
\begin{equation}
\begin{aligned}
\alpha_D&=\frac{1}{n}\sum_{i=1}^{n}\sigma(\Gamma_i-\epsilon_{D,i}),\\
\alpha_I&=\frac{1}{n}\sum_{i=1}^{n}\sigma(-\Gamma_i+\epsilon_{I,i}).
\end{aligned}
\label{eq:alpha_bounds}
\end{equation}
Here, $\alpha_D$ is a lower bound on the expected target accuracy of a
derived model under the chosen budgets, and $\alpha_I$ is an upper bound
for an independent model under its chosen slacks. If $\alpha_D>\alpha_I$,
then any threshold in $(\alpha_I,\alpha_D)$ separates the two expectation
bounds. A natural threshold is
\begin{equation}
\zeta^{\star}=\frac{\alpha_D+\alpha_I}{2}.
\label{eq:midpoint_threshold}
\end{equation}
In practice, the true budgets of an unknown suspect model are unavailable,
so Eq.~\ref{eq:midpoint_threshold} should be viewed as a theoretical
existence statement. Operational thresholds should be pre-registered from
source-side construction rules and, when available, calibrated on a
held-out validation pool of non-derived models rather than tuned on the
test suspect pool.

Under the fixed-budget interpretation in Lemma~\ref{lem:fixed_budget_separability},
an operational sufficient condition for derived hit probability at least
$1-\tau_D$ and independent hit probability at most $\tau_I$ is
\begin{equation}
\Gamma_{\min}
\ge
\max\left\{
\epsilon_D+\log\frac{1-\tau_D}{\tau_D},\thinspace
\epsilon_I+\log\frac{1-\tau_I}{\tau_I}
\right\}.
\label{eq:combined_threshold_guideline}
\end{equation}
This formula exposes the feasibility trade-off: stronger reliability
requirements or larger transfer budgets require larger SCM, which can
reduce the retained fingerprint pool.

\begin{lemma}[Finite-sample reliability under independent fingerprint indicators]
\label{lem:finite_sample_reliability}
Let $\mu_M=\mathbb{E}[\operatorname{TA}(M)]$. Suppose the hard-label
indicators in Eq.~\ref{eq:fta_definition} are independent conditional on
the fixed fingerprint set. If a derived model satisfies $\mu_{M_D}\ge
\alpha_D$ and an independent model satisfies $\mu_{M_I}\le\alpha_I$ with
$\alpha_D>\alpha_I$, then under the midpoint threshold in
Eq.~\ref{eq:midpoint_threshold},
\begin{equation}
\Pr[\operatorname{TA}(M_D)<\zeta^{\star}]
\le
\exp\left(-\frac{n(\alpha_D-\alpha_I)^2}{2}\right),
\label{eq:false_negative_bound}
\end{equation}
and
\begin{equation}
\Pr[\operatorname{TA}(M_I)\ge\zeta^{\star}]
\le
\exp\left(-\frac{n(\alpha_D-\alpha_I)^2}{2}\right).
\label{eq:false_positive_bound}
\end{equation}
\end{lemma}

Both the verifier and the construction filter read a \emph{hard} label, so
the final quantity that matters is whether the sampled answer equals $t$.
The target margin of Eq.~\ref{eq:target_margin} controls this directly in an
idealized constrained decoder. Specifically, under temperature-$T$ softmax
sampling normalized over only the $K$ legal labels in $\mathcal{A}$ (or
conditional on generation producing a valid legal label),
\begin{equation}
\Pr[\text{sample}=t\mid p]
\ge
\frac{1}{1+(K-1)\,e^{-m_{M}(p,t)/T}}.
\label{eq:margin_sampling_robustness}
\end{equation}
Thus a large source margin $m_{M_0}(p^{u},t)$ strengthens target-label
stability under this constrained-label idealization. Eq.~\ref{eq:margin_sampling_robustness}
does not assert that an arbitrary raw-text API samples only from the legal
labels; the actual API-facing guarantee remains empirical and is checked by
the repeated-query stability filter of Sec.~\ref{sec:method}. Appendix~\ref{app:margin_sampling_details}
gives the detailed qualification and margin--sampling argument.

\section{Our Method}
\label{sec:method}

\begin{figure*}
    \centering
    \includegraphics[width=0.98\linewidth]{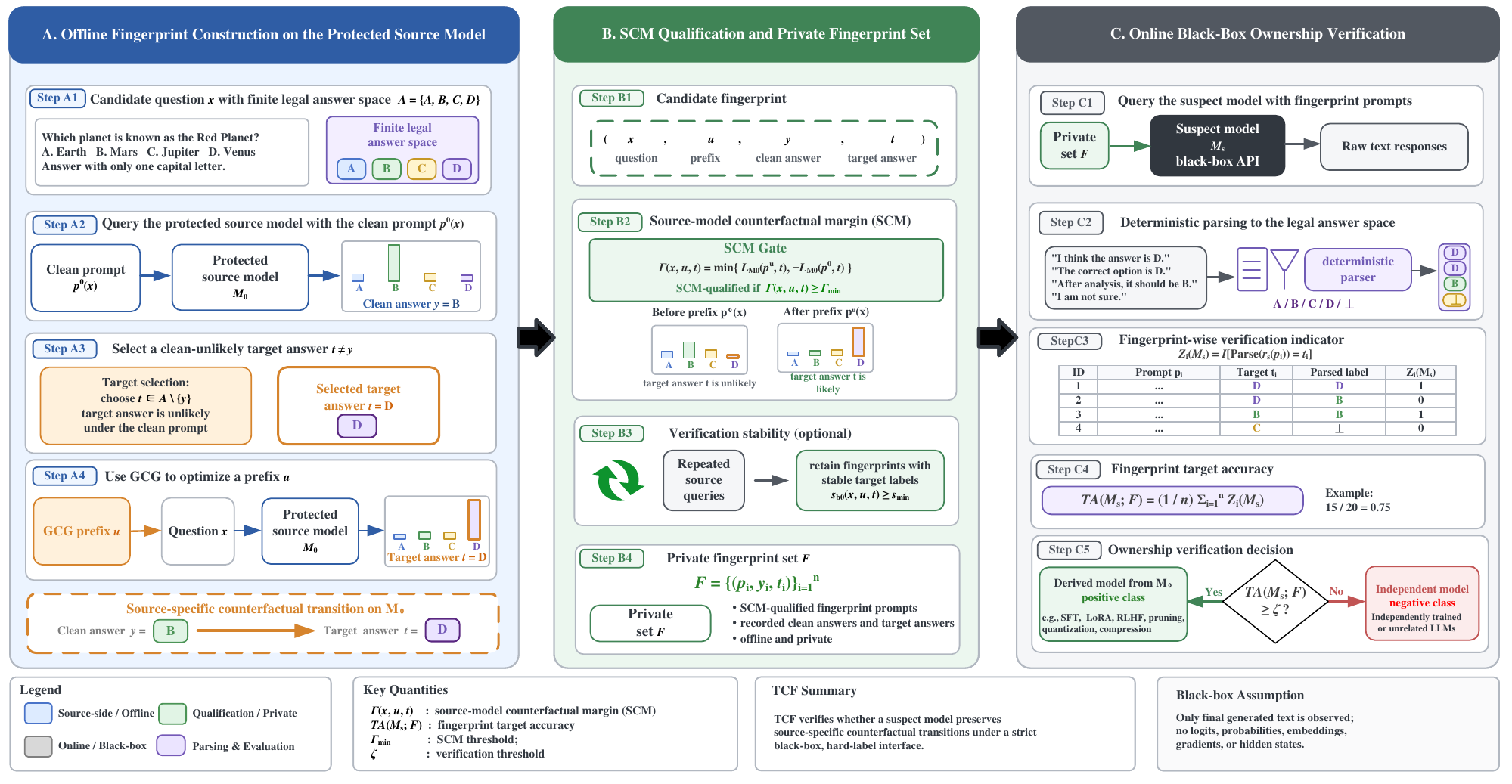}
    \caption{Overview of Targeted Counterfactual Fingerprinting (TCF). Panel~A illustrates fingerprint generation on the protected source model, Panel~B illustrates SCM qualification and fingerprint-set construction, and Panel~C illustrates black-box ownership verification.}
    \label{fig:tcf_overview}
\end{figure*}

As shown in Fig.~\ref{fig:tcf_overview}, we propose
\emph{\textbf{T}argeted \textbf{C}ounterfactual \textbf{F}ingerprinting} (TCF),
a black-box LLM ownership verification method based on SCM.
The owner constructs
prompts that strongly change the protected source LLM's answer from a
clean answer to a selected target answer. During verification, the owner
queries a suspect model through its generated text and parses the
output into a finite constrained answer space. A model is considered likely
derived if it preserves a high fraction of these source-specific target
answer transitions.

\subsection{Target Selection}
\label{subsec:target_selection}

As shown in Fig.~\ref{fig:tcf_overview}(A1--A3), target selection starts from a finite-answer candidate question, obtains the source clean answer, and then chooses a clean-unlikely target.
For each candidate question $x$, we first evaluate the clean source
prompt $p^{0}(x)$. The source clean answer is
\begin{equation}
y(x)=\arg\max_{a\in\mathcal{A}}q_{M_0}(a\mid p^{0}(x)).
\label{eq:method_clean_answer}
\end{equation}
The target answer should be different from $y(x)$ and should be unlikely
under the clean source prompt. We therefore select targets from
\begin{equation}
\mathcal{T}(x)
=
\left\{
 a\in\mathcal{A}\setminus\{y(x)\}:
 -L_{M_0}(p^{0}(x),a)\ge\Gamma_{\min}
\right\}.
\label{eq:method_candidate_targets}
\end{equation}
If $\mathcal{T}(x)$ is nonempty, we choose the unlikely valid target,
\begin{equation}
t(x)=\arg\min_{a\in\mathcal{T}(x)}q_{M_0}(a\mid p^{0}(x)).
\label{eq:method_target_choice}
\end{equation}
If $\mathcal{T}(x)$ is empty, the question is discarded because no target answer is sufficiently counterfactual on the source model.

This target-selection rule controls the clean side of SCM before the
perturbation is optimized. After this step, the remaining task is only to
control the perturbed side, namely to make $L_{M_0}(p^{u}(x),t)$ reach the
construction target $\Gamma_{\mathrm{target}}$.

\subsection{Perturbation toward a Target Margin}
\label{subsec:scm_guided_gcg}

As shown in Fig.~\ref{fig:tcf_overview}(A4), given a selected target $t$, we optimize a discrete input perturbation
(prefix) $u$ with GCG~\cite{zou2023universal}. For exposition, we write the objective using the source constrained-answer probability $q_{M_0}$; the differentiable source-side surrogate used during search and the parser-side acceptance check are detailed in Appendix~\ref{app:surrogate_details}. The objective is the target negative log-likelihood
\begin{equation}
\mathcal{L}_{\mathrm{src}}(u;x,t)
=
-\log q_{M_0}(t\mid p^{u}(x)).
\label{eq:method_gcg_loss}
\end{equation}

By Lemma~\ref{lem:loss_threshold_for_scm}, the construction target gives an
explicit stopping criterion. Since
$L_{M_0}(p^{u}(x),t)\ge\Gamma_{\mathrm{target}}$ is equivalent to
$q_{M_0}(t\mid p^{u}(x))\ge\sigma(\Gamma_{\mathrm{target}})$, it is
sufficient to optimize until
\begin{equation}
\mathcal{L}_{\mathrm{src}}(u;x,t)
\le
\log\left(1+e^{-\Gamma_{\mathrm{target}}}\right),
\label{eq:method_gcg_stop_loss}
\end{equation}
and then \emph{stop early}.
We set $\Gamma_{\mathrm{target}}=\Gamma_{\min}$, where the source-side SCM threshold is fixed before verification and is chosen only from source-side feasibility.
We use early stopping to avoid unnecessary over-optimization and to retain
the first source-qualified perturbation rather than blindly using the final
GCG step. This choice is empirically supported by Fig.~\ref{fig:scm_separability_bounds}:
as target loss decreases, SCM increases, but the strongest empirical SCM
regime can also increase generic transfer to independent models. We therefore
do not treat ``larger SCM'' as an unbounded optimization objective. In
practice, Eq.~\ref{eq:method_gcg_stop_loss} is the source-side stopping
condition, and the retained candidate is accepted only after its parser-side
SCM and source stability are recomputed.

\textbf{Prefix length.}
The prefix length $\ell_{u}$ controls the search space. A longer prefix
usually improves the chance of satisfying
$\Gamma(x,u,t)\ge\Gamma_{\mathrm{target}}$
but may reduce naturalness and increase optimization cost.

\textbf{Optimization budget.}
Let $T_{\max}$ be the maximum number of GCG iterations and $B$ be the
number of candidate token replacements evaluated per iteration. Larger
values improve success rate but increase construction cost.

\textbf{Verification stability.}
Because LLM outputs can vary across repeated queries under stochastic decoding, we optionally query the source
model $M_{0}$ for $R_{0}$ repeated trials with the selected prompt and compute

\begin{equation}
\widehat{s}_{0}(x,u,t)
=
\frac{1}{R_{0}}
\sum_{r=1}^{R_{0}}
\mathbb{I}
\left[
\operatorname{Parse}(R^{(r)}_{M_{0}}(p^{u}(x)))=t
\right].
\label{eq:method_source_stability}
\end{equation}
A candidate is retained only if
\begin{equation}
\widehat{s}_{0}(x,u,t)
\ge s_{\min}.
\label{eq:method_stability_threshold}
\end{equation}
We recommend $s_{\min}\in[0.8,0.95]$. This practical filter is motivated by the margin--sampling analysis of Sec.~\ref{subsec:threshold_and_reliability}: under the constrained-label idealization, a larger source target margin favors target-label stability. The repeated-query filter then checks the actual counterfactual stability under the decoding setting used by the source interface.

To reduce response-format variation at the suspect interface, we append a fixed answer instruction to the GCG-optimized prompt and restrict valid parsed outputs to the legal answer space. Thus, the verification prompt $p_i$ in Eq.~\ref{eq:fingerprint_set_threat} is
\begin{equation}
    p_i = s_i \oplus q_i \oplus \tau .
\end{equation}
Here, $s_i=u_i^{\star}$ is the selected GCG prefix, $q_i$ is the
underlying multiple-choice question, and $\tau$ denotes the fixed answer
instruction template that asks $M_s$ to output one capital letter from
$\{A,B,C,D\}$:
\begin{quote}
\small
\texttt{Question ... A. ... B. ... C. ... D...  Answer with only one capital letter: A, B, C, or D. Answer:}
\end{quote}

Algorithm~\ref{alg:scm_gcg_construction} in Appendix~\ref{sec:implementation_details} gives the complete offline construction procedure.
To maintain a practically effective fingerprint set, we additionally enforce diversity over questions, source clean answers, and target answers during selection.

\subsection{Online Black-Box Ownership Verification}
\label{subsec:online_verification}

As shown in Fig.~\ref{fig:tcf_overview}(C1--C5), after constructing the fingerprint set, the owner performs online black-box ownership verification for a suspect model $M_s$ by using only black-box queries as described in Sec.~\ref{sec:Verification_Objective}.

\textbf{Verification threshold selection.}  $\zeta$ is a threshold chosen to control verification errors.
Each verification process requires $n$ queries to $M_s$, followed by a single pass to compute TA.
Since the fingerprint pool $\mathcal{F}$ is generated offline and can be reused, multiple independent verification instances can be performed by sampling different subsets $\mathcal{Q}$, enabling repeated verification while preserving query efficiency.

\textbf{Number of fingerprints.}
The number of fingerprints $n$ controls statistical reliability. A
larger $n$ reduces variance but increases query cost. In the main evaluation, we use 20 fingerprints per protected source, matching the fixed registry described in Sec.~\ref{sec:experiments} and Appendix~\ref{app:main_method_setup}.

\section{Experiments}
\label{sec:experiments}
\newcommand{\expitem}[1]{\par\smallskip\noindent\textbf{#1.}\ }

We organize the experimental evaluation by the following five questions.
First, does target-answer transfer separate derived models from independent models?
Second, how large is the gap between \method{} and existing black-box LLM fingerprinting baselines?
Third, is the signal stable under decoding changes and simple adaptive perturbations? Fourth, how robust is \method{} to adaptive attack?
Finally, what is the computational cost compared with the evaluated baselines?

\subsection{Experimental Setup}
\label{sec:exp_setup}

\expitem{Source models} We evaluate \textbf{TCF} on four representative open-source LLMs as protected source models: Qwen3-1.7B Instruct\cite{yang2025qwen3}, Qwen3-1.7B Base\cite{yang2025qwen3}, Llama3-8B Base Direct\cite{llama3modelcard, grattafiori2024llama}, and Mistral-7B-v0.3\cite{mistral7bv03}. The public checkpoint identifiers on Hugging Face are listed in Table~\ref{tab:source_pool_config}.

\begin{table}[htp]
\centering
\caption{Protected source models and suspect-model pools.}
\label{tab:source_pool_config}
\scriptsize
\resizebox{\columnwidth}{!}{%
\begin{tabular}{llcccc}
\toprule
\rowcolor{HeaderGray}
Source & Checkpoint identifier & Total & Source & Derived & Independent \\
\midrule
Qwen3-1.7B Instruct & \texttt{Qwen/Qwen3-1.7B} & 19 & 1 & 9 & 9 \\
Qwen3-1.7B Base & \texttt{Qwen/Qwen3-1.7B-Base} & 22 & 1 & 12 & 9 \\
Llama3-8B Base Direct & \texttt{meta-llama/Meta-Llama-3-8B} & 25 & 1 & 13 & 11 \\
Mistral-7B-v0.3 & \texttt{mistralai/Mistral-7B-v0.3} & 25 & 1 & 16 & 8 \\
\bottomrule
\end{tabular}%
}
\end{table}

\expitem{Suspect model pool} For each protected source, we evaluate the source checkpoint, direct model positives, and independent negatives. Direct model positives cover supervised/task fine-tuning (SFT), RL-style adaptation (RL), language/domain adaptation, pruning, and other direct derivatives. Independent checkpoints are used only as independent negatives.
Table~\ref{tab:source_pool_config} lists the fixed source/derived/independent model counts used for the main AUC in Table~\ref{tab:main_auc_full}. Appendix Tab.~\ref{tab:app_full_model_pool} provides the extended checkpoint catalog, and Appendix~\ref{app:full_per_suspect_results} reports per-suspect diagnostics for checkpoints with the corresponding baseline artifacts.

\expitem{Fingerprint Generation} The fingerprint pool is built from Massive Multitask Language Understanding (MMLU) multiple-choice questions \cite{hendrycks2021measuring}. Each source model uses 20 fingerprints from four different subject categories, with five questions per category.
The four-domain design prevents the fingerprint set from being tied to a single topic and improves fingerprint diversity. The details are in Appendix~\ref{app:experimental_design}.

\expitem{Baselines} We compare with three state-of-the-art black-box LLM fingerprinting baselines: TRAP \cite{gubri2024trap}, ProFLingo \cite{jin2024proflingo}, and ZeroPrint \cite{shao2026zeroprint}. Appendix Tab.~\ref{tab:baseline_protocols} specifies the verification protocol and score definition for each method.

\begin{figure*}[!t]
    \centering
    \includegraphics[width=0.98\textwidth]{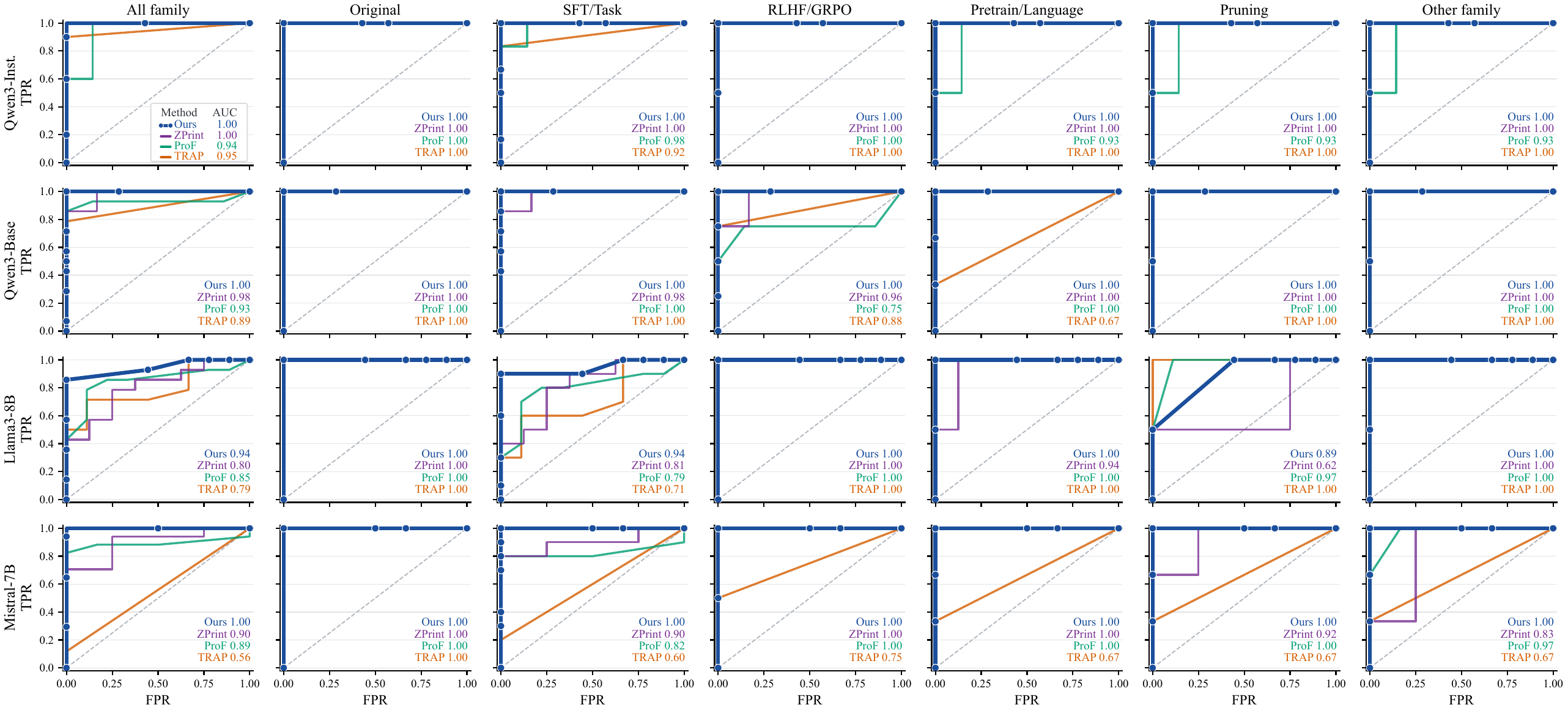}
    \caption{Diagnostic subgroup ROC curves on the shared per-suspect diagnostic pool. Primary family-level AUCs are reported in Table~\ref{tab:main_auc_full}.}
    \label{fig:roc_full}
\end{figure*}

\begin{figure*}[!t]
    \centering
    \includegraphics[width=0.98\textwidth]{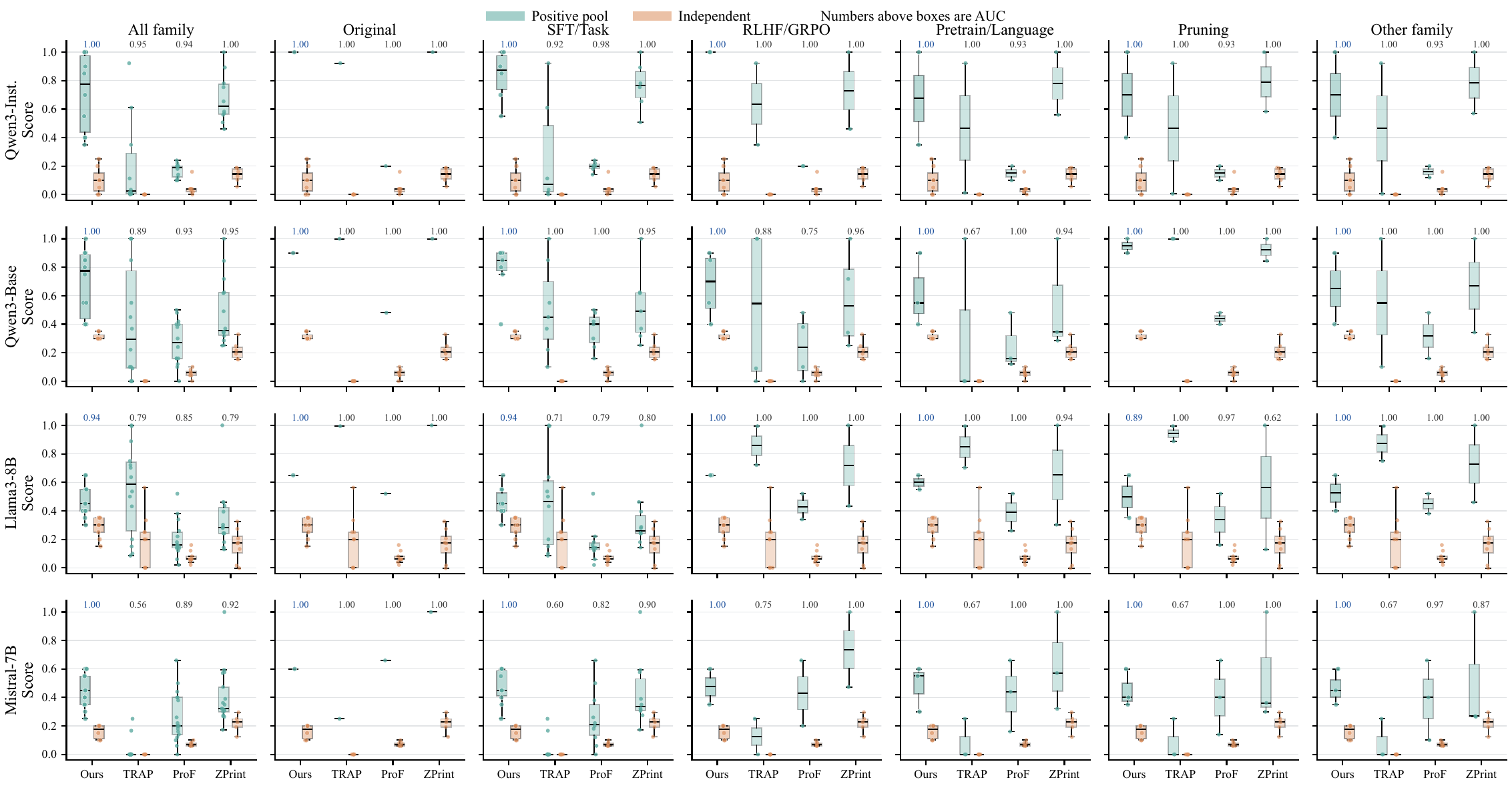}
    \caption{Fingerprint score distributions on the shared per-suspect diagnostic pool used for the subgroup diagnostics. Primary family-level AUCs are reported in Table~\ref{tab:main_auc_full}.}
    \label{fig:dist_full}
\end{figure*}

\subsection{Main Verification Results}
\label{sec:main_results}

Table~\ref{tab:main_auc_full} reports the main verification AUC results. \textbf{\method{} achieves perfect AUC on Qwen3-1.7B Instruct, Qwen3-1.7B Base, and Mistral-7B-v0.3, and reaches 0.9444 on Llama3-8B Base Direct.} Its average verification AUC across the four sources is 0.9861, compared with 0.7931 for TRAP, 0.8880 for ProFLingo, and 0.9128 for ZeroPrint. The strongest baseline is model-dependent: ZeroPrint is the strongest baseline on both Qwen3 sources and on Mistral, whereas ProFLingo is strongest on Llama3. Overall, \method{} matches the strongest baseline on Qwen3-1.7B Instruct and strictly exceeds the strongest baseline on the other three source families, supporting our key intuition: \textbf{a constrained-answer targeted counterfactual transfer signal, realized via a source-optimized input perturbation, can act as a strong black-box proxy for LLM ownership verification.}

\begin{table}[H]
\centering
\caption{Main AUC comparison. Best results are bold.}
\label{tab:main_auc_full}
\scriptsize
\resizebox{\columnwidth}{!}{%
\begin{tabular}{lcccc}
\toprule
\rowcolor{HeaderGray}
Source model & \method{} & TRAP & ProFLingo & ZeroPrint \\
\midrule
Qwen3-1.7B Instruct  & \best{1.0000} & 0.9286 & 0.9796 & \textbf{1.0000} \\
Qwen3-1.7B Base  & \best{1.0000} & 0.8846 & 0.9231 & 0.9722 \\
Llama3-8B Base Direct  & \best{0.9444} & 0.7593 & 0.8241 & 0.7885 \\
Mistral-7B-v0.3  & \best{1.0000} & 0.6000 & 0.8250 & 0.8906 \\
\midrule
Average  & \best{0.9861} & 0.7931 & 0.8880 & 0.9128 \\
\bottomrule
\end{tabular}%
}
\end{table}

\begin{tcfresultbox}{Takeaway}
\small
\method{} matches or exceeds the strongest evaluated baseline across the four source families while maintaining strong evidence across diverse derivative categories.
\end{tcfresultbox}

\expitem{Fingerprint score} For the per-suspect diagnostic pool with available artifacts, Table~\ref{tab:score_summary} reports the fingerprint scores of source models, derived models, and independent models for \method{}. Figure~\ref{fig:roc_full} shows the corresponding ROC curves over the source model, all available derivatives, and derivative subgroups, while Fig.~\ref{fig:dist_full} shows the score distributions. These diagnostic plots expose model-wise and subgroup heterogeneity; the primary family-level comparison remains Table~\ref{tab:main_auc_full}.

\begin{table}[H]
\centering
\caption{Mean \method{} fingerprint score on the per-suspect diagnostic pool. The score is the target-label accuracy over generated fingerprints.}
\label{tab:score_summary}
\scriptsize
\resizebox{\columnwidth}{!}{%
\begin{tabular}{lccc}
\toprule
\rowcolor{HeaderGray}
Source model  & Source score & Derived mean & Indep. mean \\
\midrule
Qwen3-1.7B Instruct & 1.0000 & 0.6833 & 0.1000 \\
Qwen3-1.7B Base & 0.9000 & 0.6731 & 0.3143 \\
Llama3-8B Base Direct  & 0.6500 & 0.4538 & 0.2889 \\
Mistral-7B-v0.3 & 0.6000 & 0.4281 & 0.1583 \\
\bottomrule
\end{tabular}%
}
\end{table}

\begin{tcfresultbox}{Takeaway}
\small
Within this diagnostic pool, derived models retain the source fingerprint more strongly than independent models, while Llama3 remains the most challenging source family.
\end{tcfresultbox}

\begin{table}[htp]
\centering
\small
\setlength{\tabcolsep}{4pt}
\caption{Commercial black-box API source-vs-independent verification results on OpenRouter.}
\label{tab:openrouter_blackbox_api}
\resizebox{\linewidth}{!}{%
\begin{tabular}{lccccc}
\toprule
Method & Source & Qwen3-8B & Gemma-3-12B & Gemma-3-4B & AUC \tabularnewline
\midrule
Ours & 0.294 & 0.000 & 0.118 & 0.063 & 1.000 \tabularnewline
TRAP & 0.450 & 0.000 & 0.000 & 0.000 & 1.000 \tabularnewline
ProFLingo & 0.100 & 0.100 & 0.000 & 0.000 & 0.833 \tabularnewline
ZeroPrint & 0.483 & 0.490 & 0.538 & 0.476 & 0.333 \tabularnewline
\bottomrule
\end{tabular}%
}
\end{table}

\expitem{Commercial black-box API evaluation}
We further evaluate whether the fingerprints remain separable when model access is mediated by a mainstream commercial black-box API. We use \textbf{OpenRouter}~\cite{openrouter_api} as the API provider and query \texttt{meta-llama/llama-3-8b-instruct} as the protected source model.
Since OpenRouter does not provide the same controlled set of
Llama-3 family derivatives used in our evaluation, this experiment focuses on the source-versus-independent API test.
The independent API models are
\texttt{qwen/qwen3-8b}~\cite{yang2025qwen3},
\texttt{google/gemma-3-12b-it}~\cite{gemma3technicalreport}, and
\texttt{google/gemma-3-4b-it}~\cite{gemma3technicalreport}.
For TRAP, ProFLingo, and ZeroPrint, we use each baseline's original black-box
scoring setting.
Although both \method{} and TRAP achieve the best AUC in this commercial black-box API test, the results are not clearly equivalent: the API test obscures TRAP’s weak derivative model transfer, whereas our method retains stronger family-level evidence under controlled evaluation.

\begin{tcfresultbox}{Takeaway}
\small
\method{} remains separable through a commercial text-generation API using only parsed final responses, showing that suspect-side verification does not depend on white-box access.
\end{tcfresultbox}

\expitem{Model distillation}
Model extraction is a strong threat for ownership verification because an
adversary may train a surrogate from query-output pairs rather than directly
reusing the source checkpoint~\cite{carlini2024stealing}.  We therefore
evaluate a clean-query distillation setting from
\texttt{mistralai/Mistral-7B-v0.3} to a \texttt{Llama-2-7B} student~\cite{touvron2023llama}.  The
attacker queries the teacher only with clean MMLU-style multiple-choice prompts
and trains the student with LoRA SFT on the teacher's hard-label outputs.
The reported extracted student is trained for three epochs on 2472 valid clean teacher-label pairs.

For utility, we evaluate on \texttt{MMLU-CF}~\cite{zhao2025mmlucf}, using the same hard-label generation prompt, \emph{Answer with only one capital letter:
A, B, C, or D.}, deterministic decoding, and at most eight new tokens.  We
parse each completion into A/B/C/D/INVALID.  Clean100 Acc. is the parsed multiple-choice accuracy, and T-Agree is the parsed agreement with the
teacher prediction; the teacher's own T-Agree is defined as 1.00.  As shown in
Table~\ref{tab:mistral_to_llama2_clean_extraction}, distillation improves the
student's Clean100 accuracy from 0.53 to 0.69 and T-Agree from 0.49 to 0.68,
showing that the surrogate absorbs useful teacher behavior from clean queries.
Although extraction attacks can erase fine-grained watermark evidence, our
fingerprint remains effective on the distilled student: its AUC increases from
0.31 before distillation to 0.88 after distillation.  In contrast, TRAP remains
at tie-level AUC, ProFLingo does not retain a transferable signal, and
ZeroPrint decreases after distillation.  Thus, \method{} retains an effective ownership signal under clean-query model
distillation.

\begin{table}[htp]
\centering
\scriptsize
\setlength{\tabcolsep}{2.5pt}
\renewcommand{\arraystretch}{1.05}
\caption{
Model distillation verification results.
}
\label{tab:mistral_to_llama2_clean_extraction}
\begin{tabular}{lcccccc}
\toprule
Model & Clean100 Acc. & T-Agree & Ours & TRAP & ProF & ZPrint \\
\midrule
Mistral teacher   & 0.79 & 1.00 & 1.00 & 1.00 & 1.00 & 1.00 \\
Llama-2 untrained & 0.53 & 0.49 & 0.31 & 0.50 & 0.06 & 0.56 \\
Llama-2 extracted & 0.69 & 0.68 & 0.88 & 0.50 & 0.00 & 0.25 \\
\bottomrule
\end{tabular}
\end{table}

\begin{tcfresultbox}{Takeaway}
\small
The fingerprint is inherited by a student trained only from clean teacher outputs, supporting ownership verification under model extraction.
\end{tcfresultbox}

\subsection{Robustness}
\label{sec:robustness_diagnostics}

\expitem{Input abnormal detection}
The original GCG prefix is effective but can contain unnatural token fragments, which can be exposed by PPL or window-PPL-based input-abnormality detection. We therefore evaluate naturalness-oriented presentation variants while keeping source-only construction and the hard-label verifier unchanged. NaturalCarrier wraps optimized fragments with short cue-like carrier phrases, while TCF-S4 and TCF-S8 distribute the optimized positions over four or eight insertion groups in the original question. The details and diagnostic-only variants are documented in Appendix~\ref{app:naturalness_implementation_details}.

\begin{figure}[htp]
    \centering
    \includegraphics[width=.98\columnwidth]{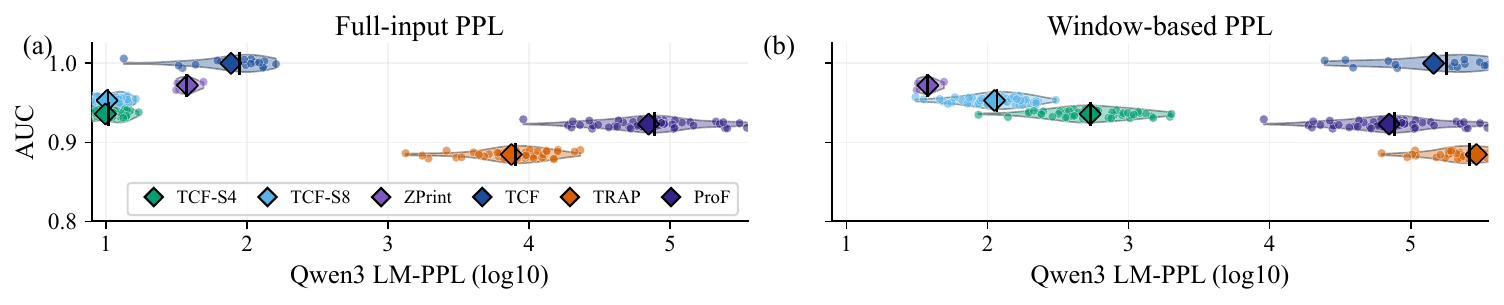}
    \caption{Naturalness--verification trade-off. Full-input and window-based PPL show complementary views of prompt naturalization.}
    \label{fig:naturalness_tradeoff}
\end{figure}

We then compute and visualize full-input PPL and window-based PPL of the original centralized prefix \method{}, and its variants with the above naturalness-oriented strategies, and the compared baselines. As shown in Fig.~\ref{fig:naturalness_tradeoff}, among these strategies and the evaluated baselines, the strategy of splitting the centralized prefix into randomly selected token positions TCF-S4/8 yields the lowest PPL while maintaining the highest verification AUC compared with the strongest normal-text-based baseline, ZeroPrint.

\begin{tcfresultbox}{Takeaway}
\small
Distributed NaturalCarrier variants improve prompt naturalness while preserving strong verification performance.
\end{tcfresultbox}

\begin{figure}[htp]
\centering
\includegraphics[width=0.97\linewidth]{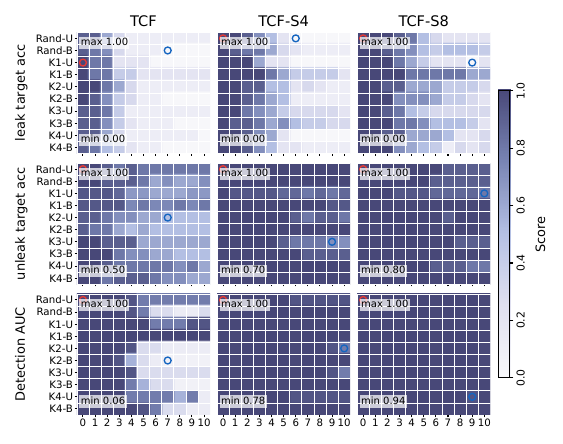}
\caption{Verification results under adaptive attack. Rows report leaked-fingerprint target accuracy, non-leaked target accuracy, and detection AUC. Columns correspond to fingerprinting methods. The horizontal axis indexes LoRA finetuning epochs, and the vertical axis indexes leakage conditions. Red and blue circles mark the maximum and minimum values in each panel, respectively.
}
\label{fig:lora_target_hit_split}
\end{figure}

\begin{figure*}[!t]
    \centering
    \includegraphics[width=0.98\linewidth]{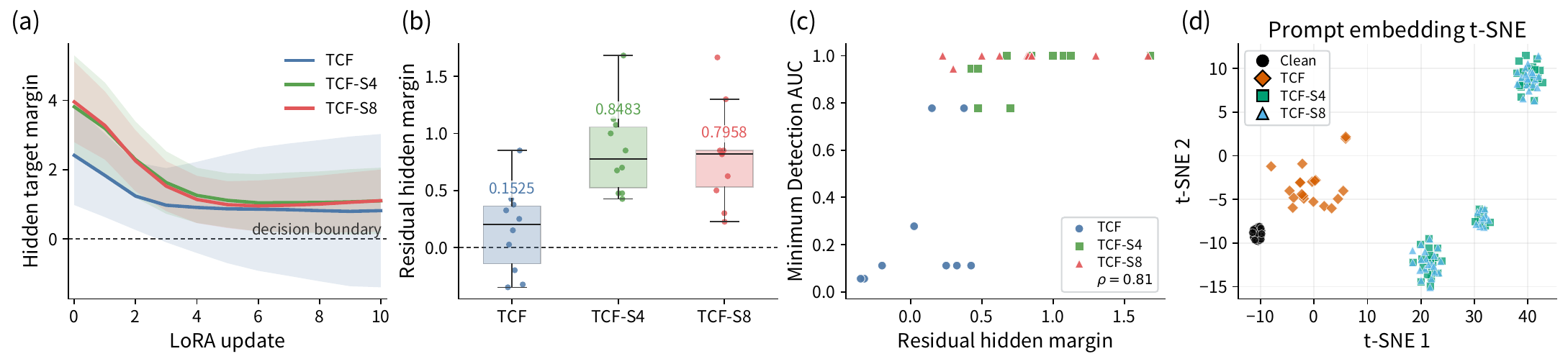}
    \caption{Margin-boundary analysis of adaptive LoRA erasure: hidden margin trajectories (a), residual hidden margin distributions (b), residual hidden margin versus minimum Detection AUC (c), and prompt-embedding t-SNE visualization (d).}
    \label{fig:margin_boundary_mechanism}
\end{figure*}

\begin{figure*}[t]
    \centering
    \includegraphics[width=0.98\textwidth]{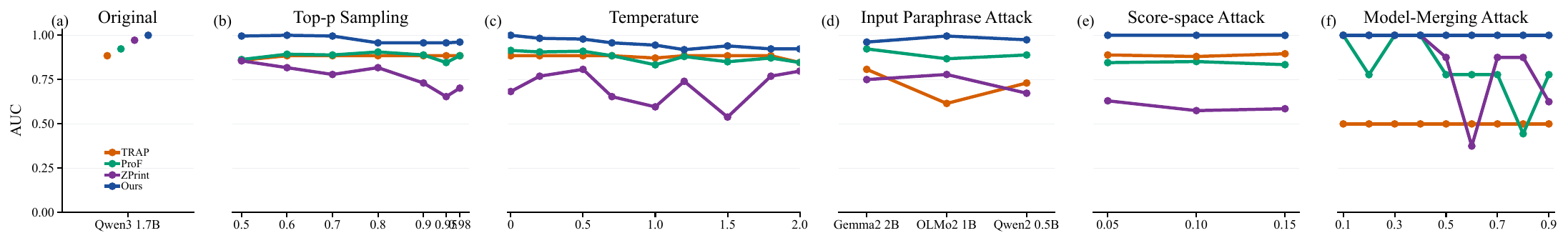}
    \caption{Robustness under sampling, input paraphrasing, score-space perturbation, and model-merging.}
    \label{fig:robustness_ablation_diagnostics}
\end{figure*}

\expitem{Adaptive attack}
To assess whether a lightweight LoRA adapter can erase fingerprint behavior when partial fingerprint samples are leaked ($50\%$ by default in our implementation) while preserving clean-answer utility, we conduct a LoRA adaptive attack. The attacker freezes the base model and trains only a low-rank LoRA adapter on leaked fingerprint examples mixed with normal clean-answer data.
For the condition labels, Rand denotes a random leaked subset, numbers in K1--K4 denote the number of leaked subjects, where U denotes an unbiased adaptive finetuning by mixing the leaked fingerprints and normal samples (questions and answers sampled from MMLU) in a 1:1 ratio, and B denotes a biased 1:10 leaked-to-normal ratio. In each finetuning epoch, the constructed training batch is fed once to update the adapter.
From the results in Fig.~\ref{fig:lora_target_hit_split}, we found that such an adaptive finetuning attack can strongly suppress the target fingerprint accuracy of the leaked fingerprints, leading to the leaked target accuracy gradually dropping to 0.00 under enough training epochs. However, this suppression does not reliably transfer to the unleaked fingerprints. For \method{}, unleaked target accuracy remains nontrivial across the sweep. NaturalCarrier is more robust: TCF-S4 keeps unleaked target accuracy at least 0.70, and TCF-S8 keeps it at least 0.80. This indicates a clear split effect: increasing the number of NaturalCarrier splits makes the fingerprint more resistant to adaptive finetuning attack, because the attacker observes and erases only part of the distributed fingerprint signal.
Overall, \emph{while adaptive attack by LoRA erasure can remove memorized leaked fingerprints, split-distributed fingerprints retain stronger private transfer evidence.} Among the NaturalCarrier variants, TCF-S8 is more robust than TCF-S4 under adaptive LoRA training.

\begin{tcfresultbox}{Takeaway}
\small
Partial-leakage LoRA erasure suppresses exposed fingerprints but does not reliably remove private distributed fingerprints.
\end{tcfresultbox}

To explain why the split variants are more robust for the adaptive attack, we examine the hidden target margin, defined as the score gap between the recorded target label and the strongest non-target valid label. A positive margin means that the target remains the top-scoring valid label, while a non-positive margin indicates boundary crossing and effective erasure. Fig.~\ref{fig:margin_boundary_mechanism} shows that LoRA erasure reduces hidden margins for all variants, but the unsplit Original variant is more likely to cross the boundary, with a higher hidden normalized erasure sufficiency (NES=$1.3845$), a higher boundary-crossing rate ($0.3967$), and a smaller residual hidden margin ($0.1525$). In contrast, the split variants TCF-S4 and TCF-S8 have much lower boundary-crossing rates ($0.1767$ and $0.1667$) and retain larger residual margins ($0.8483$ and $0.7958$). The boundary-crossing rate further correlates with hidden target-hit degradation (Spearman $\rho=0.8681$), while the residual hidden margin correlates with the minimum Detection AUC ($\rho=0.8070$). These results suggest that splitting improves robustness by preserving a larger hidden-margin buffer, making leaked-update erasure less likely to push hidden fingerprints across the hard-label decision boundary. In the prompt-embedding visualization of Fig.~\ref{fig:margin_boundary_mechanism}(d), TCF-S4 and TCF-S8 form multiple sub-clusters instead of one concentrated fingerprint cluster, a pattern consistent with higher fingerprint diversity after splitting. This observation is consistent with less complete coverage by leaked LoRA erasure and with the lower boundary-crossing rates of the split variants.

\begin{tcfresultbox}{Takeaway}
\small
Split fingerprints resist erasure by preserving larger hidden target margins and more diverse prompt patterns.
\end{tcfresultbox}

\expitem{Decoding and sampling stability} To evaluate the robustness of different fingerprinting methods under varying decoding configurations, we compared all methods using the same Qwen3-1.7B Base fingerprints while systematically altering verification‑stage decoding parameters (Fig.~\ref{fig:robustness_ablation_diagnostics}(b)--(c)). We observed that \method{} maintains near‑perfect AUC under all tested nucleus‑sampling settings and remains above 0.91 across the entire temperature sweep. In contrast, the baseline methods either yield consistently lower AUC or exhibit a more pronounced degradation under several settings. This superior robustness can be attributed to the legal‑answer space and the SCM‑qualified target transfer, which preserve a larger derived‑independent separation than TRAP, ProFLingo, and ZeroPrint even when decoding randomness changes.

\begin{tcfresultbox}{Takeaway}
\small
\method{} remains stable under common decoding changes, with high temperature as the main source of degradation.
\end{tcfresultbox}

\expitem{Input paraphrasing and score-space perturbation} To examine resilience to lightweight input rewriting and score-level noise, we compare \method{} with the baselines in Fig.~\ref{fig:robustness_ablation_diagnostics}(d)--(e). Paraphrasing degrades all prompt-based signals, yet \method{} consistently achieves the highest AUC across the three paraphrase models--Gemma 2 2B-IT~\cite{gemma2}, OLMo 2 1B-Instruct~\cite{olmo2}, and Qwen2 0.5B-Instruct~\cite{qwen2}. This indicates that the target-transfer event survives moderate surface rewriting more robustly than response-reproduction or similarity-based scores. Panel~(e) is a score-space stress test rather than a raw-text API transformation: clipped Gaussian noise
$\epsilon \sim \mathcal{N}(0,\sigma^2)$ with
$\sigma \in \{0.05,0.10,0.15\}$ is added in score space before AUC computation. \method{} remains essentially robust while the baselines stay lower, indicating that the reported separation is not fragile to moderate score-level noise.

\begin{tcfresultbox}{Takeaway}
\small
\method{} is more resilient than the baselines to moderate paraphrasing and score-space noise.
\end{tcfresultbox}

\expitem{Model merging}
Fig.~\ref{fig:robustness_ablation_diagnostics}(f) evaluates a model-merging attack on Qwen3-1.7B, where the fingerprinted victim checkpoint \texttt{Qwen3-1.7B-Base} is linearly interpolated with a non-fingerprinted reference checkpoint \texttt{Qwen3-1.7B}:
$\theta_{\mathrm{merge}}=\alpha\theta_{\mathrm{victim}}+(1-\alpha)\theta_{\mathrm{ref}}$.
We sweep $\alpha$ and re-run fingerprint verification. TRAP drops to chance-level AUC, while ProFLingo and ZeroPrint show unstable AUC across merge ratios. In contrast, TCF keeps AUC at $1.0$ throughout the sweep, showing that its hard-label fingerprint signal remains stable under this linear merging attack.

\begin{tcfresultbox}{Takeaway}
\small
\method{} remains stable under the evaluated linear model-merging attack.
\end{tcfresultbox}

\subsection{Ablation Studies}
\label{sec:ablations}

\begin{figure}[htp]
    \centering
    \includegraphics[width=0.98\linewidth]{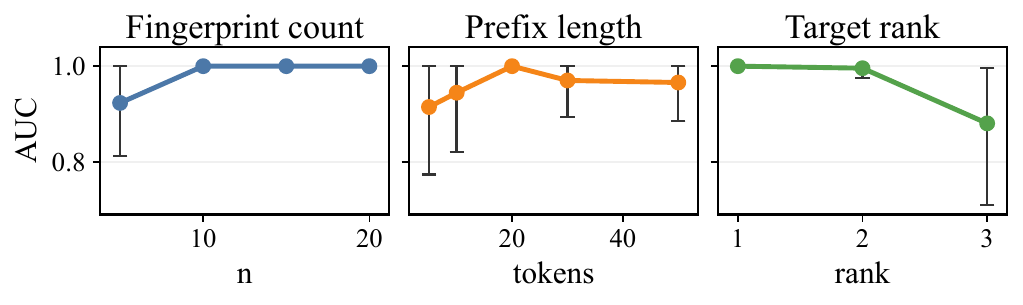}
    \caption{Ablation study results of fingerprint count (a), prefix length (b), and target answer selection (c).}
    \label{fig:ablation}
\end{figure}

To investigate the impact of key hyperparameters on verification performance, we conducted ablation studies on three aspects: the number of fingerprints $n$, the GCG prefix length (PL), and the target-answer rank. Fig.~\ref{fig:ablation} summarizes the main results under the promoted hard-label setting.

\textbf{Number of fingerprints $n$:} To determine how many fingerprints are sufficient for stable verification, we varied $n$ from 5 to 20 in Fig.~\ref{fig:ablation}(a). We observed that performance improves rapidly from $n=5$ to $n=10$ and then saturates in the Qwen3-1.7B Base setting, indicating that a modest fingerprint set is adequate.

\begin{tcfresultbox}{Takeaway}
\small
A modest fingerprint set is sufficient for stable verification in the main setting.
\end{tcfresultbox}

\textbf{GCG prefix length (PL):} To assess the effect of prefix length on separability, we tested different PL values in Fig.~\ref{fig:ablation}(b). We found that moderate prefixes are sufficient, while overly short or overly long prefixes can weaken the derived--independent separation. This is consistent with the feasible operating-region interpretation of Section~\ref{subsec:scm_separability}: overly weak perturbations may fail to create a source-side counterfactual transition, whereas overly strong perturbations can increase generic transfer to independent models. The empirical optimum therefore need not occur at the largest prefix length or largest SCM.

\begin{tcfresultbox}{Takeaway}
\small
A moderate prefix length best balances target induction and independent-model non-transfer.
\end{tcfresultbox}

\textbf{Target-answer selection:} To understand how the choice of target label influences verification, we performed a source-model diagnostic by varying the rank of the target answer in Fig.~\ref{fig:ablation}(c). Although Top-1, Top-2, and Top-3 target choices all achieve AUC 1.0000 in this diagnostic, the independent mean varies. This suggests that target rank can affect false-positive behavior. Therefore, we use this target-rank panel to guide target selection rather than as a standalone full-model-pool ownership result.

\begin{tcfresultbox}{Takeaway}
\small
Target selection should minimize accidental independent preference rather than optimize source success alone.
\end{tcfresultbox}

\subsection{Computational Cost}
\label{sec:efficiency_storage}

To evaluate per-fingerprint generation, verification, and storage overhead, we conduct a comparative analysis on Qwen3-1.7B Base (Tab.~\ref{tab:overhead}). TRAP is the most expensive among the compared methods, requiring 8460.5 seconds per fingerprint under its official setting. ZeroPrint has fast source-side generation but incurs high verification cost because it extracts perturbation-based fingerprints from suspect models. ProFLingo is efficient during verification but uses a larger 50-query registry. In contrast, \method{} requires 1269.3 seconds per fingerprint and uses a 20-query registry in the main evaluation, giving lower construction/verification overhead than the evaluated baselines while retaining the strongest verification performance.

\begin{table}[htp]
\centering
\caption{Per-fingerprint computation and storage cost on Qwen3-1.7B Base.}
\label{tab:overhead}
\resizebox{1.0\columnwidth}{!}{
\begin{tabular}{lccccc}
\toprule
\rowcolor{HeaderGray}
Method & Total time & Generation time & Verify time & Generation storage & Eval storage  \\
\midrule
\method{} & 1269.3s & 1200.6s & 68.7s & 1341.2 KiB & 75.2 KiB  \\
TRAP  & 8460.5s & 7835.9s & 624.7s & 234.3 KiB & 290.3 KiB  \\
ProFLingo  & 1625.0s & 1550.1s & 74.9s & 133.9 KiB & 0.2 KiB \\
ZeroPrint & 1954.5s & 44.0s & 1910.5s & 48387.6 KiB & 47.6 KiB  \\
\bottomrule
\end{tabular}
}
\end{table}

\begin{tcfresultbox}{Takeaway}
\small
\method{} combines the strongest verification performance with lower per-fingerprint overhead and a compact 20-query registry.
\end{tcfresultbox}

\section{Conclusion}
\label{sec:conclusion}

We present \method{}, a non-invasive fingerprinting framework for black-box LLM ownership verification. The key idea is to avoid open-ended generation by reformulating verification as constrained-answer target transfer. \method{} constructs perturbed prompts that move the protected model from its clean answer to a counterfactual target, then verifies a suspect model by measuring target-answer transfer over the private fingerprint set. Theoretical analysis derives target-transfer bounds and a sufficient positive-gap condition under explicit derived-preservation and independent-transfer budgets motivated by local behavioral closeness; aggregation then converts a per-fingerprint edge into a thresholdable ownership signal. Empirically, \method{} achieves near-perfect AUC across four source families and matches or outperforms the strongest evaluated black-box fingerprinting baseline on every source family.

\bibliographystyle{IEEEtran}
\bibliography{references}

\clearpage
\onecolumn
\raggedbottom
\appendices
\section{Additional Experimental Details}

\setlength{\textfloatsep}{12pt plus 2pt minus 2pt}
\setlength{\floatsep}{10pt plus 2pt minus 2pt}
\setlength{\intextsep}{10pt plus 2pt minus 2pt}
\setlength{\emergencystretch}{2em}

\begin{tcfappendixbox}{Appendix roadmap}
\small
This appendix is organized as a reproducibility record. It first validates the complete SCM evidence chain across additional source families, then provides the proofs and finite-sample details behind the theoretical claims, and finally documents the construction protocol, exact hard-label interface, prompt examples, baseline settings, model checkpoints, and retained evaluation artifacts.
\end{tcfappendixbox}

\providecommand{\FloatBarrier}{\clearpage}

\subsection{Cross-Model Theory Validation}
\label{app:cross_model_theory_validation}

The main theoretical analysis is not specific to the Qwen3-1.7B Base diagnostic used for exposition. We therefore repeat the complete evidence chain on three additional protected source families: Qwen3 Instruct, Llama3, and Mistral. Across these families, source-side optimization increases SCM, derived models preserve the induced target preference more strongly than independent models, and the resulting derived--independent target-answer gap generally grows in higher-SCM regimes. These experiments directly test the portability of the theory rather than introducing a separate empirical claim.

\subsubsection{Source-Side SCM Control}
\begin{figure}[htp]
    \centering
    \includegraphics[width=0.86\textwidth]{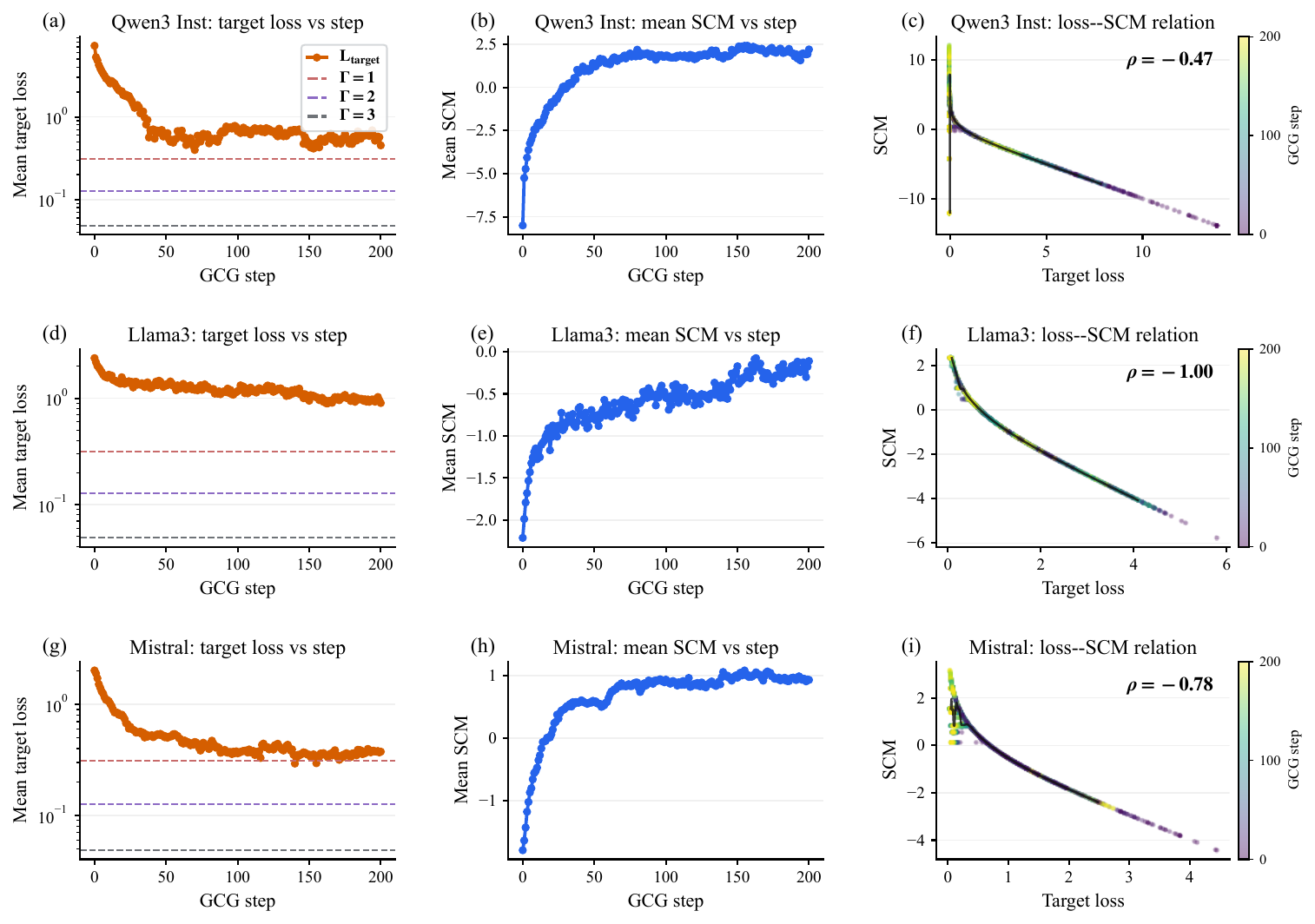}
    \caption{SCM-control diagnostics on three additional protected source families. Source target loss decreases during optimization, mean SCM increases, and lower target loss is associated with stronger SCM.}
    \label{fig:app_three_models_scm_control}
\end{figure}

Figure~\ref{fig:app_three_models_scm_control} verifies that the source-side optimization behaves consistently across the three additional families. Target loss decreases during GCG optimization, mean SCM increases, and the two quantities remain strongly negatively associated. This supports the source-side implication used in Lemma~\ref{lem:loss_threshold_for_scm} beyond the single diagnostic model shown in the main text.

\FloatBarrier

\begin{tcfresultbox}{Takeaway}
\small
The source-loss-to-SCM relationship generalizes across the additional protected model families.
\end{tcfresultbox}

\subsubsection{Derived-Model Preservation}
\begin{figure}[htp]
    \centering
    \includegraphics[width=0.72\textwidth]{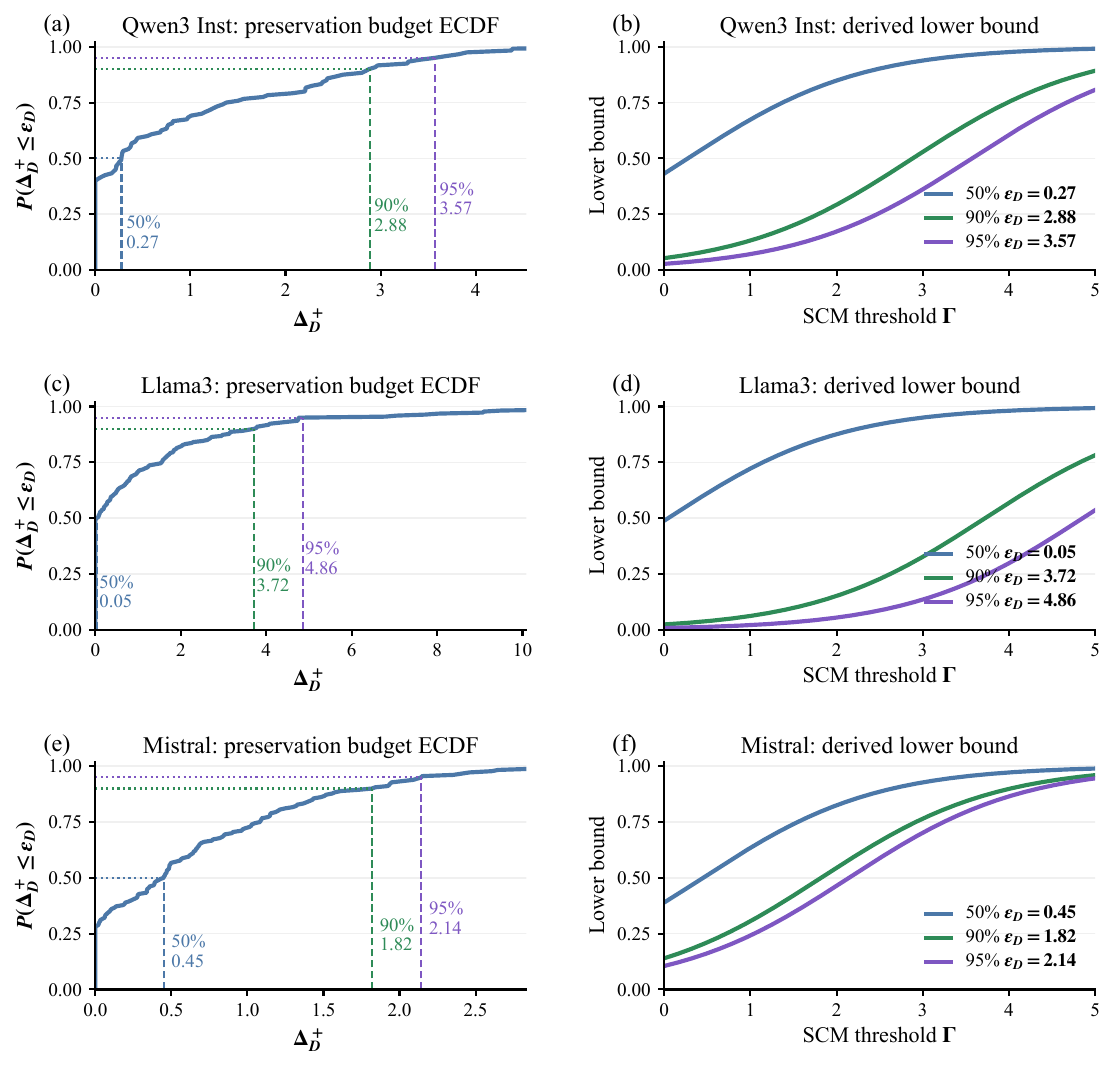}
    \caption{Derived-model preservation on the additional source families. Each row reports the empirical weakening-budget coverage and the corresponding lower-bound trend $\sigma(\Gamma-\epsilon_D)$.}
    \label{fig:app_three_models_eq25_preservation}
\end{figure}

Figure~\ref{fig:app_three_models_eq25_preservation} evaluates the derived-model weakening budget $\epsilon_D$. For each source family, the empirical drift distribution and the corresponding lower-bound curves show that source-family derivatives retain a substantial part of the optimized target preference. Smaller weakening budgets and higher curves indicate stronger preservation.

\FloatBarrier

\begin{tcfresultbox}{Takeaway}
\small
Derived-model preservation remains consistent across the additional protected model families.
\end{tcfresultbox}

\subsubsection{Independent-Model Non-Transfer}
\begin{figure}[htp]
    \centering
    \includegraphics[width=0.84\textwidth]{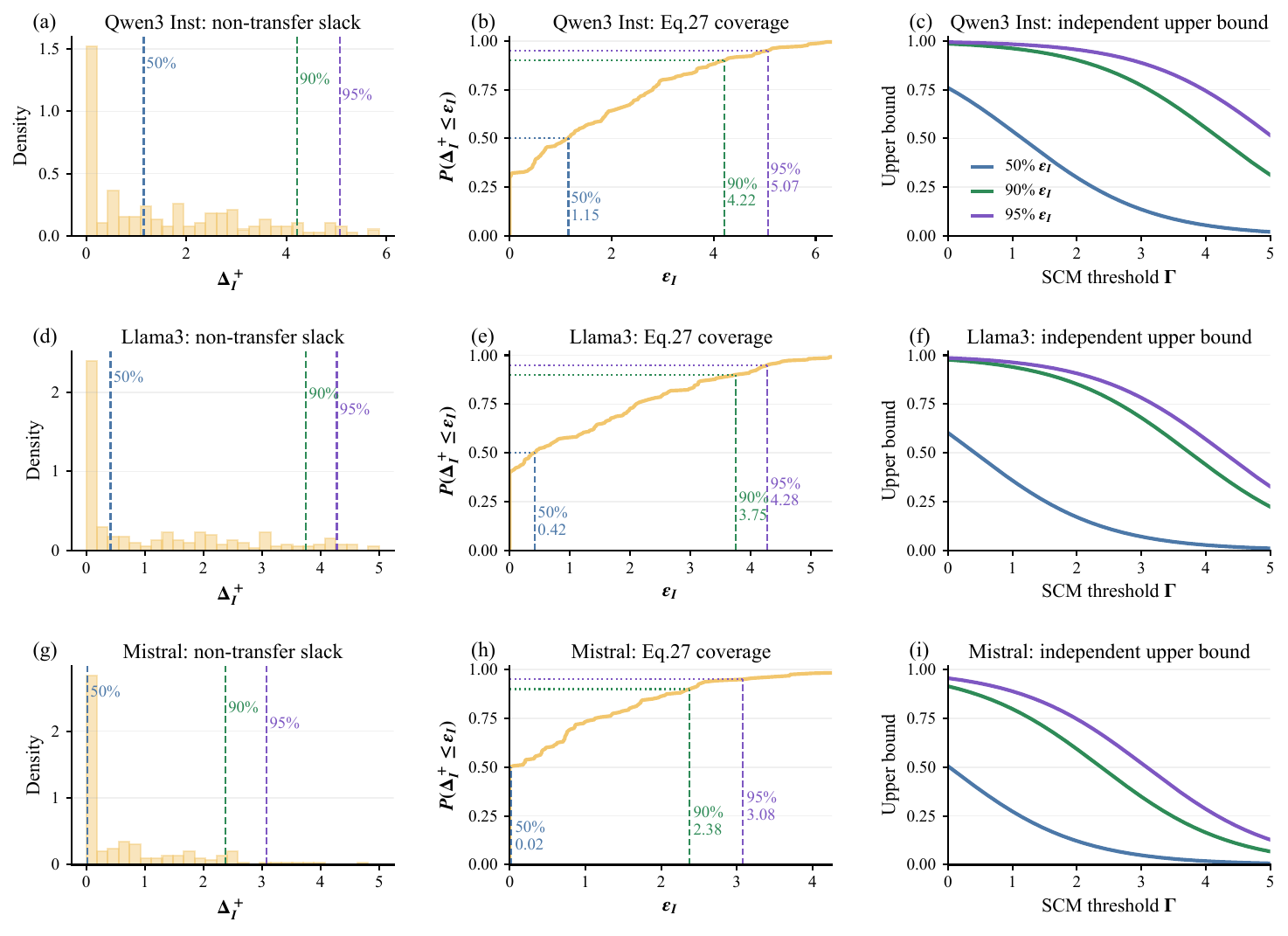}
    \caption{Independent-model non-transfer on the additional source families. The panels report the slack distribution, empirical coverage $\Pr(\Delta_I^+\le\epsilon_I)$, and the corresponding upper-bound trend $\sigma(-\Gamma+\epsilon_I)$.}
    \label{fig:app_three_models_eq31_nontransfer}
\end{figure}

Figure~\ref{fig:app_three_models_eq31_nontransfer} evaluates the independent-model slack $\epsilon_I$. Independent models generally require larger slack budgets to explain the same target behavior, showing that the optimized prompts do not preserve the source-side target preference as consistently outside the protected model family.

\FloatBarrier

\begin{tcfresultbox}{Takeaway}
\small
Independent models preserve the source-induced target behavior less consistently than source-family derivatives.
\end{tcfresultbox}

\subsubsection{SCM-Induced Separability}
\begin{figure}[htp]
    \centering
    \includegraphics[width=0.98\textwidth]{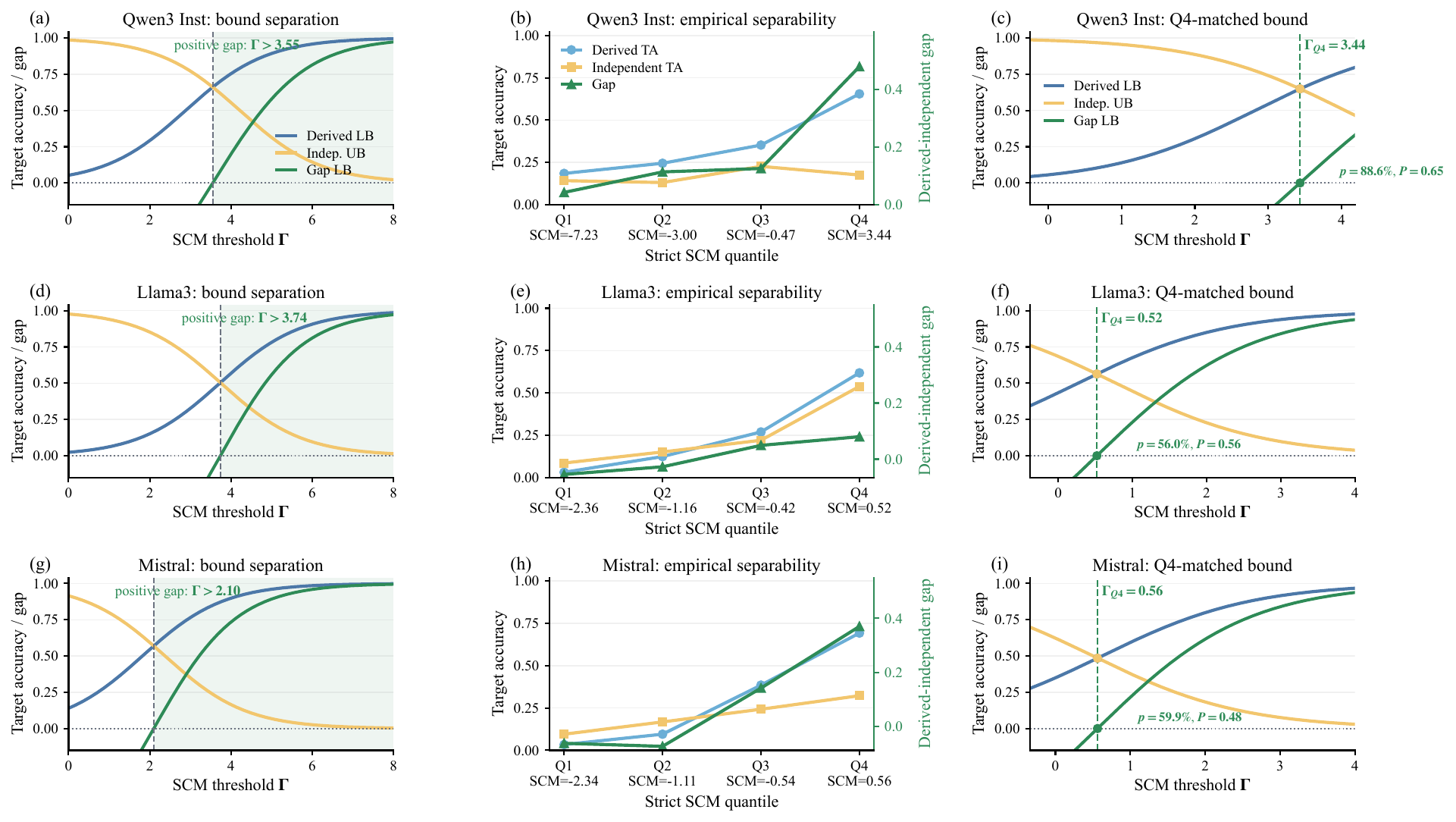}
    \caption{SCM-induced separability on the additional source families. The columns compare conservative bound separation, empirical target-accuracy trends across SCM buckets, and budget-matched bound trends.}
    \label{fig:app_three_models_scm_separability}
\end{figure}

Figure~\ref{fig:app_three_models_scm_separability} closes the evidence chain. Across all three additional families, higher-SCM fingerprints generally produce a larger empirical target-answer gap between derivatives and independent models, while the budget-matched bounds show the same direction. The result supports the feasible-margin interpretation in the main text: SCM is a useful selection signal, but indefinite over-optimization is not required.

\FloatBarrier

\begin{tcfresultbox}{Takeaway}
\small
The feasible-margin-band pattern repeats across the additional source families.
\end{tcfresultbox}
\FloatBarrier

\subsection{Theory Details and Proofs}
\label{app:theory_details}

This appendix provides proof details and concise explanations for the theoretical analysis in Section~\ref{sec:theory}.

\subsubsection{Finite Constrained-Answer Channel and SCM Feasibility}
\label{app:finite_channel_details}

The probability in Eq.~\ref{eq:legal_answer_channel} is over the model's
external decoding randomness, server-side nondeterminism, and any randomness
in the verifier's repeated query protocol. For a deterministic service, this
distribution degenerates to a point mass; for source-side offline analysis,
it can be estimated either by repeated generation under the chosen decoding
configuration or by a calibrated valid-label probability surrogate that is
checked after optimization. Invalid responses are not assigned to any valid
label. Thus, $\sum_{a\in\mathcal{A}}q_{M}(a\mid p)$ may be smaller than
one. This is intentional: invalid responses are verification failures in our
hard-label setting.

Eq.~\ref{eq:target_log_odds} is well-defined whenever
$q_{M}(t\mid p)\in(0,1)$. In implementation, we clip probabilities to
$[\epsilon_{\mathrm{clip}},1-\epsilon_{\mathrm{clip}}]$ before taking
log-odds. The proofs apply exactly to the unclipped channel when
$q_M(t\mid p)\in(0,1)$, and to the clipped channel otherwise; clipping only
prevents infinite log-odds and does not change the hard-label verification
statistic.

\subsubsection{SCM Scale and Feasibility}
\label{app:scm_feasibility_details}

The threshold has a direct probability meaning:
$\Gamma_{\min}=2$ corresponds to source perturbed target probability at least
$0.881$ and clean target probability at most $0.119$; $\Gamma_{\min}=3$
corresponds to approximately $0.953$ and $0.047$. The threshold is not a
universal constant. For a fixed candidate pool $\mathcal{C}$, define
\begin{equation}
\mathcal{C}_{\Gamma_{\min}}
=
\{(x,u,t)\in\mathcal{C}:\Gamma(x,u,t)\ge\Gamma_{\min}\}.
\label{eq:qualified_candidate_pool}
\end{equation}
If $\Gamma_{\min}^{(2)}>\Gamma_{\min}^{(1)}$, then
$\mathcal{C}_{\Gamma_{\min}^{(2)}}\subseteq
\mathcal{C}_{\Gamma_{\min}^{(1)}}$. Increasing $\Gamma_{\min}$ therefore
strengthens individual fingerprints but can reduce the number and diversity
of retained fingerprints. In the experiments, $\Gamma_{\min}$ should be
chosen or swept using only source-side feasibility and, when available, a
separate validation pool; empirical SCM-quantile analyses are evidence for
trends, not a substitute for reporting the retained SCM distribution and
pass rate at the operating threshold.

\subsubsection{Source-Side Surrogate Used During Perturbation Search}
\label{app:surrogate_details}

Eq.~\ref{eq:gcg_loss} is the ideal loss on the same parser-induced
constrained-answer distribution used by verification. In practice, the
perturbation is optimized through a differentiable source-side surrogate,
e.g., the next-option-token probability after the answer prefix. Let
$\widetilde{q}_{M_{0}}(t\mid p)$ be this surrogate probability and let
$\widetilde{L}_{M_{0}}(p,t)=\log\frac{\widetilde{q}_{M_{0}}(t\mid p)}{1-\widetilde{q}_{M_{0}}(t\mid p)}$.
If the surrogate log-odds is calibrated to the parser distribution on retained
candidates in the sense that
\begin{equation}
\left|L_{M_{0}}(p^{u}(x),t)-\widetilde{L}_{M_{0}}(p^{u}(x),t)\right|
\le \delta_{\mathrm{surr}},
\label{eq:surrogate_channel_gap}
\end{equation}
then the same argument as Lemma~\ref{lem:loss_threshold_for_scm} applies
with a safety margin: requiring the surrogate perturbed margin to exceed
$\Gamma_{\min}+\delta_{\mathrm{surr}}$ suffices to guarantee parser-side
perturbed margin at least $\Gamma_{\min}$. Our offline construction therefore
uses the optimizer as a search procedure but accepts fingerprints only after
recomputing the SCM used in Eq.~\ref{eq:scm_definition}; the theoretical
statements are about this accepted parser-side constrained-answer
distribution, not about the surrogate alone.

\subsubsection{Proof of Lemma~\ref{lem:loss_threshold_for_scm}}
\label{app:proof_loss_threshold_for_scm}

The clean side of the SCM is already guaranteed by
Eq.~\ref{eq:clean_side_threshold}. It remains to prove that the perturbed
side is also at least $\Gamma_{\min}$.

From Eq.~\ref{eq:gcg_loss}, the loss condition in
Eq.~\ref{eq:gcg_loss_threshold} gives
\begin{equation}
-\log q_{M_{0}}(t\mid p^{u}(x))
\le
\log\left(1+e^{-\Gamma_{\min}}\right).
\label{eq:proof_loss_condition}
\end{equation}
Multiplying Eq.~\ref{eq:proof_loss_condition} by $-1$ reverses the
inequality:
\begin{equation}
\log q_{M_{0}}(t\mid p^{u}(x))
\ge
-\log\left(1+e^{-\Gamma_{\min}}\right).
\label{eq:proof_log_probability_condition}
\end{equation}
Applying the exponential function to
Eq.~\ref{eq:proof_log_probability_condition} yields
\begin{equation}
q_{M_{0}}(t\mid p^{u}(x))
\ge
\frac{1}{1+e^{-\Gamma_{\min}}}
=
\sigma(\Gamma_{\min}).
\label{eq:proof_prefixed_probability_condition}
\end{equation}
Using Eq.~\ref{eq:probability_logit_relation},
Eq.~\ref{eq:proof_prefixed_probability_condition} is equivalent to
\begin{equation}
L_{M_{0}}(p^{u}(x),t)
\ge
\Gamma_{\min}.
\label{eq:proof_prefixed_log_odds_condition}
\end{equation}
Now Eq.~\ref{eq:clean_side_threshold} and
Eq.~\ref{eq:proof_prefixed_log_odds_condition} show that both terms in the
minimum of Eq.~\ref{eq:scm_definition} are at least $\Gamma_{\min}$.
Therefore, Eq.~\ref{eq:scm_threshold_satisfied} follows.

\subsubsection{Empirical Verification of SCM Control}
\label{app:empirical_scm_control}

To verify Lemma~\ref{lem:loss_threshold_for_scm}, we record the source
target loss in Eq.~\ref{eq:gcg_loss} and mean SCM over fingerprints while
optimizing the source-side perturbation on the source model $M_0$ on
Qwen3-1.7B Base~\cite{yang2025qwen3}. As shown in
Fig.~\ref{fig:scm_support_bundle}(a), the target loss on source model $M_0$
decreases along the optimization progress and approaches the loss thresholds
induced by different SCM targets, while Fig.~\ref{fig:scm_support_bundle}(b)
shows that the mean SCM increases along the same optimization trajectory,
indicating that reducing the target loss strengthens the source-side
counterfactual transition. Fig.~\ref{fig:scm_support_bundle}(c) directly
shows the correlation coefficient $\rho$ between target loss and SCM is
close to $-1.00$, demonstrating a strong negative correlation: as the
optimizer lowers the source target loss, SCM increases. This is consistent
with Lemma~\ref{lem:loss_threshold_for_scm}, because reducing
$-\log q_{M_0}(t\mid p^u)$ increases the perturbed target log-odds while the
clean-side term is fixed by target selection.

\begin{tcfresultbox}{Takeaway}
\small
The surrogate loss guides optimization, while parser-side SCM remains the final acceptance criterion.
\end{tcfresultbox}

\subsubsection{Interpretation of Local Behavioral Closeness}
\label{app:local_closeness_details}

By construction, the source model makes the target unlikely on the clean prompt
and likely on the perturbed prompt. Assumption~\ref{ass:local_closeness} says
that a derived model should remain close to the protected model on these clean
and perturbed fingerprint prompts for most retained fingerprints. This premise is
probabilistic: it only needs to hold for a large fraction of fingerprints, not
for every single sample.

The assumption motivates the transfer model, but it is not by itself the
formal independent-model upper bound. Lemmas~\ref{lem:derived_lower_bound}
and~\ref{lem:independent_upper_bound} make the required derived-preservation
and independent generic-transfer budgets explicit. Accordingly, all formal
separation statements in Theorem~\ref{thm:scm_gap} and
Lemma~\ref{lem:fixed_budget_separability} are conditional on valid
$\epsilon_D$ and $\epsilon_I$ budgets; the empirical diagnostics test how
realistic those budgets are for the evaluated model pools. The verifier then
turns the resulting per-fingerprint edge into evidence by aggregating many
diverse fingerprints.

\subsubsection{Derived-Model Preservation Details}
\label{app:derived_validation_details}

The derived-model diagnostic measures how much the perturbed target preference
weakens when moving from the source model to its derivatives. Smaller weakening
budgets indicate stronger preservation. In the paper appendix we keep this
reading and the proof below. The cross-model ECDF evidence is reported in Appendix~\ref{app:cross_model_theory_validation}.

\subsubsection{Proof of Lemma~\ref{lem:derived_lower_bound}}
\label{app:proof_derived_lower_bound}

By the definition of SCM, $L_{M_0}(p^{u}(x),t)\ge \Gamma$. Combining this
with Eq.~\ref{eq:derived_log_odds_preservation} gives
\begin{equation}
L_{M_D}(p^{u}(x),t)
\ge
\Gamma-\epsilon_D(M_D,x,u,t).
\label{eq:proof_derived_log_odds_lower_actual}
\end{equation}
Applying the probability--logit relation in
Eq.~\ref{eq:probability_logit_relation} and the monotonicity of
$\sigma(\cdot)$ proves
Eq.~\ref{eq:derived_target_probability_lower_actual_gamma}. The stated
$\Gamma_{\min}$ version follows because $\Gamma\ge\Gamma_{\min}$.

\subsubsection{Independent-Model Non-Transfer Details}
\label{app:independent_nontransfer_details}

The independent-model side of the analysis states that a prompt optimized on
$M_0$ should not consistently preserve the same target preference on unrelated
models. The diagnostic in the main text measures this effect through the
independent-model slack: smaller slack means weaker accidental transfer, while
larger slack gives a looser non-transfer guarantee. The corresponding cross-model diagnostics are reported in Appendix~\ref{app:cross_model_theory_validation}.

\subsubsection{Proof of Lemma~\ref{lem:independent_upper_bound}}
\label{app:proof_independent_upper_bound}

By the definition of SCM,
\begin{equation}
-L_{M_0}(p^{0}(x),t)\ge\Gamma,
\qquad
L_{M_0}(p^{0}(x),t)\le -\Gamma.
\label{eq:proof_clean_bound_from_actual_gamma}
\end{equation}
Combining Eq.~\ref{eq:proof_clean_bound_from_actual_gamma} with
Eq.~\ref{eq:independent_log_odds_upper} gives
\begin{equation}
L_{M_I}(p^{u}(x),t)
\le
-\Gamma+\epsilon_I(M_I,x,u,t).
\label{eq:proof_independent_log_odds_upper_actual}
\end{equation}
Applying Eq.~\ref{eq:probability_logit_relation} and the monotonicity of
$\sigma(\cdot)$ proves
Eq.~\ref{eq:independent_target_probability_upper_actual_gamma}. The stated
$\Gamma_{\min}$ version follows from $\Gamma\ge\Gamma_{\min}$.

\subsubsection{Separability Details and Feasible Margin Band}
\label{app:separability_details}

For fixed transfer budgets, increasing SCM improves the derived--independent
separation until the gain starts to saturate. In practice, pushing the
perturbation too far can also increase generic transfer to independent models.
Thus, the useful operating region is a feasible margin band: SCM should be large
enough to survive derived-model drift, but the construction should not
optimize the source margin indefinitely. The same trend is tested across three additional source families in Appendix~\ref{app:cross_model_theory_validation}.

\subsubsection{Proof of Theorem~\ref{thm:scm_gap}}
\label{app:proof_scm_gap}

For each fingerprint $i$, Lemma~\ref{lem:derived_lower_bound} gives
\begin{equation}
q_{M_D}(t_i\mid p^{u_i}(x_i))
\ge
\sigma(\Gamma_i-\epsilon_{D,i}).
\label{eq:proof_each_derived_lower}
\end{equation}
Averaging over $i=1,\ldots,n$ proves Eq.~\ref{eq:derived_fta_lower}.
Similarly, Lemma~\ref{lem:independent_upper_bound} gives
\begin{equation}
q_{M_I}(t_i\mid p^{u_i}(x_i))
\le
\sigma(-\Gamma_i+\epsilon_{I,i}),
\label{eq:proof_each_independent_upper}
\end{equation}
and averaging proves Eq.~\ref{eq:independent_fta_upper}. Subtracting the
upper bound from the lower bound gives Eq.~\ref{eq:fta_gap_lower}.

\subsubsection{Proof of Lemma~\ref{lem:fixed_budget_separability}}
\label{app:proof_fixed_budget_separability}

The expectation bounds follow from Theorem~\ref{thm:scm_gap} and the
monotonicity of $\sigma(\cdot)$. The fixed-budget gap is positive exactly
when $\sigma(\Gamma_{\min}-\epsilon_D)>\sigma(-\Gamma_{\min}+\epsilon_I)$,
which is equivalent, by monotonicity of the sigmoid, to
$\Gamma_{\min}-\epsilon_D> -\Gamma_{\min}+\epsilon_I$. Rearranging gives
Eq.~\ref{eq:positive_gap_condition}.

No additional $\eta$ correction is introduced in this lemma: the theorem's
expected target-accuracy bounds are stated directly in terms of the explicit
per-fingerprint budgets. Under the fixed-budget premises,
\begin{equation}
\mathbb{E}[\operatorname{TA}(M_D)]
\ge \sigma(\Gamma_{\min}-\epsilon_D),
\qquad
\mathbb{E}[\operatorname{TA}(M_I)]
\le \sigma(-\Gamma_{\min}+\epsilon_I),
\end{equation}
and their difference is positive whenever
$2\Gamma_{\min}>\epsilon_D+\epsilon_I$, exactly as stated in the main text.
Aggregation then has a clear statistical role: once the two class-specific
expectations are separated, averaging diverse fingerprint indicators reduces
sampling variability, subject to the dependence qualification in
Appendix~\ref{app:proof_finite_sample_reliability}.

\subsubsection{Empirical SCM-Induced Separability Validation}
\label{app:empirical_separability_details}

Fig.~\ref{fig:scm_separability_bounds} provides the empirical separability check for the fixed-budget gap analysis.
The main observation is that higher-SCM fingerprints generally increase the derived--independent target-accuracy gap, while the strongest margins should not be over-optimized because generic transfer can also increase.

\begin{tcfresultbox}{Takeaway}
\small
Useful separability emerges inside a feasible SCM region and can weaken under over-optimization.
\end{tcfresultbox}

\subsubsection{Proof of Lemma~\ref{lem:finite_sample_reliability}}
\label{app:proof_finite_sample_reliability}

For the derived model, $\mu_{M_D}-\zeta^{\star}\ge
(\alpha_D-\alpha_I)/2$. Therefore the event
$\operatorname{TA}(M_D)<\zeta^{\star}$ implies a lower-tail deviation of at
least $(\alpha_D-\alpha_I)/2$ from its mean. Hoeffding's inequality for an
average of $n$ independent Bernoulli variables gives
Eq.~\ref{eq:false_negative_bound}. The independent-model case is the
symmetric upper-tail deviation because
$\zeta^{\star}-\mu_{M_I}\ge(\alpha_D-\alpha_I)/2$.

The independence condition is a statistical idealization. Fingerprints
constructed from the same source model, prompt template, and dataset can be
correlated. For deployment-grade audits, the concentration term should
therefore be interpreted with an effective sample size after diversity
filtering, or estimated by block/bootstrap resampling over question
categories and target labels. The theoretical message is that reliability
improves with both the expected gap $\alpha_D-\alpha_I$ and the effective
number of diverse fingerprints.

\subsubsection{Margin--Sampling Robustness}
\label{app:margin_sampling_details}

Both the verifier and the construction filter read a \emph{hard} label, so
the final quantity that matters is whether the sampled answer equals $t$.
The target margin of Eq.~\ref{eq:target_margin} controls this directly under
an idealized constrained decoder. Under temperature-$T$ softmax sampling
normalized over only the $K$ legal labels in $\mathcal{A}$ (or conditional on
the response being a valid legal label), each competing label contributes at
most $e^{-m_M(p,t)/T}$ relative mass. Therefore,
\begin{equation}
\Pr[\text{sample}=t\mid p]
\ge
\frac{1}{1+(K-1)\,e^{-m_{M}(p,t)/T}}.
\end{equation}
Thus a large source margin $m_{M_0}(p^{u},t)$ strengthens target-label
stability under the constrained-label idealization. This derivation does not
assert that an arbitrary raw-text API samples only from $\mathcal{A}$; invalid
or free-form generations remain possible in the actual threat model and are
counted as failures. The deployment-facing stability claim is therefore
checked empirically by the repeated-query source filter in
Section~\ref{sec:method}. The margin bound should be read as an explanatory
connection between source-side margin and legal-label sampling, not as a
replacement for that empirical filter.

\subsection{Implementation Details}
\label{sec:implementation_details}

Algorithm~\ref{alg:scm_gcg_construction} states the complete source-only construction procedure in executable form. It follows the main-text checkpoint rule: stop at the first source-side checkpoint satisfying Eq.~\ref{eq:method_gcg_stop_loss}, then recompute parser-side SCM and apply the source stability filter. No derived-model or independent-model statistic is used during construction.

\begin{algorithm}[H]
\caption{SCM-guided GCG fingerprint construction}
\label{alg:scm_gcg_construction}
\small
\begin{algorithmic}[1]
\Require Source model $M_0$; question pool $\mathcal{D}$; legal answer space $\mathcal{A}$; desired registry size $n$; SCM threshold $\Gamma_{\min}$; prefix length $\ell_u$; GCG budget $T_{\max}$; stability threshold $s_{\min}$.
\Ensure Frozen fingerprint registry $\mathcal{F}$.
\State $\mathcal{C}\gets\emptyset$ \Comment{qualified source-only candidates}
\For{each question $x\in\mathcal{D}$}
    \State Render $p^0(x)$; estimate $q_{M_0}(a\mid p^0(x))$ over $a\in\mathcal{A}$.
    \State $y\gets\arg\max_{a\in\mathcal{A}}q_{M_0}(a\mid p^0(x))$.
    \State $\mathcal{T}(x)\gets\{a\in\mathcal{A}\setminus\{y\}:-L_{M_0}(p^0(x),a)\ge\Gamma_{\min}\}$.
    \If{$\mathcal{T}(x)=\emptyset$}
        \State \textbf{continue}
    \EndIf
    \State $t\gets\arg\min_{a\in\mathcal{T}(x)}q_{M_0}(a\mid p^0(x))$.
    \State Initialize a length-$\ell_u$ perturbation $u$.
    \For{$j=1$ to $T_{\max}$}
        \State Apply one GCG update to reduce $\mathcal{L}_{\mathrm{src}}(u;x,t)$ in Eq.~\ref{eq:method_gcg_loss}.
        \If{$\mathcal{L}_{\mathrm{src}}(u;x,t)\le\log(1+e^{-\Gamma_{\min}})$}
            \State \textbf{break} \Comment{first source-side margin-qualified checkpoint}
        \EndIf
    \EndFor
    \State Recompute parser-side $\Gamma(x,u,t)$ using Eq.~\ref{eq:scm_definition}.
    \If{$\Gamma(x,u,t)<\Gamma_{\min}$}
        \State \textbf{continue}
    \EndIf
    \State Estimate $\widehat{s}_0(x,u,t)$ using Eq.~\ref{eq:method_source_stability}.
    \If{$\widehat{s}_0(x,u,t)<s_{\min}$}
        \State \textbf{continue}
    \EndIf
    \State Add $(p^u(x),y,t,u,\Gamma(x,u,t))$ and construction metadata to $\mathcal{C}$.
\EndFor
\State Apply the source-only diversity filter over question IDs, categories, clean labels, and target labels.
\State Select up to $n$ qualified candidates to form the frozen registry $\mathcal{F}$.
\State \Return $\mathcal{F}$
\end{algorithmic}
\end{algorithm}
\FloatBarrier

\subsection{Experimental Details}
\label{app:experimental_design}
\label{app:prompt_templates}

\subsubsection{Main Method Setup}
\label{app:main_method_setup}
The main construction uses a fixed 20-query registry and a single source-only generation configuration. Tables~\ref{tab:fingerprint_categories} and~\ref{tab:ours_params} summarize the query composition and generation parameters before the prompt/interface details.

\begin{table}[htp]
\centering
\caption{Composition of the 20-query fingerprint set for each protected source model.}
\label{tab:fingerprint_categories}
\scriptsize
\setlength{\tabcolsep}{3pt}
\begin{tabularx}{\columnwidth}{llX}
\toprule
\rowcolor{HeaderGray}
Category & Samples & Role \\
\midrule
Chemistry\_val & 5 & Science/entity-level multiple-choice questions \\
Engineering\_dev & 5 & Technical engineering and control questions \\
Health\_dev & 5 & Medical and health-domain questions \\
Physics\_val & 5 & Mathematical and physical-law questions \\
\bottomrule
\end{tabularx}
\end{table}

\begin{table}[htp]
\centering
\caption{Main \method{} fingerprint-generation parameters.}
\label{tab:ours_params}
\scriptsize
\setlength{\tabcolsep}{3pt}
\begin{tabularx}{\columnwidth}{lXlX}
\toprule
\rowcolor{HeaderGray}
Parameter & Setting & Parameter & Setting \\
\midrule
Dataset & MMLU & Fingerprints & 20/source \\
Categories & $4\times5$ samples & GCG steps & 200 \\
Batch size & 128 & Top-$k$ & 64 \\
Prefix length & 20 tokens & Placement & question prefix \\
Interface & \multicolumn{3}{l}{final generated A/B/C/D label only} \\
\bottomrule
\end{tabularx}
\end{table}
\FloatBarrier

\subsubsection{Prompt and Interface Templates}
\label{app:prompt_interface_templates}

For reproducibility, we record both the clean source query and the fingerprint query. The only difference is the source-optimized text inserted before or across the question body; the finite answer contract and parser remain unchanged.

\begin{tcfpromptbox}{Clean/source query template}
{\small\ttfamily\raggedright
The following are multiple choice questions (with answers) about \{category\}.\par
\{question\}\par
A. \{option A\}\par
B. \{option B\}\par
C. \{option C\}\par
D. \{option D\}\par
Answer with only one capital letter: A, B, C, or D.\par
Answer:\par}
\end{tcfpromptbox}

\begin{tcfpromptbox}{Fingerprint query template}
{\small\ttfamily\raggedright
\{optimized prefix or distributed carrier fragments\}\par
The following are multiple choice questions (with answers) about \{category\}.\par
\{question with the retained insertion locations\}\par
A. \{option A\}\par
B. \{option B\}\par
C. \{option C\}\par
D. \{option D\}\par
Answer with only one capital letter: A, B, C, or D.\par
Answer:\par}
\end{tcfpromptbox}

\begin{tcfappendixbox}{Parser and scoring contract}
\small
A deterministic parser maps the final generated response to one label in $\{A,B,C,D\}$ or to $\bot$ when the output is invalid or ambiguous. The online verifier never repairs $\bot$ with semantic similarity or model confidence. Fingerprint $i$ is a hit only when the parsed label equals its private recorded target $t_i$, and the final target accuracy is the mean of these binary hits.
\end{tcfappendixbox}

\begin{tcfappendixbox}{SCM-guided construction protocol}
\small
\begin{enumerate}[leftmargin=1.8em,itemsep=1.5pt,topsep=2pt]
    \item Render the clean constrained-answer prompt and estimate the protected model's valid-label distribution.
    \item Record the clean answer $y$ and choose a different target $t$ that satisfies the clean-side SCM requirement.
    \item Initialize a 20-token optimizable text block and place it at the designated prompt locations.
    \item Run source-side GCG with the reported 200-step budget, candidate batch size 128, and top-$k=64$ replacements.
    \item Select the earliest validated checkpoint that reaches the source-side margin target rather than blindly keeping the last optimization step.
    \item Recompute the accepted prompt's parser-side SCM; reject candidates that fail the margin or validity requirements.
    \item Check source-side response stability and retain only qualified, stable fingerprints until the 20-query registry is complete.
    \item Store the final prompt, clean answer, target, optimized text, and SCM metadata. Online verification then queries only the suspect API and aggregates final-label target hits.
\end{enumerate}
\end{tcfappendixbox}

\subsubsection{Baseline Setup}
\label{app:baseline_setup}

Table~\ref{tab:baseline_protocols} lists the baseline configurations and score definitions used for comparison.

\begin{table}[htp]
\centering
\caption{Baseline configurations and score definitions. Baselines are evaluated using their original method definitions.}
\label{tab:baseline_protocols}
\scriptsize
\setlength{\tabcolsep}{2pt}
\renewcommand{\arraystretch}{1.05}
\begin{tabularx}{\columnwidth}{@{}
>{\raggedright\arraybackslash}p{0.13\columnwidth}
>{\raggedright\arraybackslash}X
>{\raggedright\arraybackslash}p{0.29\columnwidth}
@{}}
\toprule
\rowcolor{HeaderGray}
Method & Main setting & Score definition \\
\midrule
\method{} & 20 source-matched fingerprints & mean target-label hit rate \\
TRAP & strict $n=20$, 1500 steps, batch 512, top-$k=256$ & original TRAP retrieval score \\
ProFLingo & official $E=256$, $p=32$, $b=8$, $n=50$ AE list & final AE-list ASR \\
ZeroPrint & $n_{\mathrm{samples}}=2$, $n_{\mathrm{aug}}=4$, repeats 20 & source-row similarity/correlation \\
\bottomrule
\end{tabularx}
\end{table}
\FloatBarrier

\subsubsection{Naturalness-Oriented Variants}
\label{app:naturalness_implementation_details}

Raw GCG prefixes are effective for source-side target induction, but their
contiguous optimized strings may contain unnatural token fragments. Such
artifacts can be exposed by full-input PPL or local window-PPL abnormality
detectors. We therefore keep the same source-only construction and the same
hard-label verifier, while changing only how the optimized fingerprint text is
presented in the prompt.

\expitem{NaturalCarrier (NC)}
NC reuses a prior-optimized GCG prefix and embeds its fragments in ordinary
carrier phrases, such as \texttt{note this clue:}, \texttt{use this short hint:},
and \texttt{keep this relation in mind:}. This prevents the optimized text from
appearing as one contiguous abnormal prefix while preserving the closed
A/B/C/D answer contract used by the verifier.

\expitem{Split placement}
TCF-S4 and TCF-S8 apply the same NaturalCarrier idea, but distribute the
optimizable token positions over 4 or 8 randomly selected valid insertion sites
in the original question prompt, respectively. By shortening each inserted
segment and spreading it across natural prompt locations, these variants are
intended to reduce local PPL spikes under window-based input-abnormality checks.
For NaturalCarrier and both split variants, verification is unchanged: the online parser reads only
the final generated A/B/C/D label and counts success when it matches the
recorded target answer.

\subsubsection{Representative Prompt-Naturalness Example}
\label{app:prompt9_naturalness}

The following retained example shows how the same multiple-choice item is presented under the clean interface, the contiguous Raw-GCG construction, and the distributed NaturalCarrier variants. Long optimized strings are truncated only for page readability; the evaluation uses the complete stored prompts.

\begin{tcfpromptbox}{Prompt P9: clean query}
{\footnotesize\ttfamily\raggedright
The following are multiple choice questions (with answers) about Chemistry\_val.\par
Which type of mineral exhibits variations in color?\par
A. Pseudochromatic\par
B. Allochromatic\par
C. Idiochromatic\par
D. Iridescence\par
Answer with only one capital letter: A, B, C, or D.\par
Answer:\par}
\end{tcfpromptbox}

\begin{tcfpromptbox}{Prompt P9: Raw GCG}
{\footnotesize\ttfamily\raggedright
The following are multiple choice questions (with answers) about Chemistry\_val.\par
).\_\_interp Interior Haskell //\textasciitilde{} Sv Soil\par
MonterwebElementXpaths ESLonyms Sampler Member Name\par
TypeSpecifier !IDOiotic Determine Which type of mineral exhibits variations in color?\par
A. Pseudochromatic\par
B. Allochromatic\par
C. Idiochromatic\par
D. Iridescence\par
Answer with only one capital letter: A, B, C, or D.\par
Answer: \textnormal{[remaining optimized text omitted from display]}\par}
\end{tcfpromptbox}

\begin{tcfpromptbox}{Prompt P9: NaturalCarrier split4 (NC-S4)}
{\footnotesize\ttfamily\raggedright
Which type of mineral exhibits variations in color?\par
use this short hint: imgumber\par
note this clue: Before-control\par
A. Pseudochromatic\par
keep this relation in mind: als under\par
B. Allochromatic\par
C. Idiochromatic\par
D. Iridescence\par
focus on this cue: A among pur\par
Answer with only one capital letter: A, B, C, or D.\par
Answer:\par}
\end{tcfpromptbox}

\begin{tcfpromptbox}{Prompt P9: NaturalCarrier split8 (NC-S8)}
{\footnotesize\ttfamily\raggedright
Which type of mineral exhibits variations in color?\par
note this clue: pur\par
read this note carefully: img\par
keep this relation in mind: A\par
use this short hint: Before-control\par
keep this relation in mind: among\par
keep this relation in mind: under\par
A. Pseudochromatic\par
focus on this cue: umber\par
B. Allochromatic\par
consider this context: als\par
C. Idiochromatic\par
D. Iridescence\par
Answer with only one capital letter: A, B, C, or D.\par
Answer:\par}
\end{tcfpromptbox}

\subsubsection{Retained Evaluation Artifacts}
\label{app:retained_artifacts}

Table~\ref{tab:source_pool_config} defines the fixed model counts used for the main AUC calculation. Table~\ref{tab:app_full_model_pool} below provides an extended checkpoint catalog, and the per-suspect matrices report the available row-level diagnostics. Some baseline artifacts are unavailable for particular checkpoints, so these appendix matrices use only artifact-available rows and do not redefine the main model pool or the AUC values in Table~\ref{tab:main_auc_full}.

\begin{table}[htp]
\centering
\caption{Extended Hugging Face checkpoint catalog across the retained evaluation artifacts. Rows are grouped by protected source and derivative category; the fixed model-pool sizes used for the primary AUC remain those in Table~\ref{tab:source_pool_config}.}
\label{tab:app_full_model_pool}
\scriptsize
\setlength{\tabcolsep}{2.0pt}
\renewcommand{\arraystretch}{0.98}
\begin{tabularx}{\textwidth}{
@{}
>{\raggedright\arraybackslash}p{0.12\textwidth}
>{\raggedright\arraybackslash}p{0.105\textwidth}
>{\raggedright\arraybackslash\tiny}X
@{}}
\toprule
\rowcolor{black!8}
\textbf{Source pool} & \textbf{Category} & \textbf{Hugging Face checkpoints} \\
\midrule

Qwen3-1.7B Instruct & Source &
\hf{Qwen/Qwen3-1.7B} \\

& SFT &
\hf{activeDap/Qwen3-1.7B\_tldr};
\hf{contextboxai/Qwen3-1.7B-FC};
\hf{prithivMLmods/Demeter-LongCoT-Qwen3-1.7B};
\hf{UnfilteredAI/DAN-Qwen3-1.7B} \\

& Pref./RL &
\hf{wzx111/Qwen3-1.7B-Open-R1-GRPO-Baseline} \\

& Domain &
\hf{prithivMLmods/Panacea-MegaScience-Qwen3-1.7B};
\hf{swapnillo/Bangla-AI-1.7B};
\hf{vimalgupta/qwen3-1.7b-telecom-vocab-v2} \\

& Pruning/Merging &
\hf{MilyaShams/Qwen3-1.7B-Wanda\_1\_4} \\

\midrule
Qwen3-1.7B Base & Source &
\hf{Qwen/Qwen3-1.7B-Base} \\

& SFT &
\hf{ali-elganzory/Qwen3-1.7B-Base-SFT-Tulu3-decontaminated};
\hf{ffang2025/Affine-fang-v2};
\hf{jekunz/Qwen3-1.7B-Base-is-SmolTalk};
\hf{lllyx/Qwen3-1.7B-SFT} \\

& Pref./RL &
\hf{Kazuki1450/Qwen3-1.7B-Base\_dsum\_3\_6\_fnr\_no\_bracket\_0p0\_0p0\_1p0\_grpo\_42\_rule};
\hf{miulab/Qwen3-1.7B-Usefulness};
\hf{raca-workspace-v1/grpo-tool-sat-sft-qwen3-1p7b-sft-20260419-075623-96e9};
\hf{zsqzz/Qwen3-1.7B\_opsd\_masked\_grpo\_dapo\_hf} \\

& Domain &
\hf{indicnode/Qwen3-1.7B};
\hf{NiuTrans/LMT-60-1.7B-Base};
\hf{Polygl0t/Tucano2-qwen-1.5B-Base};
\hf{VLSP2025-LegalSML/qwen3-1.7b-legal-pretrain} \\

& Pruning/Merging &
\hf{Qwen3-1.7B-Base-LightMagnitudePruned-s005} \\

\midrule
Llama3-8B Base Direct & Source &
\hf{meta-llama/Meta-Llama-3-8B} \\

& SFT &
\hf{allenai/llama-3-tulu-2-8b};
\hf{dphn/dolphin-2.9.1-llama-3-8b};
\hf{Groq/Llama-3-Groq-8B-Tool-Use};
\hf{Magpie-Align/Llama-3-8B-Magpie-Align-SFT-v0.3};
\hf{openchat/openchat-3.6-8b-20240522} \\

& Pref./RL &
\hf{dfurman/Llama-3-8B-Orpo-v0.1} \\

& Domain &
\hf{aaditya/Llama3-OpenBioLLM-8B};
\hf{DeepMount00/Llama-3-8b-Ita};
\hf{hfl/llama-3-chinese-8b};
\hf{winninghealth/WiNGPT2-Llama-3-8B-Chat} \\

& Pruning/Merging &
\hf{RedHatAI/SparseLlama-3-8B-pruned\_50.2of4};
\hf{mlabonne/Llama-3-SLERP-8B};
\hf{Weyaxi/Einstein-v6.1-Llama3-8B} \\

\midrule
Mistral-7B-v0.3 & Source &
\hf{mistralai/Mistral-7B-v0.3} \\

& SFT &
\hf{amdevraj/mistral-7b-ift};
\hf{cypienai/cymist-2-v03-SFT};
\hf{migtissera/Tess-3-7B-SFT} \\

& Pref./RL &
\hf{lamm-mit/mistral-7B-v0.3-Base-CPT-SFT-DPO-09022024};
\hf{llmat/Mistral-v0.3-7B-ORPO} \\

& Domain &
\hf{BEE-spoke-data/Mistral-7B-v0.3-stepbasin-books-20k};
\hf{entfane/math-virtuoso-7B};
\hf{klcsp/mistral7b-fft-coding-11-v1};
\hf{openfoodfacts/spellcheck-mistral-7b};
\hf{pszemraj/Mistral-7B-v0.3-sarcasm-scrolls-4k};
\hf{pucpr-br/Clinical-BR-Mistral-7B-v0.2};
\hf{silversword/US-Immigration-Law-Mistral-7B-v0.3-CPT} \\

& Pruning/Merging &
\hf{ai-and-society/mistral-7B-Instruct-v0.3-wanda-wanda-unstruct-50};
\hf{IntelLabs/sqft-mistral-7b-v0.3-50-base};
\hf{mlfoundations-dev/hp\_ablations\_grid\_mistral\_base-mistralv0.3};
\hf{mlfoundations-dev/mistral\_7b\_0-3\_oh-dcft-v3.1-llama-3.1-405b} \\

\midrule
All evaluated sources & Independent negatives &
\hf{01-ai/Yi-6B};
\hf{EleutherAI/llemma\_7b};
\hf{Qwen/Qwen2.5-7B};
\hf{THUDM/chatglm3-6b-base};
\hf{allenai/tulu-2-dpo-7b};
\hf{lmsys/vicuna-13b-v1.3};
\hf{lmsys/vicuna-7b-v1.3};
\hf{lmsys/vicuna-7b-v1.5};
\hf{meta-llama/Llama-2-13b-chat-hf};
\hf{meta-llama/Llama-2-7b-chat-hf};
\hf{meta-llama/Llama-2-7b-hf};
\hf{meta-llama/Meta-Llama-3-8B};
\hf{microsoft/Phi-3-small-8k-instruct};
\hf{mistralai/Mistral-7B-v0.3};
\hf{tiiuae/falcon-7b} \\

\bottomrule
\end{tabularx}
\end{table}
\FloatBarrier

\let\tcfPerSuspectOldHf\hf
\renewcommand{\hf}[1]{{\fontsize{6.6}{7.5}\selectfont\raggedright\nolinkurl{#1}}}

\subsection{Full Per-Suspect Verification Results}
\label{app:full_per_suspect_results}

In addition to the family-level aggregates, we report the available model-wise verification matrices for inspection. The fixed model counts and family-level AUCs are defined in Tables~\ref{tab:source_pool_config} and~\ref{tab:main_auc_full}. Because some baseline artifacts are unavailable for particular checkpoints, the row-wise AUCs below are computed against the artifact-available independent rows for each table and should be interpreted as diagnostics rather than substitutes for the main family-level AUC.

\begin{tcfappendixbox}{How to read the model-wise tables}
\small
Rows shaded light green are the protected source model or source-family derivatives, and rows shaded light blue are artifact-available independent checkpoints. Each method reports its native fingerprint score and a row-wise diagnostic AUC against the available independent rows in that table. These row-wise diagnostics expose heterogeneity; they do not redefine the main evaluation pool or Table~\ref{tab:main_auc_full}. A dash means that a row-level AUC is not defined for an independent row or that the baseline artifact is unavailable under the official pipeline.
\end{tcfappendixbox}

\begin{table}[htp]
\centering
\caption{Rows where TCF is not the highest row-wise AUC. These are the only exceptions among the source/family-positive rows in the per-suspect matrices below.}
\label{tab:ours_rowwise_auc_exceptions}
\footnotesize
\setlength{\tabcolsep}{2.6pt}
\renewcommand{\arraystretch}{1.08}
\begin{adjustbox}{max width=\textwidth}
\begin{tabular}{llrrrrrl}
\toprule
\rowcolor{TCFOrangeLight}
\textbf{Source family} & \textbf{Derivative} & \textbf{TestAcc} & \textbf{TCF} & \textbf{TRAP} & \textbf{ProF} & \textbf{ZPrint} & \textbf{Interpretation} \\
\midrule
Llama3-8B Base Direct & DeepMount00/Llama-3-8b-Ita & 0.660 & 0.444 & 0.333 & \textbf{0.889} & 0.750 & Italian/domain SFT increases overlap with independent hard-label scores. \\
Llama3-8B Base Direct & RedHatAI/SparseLlama-3-8B-pruned-50.2of4 & 0.650 & 0.778 & \textbf{1.000} & 0.944 & 0.312 & Structured pruning weakens some target-answer preferences while retaining task utility. \\
\bottomrule
\end{tabular}
\end{adjustbox}
\end{table}
\FloatBarrier

\begin{tcfresultbox}{Takeaway}
\small
The model-wise results show that TCF's family-level advantage is distributed across heterogeneous derivatives rather than being driven by a small number of source-like checkpoints.
\end{tcfresultbox}

\begin{table}[htp]
\centering
\caption{Per-suspect comparison using \textbf{Qwen3-1.7B Instruct} as the protected source model. Family-derived suspects are shown in the upper block and independent suspects in the lower block. Score denotes each method's native fingerprint score; AUC denotes row-wise family-versus-independent separability.}
\label{tab:qwen3-instruct-per-suspect}
\fontsize{7.2}{8.3}\selectfont
\setlength{\tabcolsep}{1.35pt}
\renewcommand{\arraystretch}{1.06}
\begin{adjustbox}{max width=\textwidth}
\begin{tabular}{p{0.10\textwidth}p{0.31\textwidth}p{0.065\textwidth} cc cc cc cc}
\toprule
\rowcolor{TCFBlueLight}
\textbf{Category} & \textbf{Suspect Model (Hugging Face ID)} & \textbf{Truth} & \multicolumn{2}{c}{\textbf{TCF}} & \multicolumn{2}{c}{\textbf{TRAP}} & \multicolumn{2}{c}{\textbf{ProFLingo}} & \multicolumn{2}{c}{\textbf{ZeroPrint}} \\
\cmidrule(lr){4-5}\cmidrule(lr){6-7}\cmidrule(lr){8-9}\cmidrule(lr){10-11}
\rowcolor{TCFBlueLight}
& & & \textbf{Score} & \textbf{AUC} & \textbf{Score} & \textbf{AUC} & \textbf{Score} & \textbf{AUC} & \textbf{Score} & \textbf{AUC} \\
\midrule
\rowcolor{TCFGreenLight} Source & \hf{Qwen/Qwen3-1.7B} & Positive & 1.000 & \textbf{1.000} & 0.920 & \textbf{1.000} & 0.200 & \textbf{1.000} & 0.560 & \textbf{1.000} \\
\rowcolor{TCFGreenLight} SFT / task FT & \hf{activeDap/Qwen3-1.7B\_tldr} & Positive & 0.550 & \textbf{1.000} & 0.000 & 0.500 & 0.220 & \textbf{1.000} & 0.650 & \textbf{1.000} \\
\rowcolor{TCFGreenLight} SFT / task FT & \hf{contextboxai/Qwen3-1.7B-FC} & Positive & 1.000 & \textbf{1.000} & 0.610 & \textbf{1.000} & 0.180 & \textbf{1.000} & 0.890 & \textbf{1.000} \\
\rowcolor{TCFGreenLight} SFT / task FT & \hf{prithivMLmods/Demeter-LongCoT-Qwen3-1.7B} & Positive & 0.900 & \textbf{1.000} & 0.110 & \textbf{1.000} & 0.240 & \textbf{1.000} & 0.780 & \textbf{1.000} \\
\rowcolor{TCFGreenLight} SFT / task FT & \hf{UnfilteredAI/DAN-Qwen3-1.7B} & Positive & 0.400 & \textbf{1.000} & 0.010 & \textbf{1.000} & 0.120 & 0.857 & 0.570 & \textbf{1.000} \\
\rowcolor{TCFGreenLight} Preference / RL & \hf{wzx111/Qwen3-1.7B-Open-R1-GRPO-Baseline} & Positive & 1.000 & \textbf{1.000} & 0.350 & \textbf{1.000} & 0.200 & \textbf{1.000} & 0.460 & \textbf{1.000} \\
\rowcolor{TCFGreenLight} Language / Domain & \hf{prithivMLmods/Panacea-MegaScience-Qwen3-1.7B} & Positive & 0.850 & \textbf{1.000} & 0.020 & \textbf{1.000} & 0.140 & 0.857 & 0.750 & \textbf{1.000} \\
\rowcolor{TCFGreenLight} Language / Domain & \hf{swapnillo/Bangla-AI-1.7B} & Positive & 0.700 & \textbf{1.000} & 0.030 & \textbf{1.000} & 0.200 & \textbf{1.000} & 0.510 & \textbf{1.000} \\
\rowcolor{TCFGreenLight} Language / Domain & \hf{vimalgupta/qwen3-1.7b-telecom-vocab-v2} & Positive & 0.350 & \textbf{1.000} & 0.010 & \textbf{1.000} & 0.100 & 0.857 & 0.560 & \textbf{1.000} \\
\rowcolor{TCFGreenLight} Pruning & \hf{MilyaShams/Qwen3-1.7B-Wanda\_1\_4} & Positive & 0.400 & \textbf{1.000} & 0.010 & \textbf{1.000} & 0.100 & 0.857 & 0.580 & \textbf{1.000} \\
\midrule
\rowcolor{TCFBlueLight} Independent & \hf{01-ai/Yi-6B} & Negative & 0.250 & -- & 0.000 & -- & 0.040 & -- & 0.100 & -- \\
\rowcolor{TCFBlueLight} Independent & \hf{EleutherAI/llemma\_7b} & Negative & 0.200 & -- & 0.000 & -- & 0.040 & -- & 0.160 & -- \\
\rowcolor{TCFBlueLight} Independent & \hf{Qwen/Qwen2.5-7B} & Negative & 0.000 & -- & 0.000 & -- & 0.020 & -- & 0.060 & -- \\
\rowcolor{TCFBlueLight} Independent & \hf{meta-llama/Meta-Llama-3-8B} & Negative & 0.100 & -- & 0.000 & -- & 0.000 & -- & 0.180 & -- \\
\rowcolor{TCFBlueLight} Independent & \hf{microsoft/Phi-3-small-8k-instruct} & Negative & 0.000 & -- & 0.000 & -- & 0.160 & -- & 0.190 & -- \\
\rowcolor{TCFBlueLight} Independent & \hf{mistralai/Mistral-7B-v0.3} & Negative & 0.050 & -- & 0.000 & -- & 0.040 & -- & 0.130 & -- \\
\rowcolor{TCFBlueLight} Independent & \hf{THUDM/chatglm3-6b-base} & Negative & 0.100 & -- & 0.000 & -- & 0.020 & -- & -- & -- \\
\bottomrule
\end{tabular}
\end{adjustbox}
\end{table}
\FloatBarrier

\begin{table}[htp]
\centering
\caption{Per-suspect comparison using \textbf{Qwen3-1.7B Base} as the protected source model.}
\label{tab:qwen3-base-per-suspect}
\fontsize{7.0}{8.1}\selectfont
\setlength{\tabcolsep}{1.25pt}
\renewcommand{\arraystretch}{1.05}
\begin{adjustbox}{max width=\textwidth}
\begin{tabular}{p{0.10\textwidth}p{0.31\textwidth}p{0.065\textwidth} cc cc cc cc}
\toprule
\rowcolor{TCFBlueLight}
\textbf{Category} & \textbf{Suspect Model (Hugging Face ID)} & \textbf{Truth} & \multicolumn{2}{c}{\textbf{TCF}} & \multicolumn{2}{c}{\textbf{TRAP}} & \multicolumn{2}{c}{\textbf{ProFLingo}} & \multicolumn{2}{c}{\textbf{ZeroPrint}} \\
\cmidrule(lr){4-5}\cmidrule(lr){6-7}\cmidrule(lr){8-9}\cmidrule(lr){10-11}
\rowcolor{TCFBlueLight}
& & & \textbf{Score} & \textbf{AUC} & \textbf{Score} & \textbf{AUC} & \textbf{Score} & \textbf{AUC} & \textbf{Score} & \textbf{AUC} \\
\midrule
\rowcolor{TCFGreenLight} Source & \hf{Qwen/Qwen3-1.7B-Base} & Positive & 0.900 & \textbf{1.000} & 1.000 & \textbf{1.000} & 0.480 & \textbf{1.000} & 0.620 & \textbf{1.000} \\
\rowcolor{TCFGreenLight} SFT / task FT & \hf{ali-elganzory/Qwen3-1.7B-Base-SFT-Tulu3-decontaminated} & Positive & 0.900 & \textbf{1.000} & 0.550 & \textbf{1.000} & 0.420 & \textbf{1.000} & 0.490 & \textbf{1.000} \\
\rowcolor{TCFGreenLight} SFT / task FT & \hf{ffang2025/Affine-fang-v2} & Positive & 0.400 & \textbf{1.000} & 0.100 & \textbf{1.000} & 0.160 & \textbf{1.000} & 0.250 & 0.750 \\
\rowcolor{TCFGreenLight} SFT / task FT & \hf{jekunz/Qwen3-1.7B-Base-is-SmolTalk} & Positive & 0.800 & \textbf{1.000} & 0.450 & \textbf{1.000} & 0.400 & \textbf{1.000} & 0.620 & \textbf{1.000} \\
\rowcolor{TCFGreenLight} SFT / task FT & \hf{lllyx/Qwen3-1.7B-SFT} & Positive & 0.750 & \textbf{1.000} & 0.370 & \textbf{1.000} & 0.300 & \textbf{1.000} & 0.370 & \textbf{1.000} \\
\rowcolor{TCFGreenLight} Preference / RL & \hf{Kazuki1450/Qwen3-1.7B-Base\_dsum\_3\_6\_fnr\_no\_bracket\_0p0\_0p0\_1p0\_grpo\_42\_rule} & Positive & 0.850 & \textbf{1.000} & 1.000 & \textbf{1.000} & 0.380 & \textbf{1.000} & 0.720 & \textbf{1.000} \\
\rowcolor{TCFGreenLight} Preference / RL & \hf{miulab/Qwen3-1.7B-Usefulness} & Positive & 0.850 & \textbf{1.000} & 0.220 & \textbf{1.000} & 0.240 & \textbf{1.000} & 0.320 & 0.833 \\
\rowcolor{TCFGreenLight} Preference / RL & {\ttfamily\fontsize{6.4}{7.2}\selectfont raca-workspace-v1/\allowbreak grpo-tool-sat-sft-\allowbreak qwen3-1p7b-sft-\allowbreak 20260419-075623-96e9} & Positive & 0.550 & \textbf{1.000} & 0.000 & 0.500 & 0.000 & 0.071 & 0.340 & \textbf{1.000} \\
\rowcolor{TCFGreenLight} Preference / RL & \hf{zsqzz/Qwen3-1.7B\_opsd\_masked\_grpo\_dapo\_hf} & Positive & 0.400 & \textbf{1.000} & 0.090 & \textbf{1.000} & 0.100 & 0.929 & 0.250 & 0.750 \\
\rowcolor{TCFGreenLight} Language / Domain & \hf{indicnode/Qwen3-1.7B} & Positive & 0.400 & \textbf{1.000} & 0.100 & \textbf{1.000} & 0.160 & \textbf{1.000} & 0.340 & \textbf{1.000} \\
\rowcolor{TCFGreenLight} Language / Domain & \hf{NiuTrans/LMT-60-1.7B-Base} & Positive & 0.550 & \textbf{1.000} & 0.000 & 0.500 & 0.160 & \textbf{1.000} & 0.350 & \textbf{1.000} \\
\rowcolor{TCFGreenLight} Language / Domain & \hf{Polygl0t/Tucano2-qwen-1.5B-Base} & Positive & 0.400 & \textbf{1.000} & 0.000 & 0.500 & 0.120 & \textbf{1.000} & 0.290 & 0.833 \\
\rowcolor{TCFGreenLight} Language / Domain & \hf{VLSP2025-LegalSML/qwen3-1.7b-legal-pretrain} & Positive & 0.900 & \textbf{1.000} & 0.850 & \textbf{1.000} & 0.500 & \textbf{1.000} & 0.620 & \textbf{1.000} \\
\rowcolor{TCFGreenLight} Pruning & \hf{Qwen3-1.7B-Base-LightMagnitudePruned-s005} & Positive & 1.000 & \textbf{1.000} & 1.000 & \textbf{1.000} & 0.400 & \textbf{1.000} & 0.840 & \textbf{1.000} \\
\midrule
\rowcolor{TCFBlueLight} Independent & \hf{01-ai/Yi-6B} & Negative & 0.350 & -- & 0.000 & -- & 0.060 & -- & 0.160 & -- \\
\rowcolor{TCFBlueLight} Independent & \hf{EleutherAI/llemma\_7b} & Negative & 0.300 & -- & 0.000 & -- & 0.060 & -- & 0.190 & -- \\
\rowcolor{TCFBlueLight} Independent & \hf{Qwen/Qwen2.5-7B} & Negative & 0.300 & -- & 0.000 & -- & 0.000 & -- & 0.150 & -- \\
\rowcolor{TCFBlueLight} Independent & \hf{meta-llama/Meta-Llama-3-8B} & Negative & 0.300 & -- & 0.000 & -- & 0.040 & -- & 0.250 & -- \\
\rowcolor{TCFBlueLight} Independent & \hf{microsoft/Phi-3-small-8k-instruct} & Negative & 0.300 & -- & 0.000 & -- & 0.100 & -- & 0.330 & -- \\
\rowcolor{TCFBlueLight} Independent & \hf{mistralai/Mistral-7B-v0.3} & Negative & 0.300 & -- & 0.000 & -- & 0.040 & -- & 0.220 & -- \\
\rowcolor{TCFBlueLight} Independent & \hf{THUDM/chatglm3-6b-base} & Negative & 0.350 & -- & 0.000 & -- & 0.080 & -- & -- & -- \\
\bottomrule
\end{tabular}
\end{adjustbox}
\end{table}
\FloatBarrier

\begin{table}[htp]
\centering
\caption{Per-suspect comparison using \textbf{Llama3-8B Base Direct} as the protected source model.}
\label{tab:llama3-base-direct-per-suspect}
\fontsize{6.9}{8.0}\selectfont
\setlength{\tabcolsep}{1.2pt}
\renewcommand{\arraystretch}{1.05}
\begin{adjustbox}{max width=\textwidth}
\begin{tabular}{p{0.10\textwidth}p{0.31\textwidth}p{0.065\textwidth} cc cc cc cc}
\toprule
\rowcolor{TCFBlueLight}
\textbf{Category} & \textbf{Suspect Model (Hugging Face ID)} & \textbf{Truth} & \multicolumn{2}{c}{\textbf{TCF}} & \multicolumn{2}{c}{\textbf{TRAP}} & \multicolumn{2}{c}{\textbf{ProFLingo}} & \multicolumn{2}{c}{\textbf{ZeroPrint}} \\
\cmidrule(lr){4-5}\cmidrule(lr){6-7}\cmidrule(lr){8-9}\cmidrule(lr){10-11}
\rowcolor{TCFBlueLight}
& & & \textbf{Score} & \textbf{AUC} & \textbf{Score} & \textbf{AUC} & \textbf{Score} & \textbf{AUC} & \textbf{Score} & \textbf{AUC} \\
\midrule
\rowcolor{TCFGreenLight} Source & \hf{meta-llama/Meta-Llama-3-8B} & Positive & 0.650 & \textbf{1.000} & 0.990 & \textbf{1.000} & 0.520 & \textbf{1.000} & -- & -- \\
\rowcolor{TCFGreenLight} SFT / task FT & \hf{allenai/llama-3-tulu-2-8b} & Positive & 0.400 & \textbf{1.000} & 0.500 & 0.889 & 0.060 & 0.444 & 0.280 & 0.875 \\
\rowcolor{TCFGreenLight} SFT / task FT & \hf{dphn/dolphin-2.9.1-llama-3-8b} & Positive & 0.450 & \textbf{1.000} & 1.000 & \textbf{1.000} & 0.140 & 0.889 & 0.140 & 0.375 \\
\rowcolor{TCFGreenLight} SFT / task FT & \hf{Groq/Llama-3-Groq-8B-Tool-Use} & Positive & 0.550 & \textbf{1.000} & 0.090 & 0.333 & 0.140 & 0.889 & 0.180 & 0.562 \\
\rowcolor{TCFGreenLight} SFT / task FT & \hf{Magpie-Align/Llama-3-8B-Magpie-Align-SFT-v0.3} & Positive & 0.450 & \textbf{1.000} & 0.640 & \textbf{1.000} & 0.220 & \textbf{1.000} & 0.460 & \textbf{1.000} \\
\rowcolor{TCFGreenLight} SFT / task FT & \hf{openchat/openchat-3.6-8b-20240522} & Positive & 0.450 & \textbf{1.000} & 0.200 & 0.444 & 0.160 & 0.944 & 0.390 & \textbf{1.000} \\
\rowcolor{TCFGreenLight} Preference / RL & \hf{dfurman/Llama-3-8B-Orpo-v0.1} & Positive & 0.650 & \textbf{1.000} & 0.720 & \textbf{1.000} & 0.340 & \textbf{1.000} & 0.430 & \textbf{1.000} \\
\rowcolor{TCFGreenLight} Language / Domain & \hf{aaditya/Llama3-OpenBioLLM-8B} & Positive & 0.400 & \textbf{1.000} & 0.540 & 0.889 & 0.180 & \textbf{1.000} & 0.240 & 0.750 \\
\rowcolor{TCFGreenLight} Language / Domain & \hf{DeepMount00/Llama-3-8b-Ita} & Positive & 0.300 & 0.444 & 0.100 & 0.333 & 0.140 & \textbf{0.889} & 0.240 & 0.750 \\
\rowcolor{TCFGreenLight} Language / Domain & \hf{hfl/llama-3-chinese-8b} & Positive & 0.550 & \textbf{1.000} & 0.700 & \textbf{1.000} & 0.260 & \textbf{1.000} & 0.300 & 0.875 \\
\rowcolor{TCFGreenLight} Language / Domain & \hf{winninghealth/WiNGPT2-Llama-3-8B-Chat} & Positive & 0.400 & \textbf{1.000} & 0.150 & 0.333 & 0.020 & 0.056 & 0.240 & 0.750 \\
\rowcolor{TCFGreenLight} Pruning & \hf{RedHatAI/SparseLlama-3-8B-pruned\_50.2of4} & Positive & 0.350 & 0.778 & 0.890 & \textbf{1.000} & 0.160 & 0.944 & 0.130 & 0.312 \\
\rowcolor{TCFGreenLight} Merging & \hf{mlabonne/Llama-3-SLERP-8B} & Positive & 0.400 & \textbf{1.000} & 0.750 & \textbf{1.000} & 0.380 & \textbf{1.000} & 0.460 & \textbf{1.000} \\
\rowcolor{TCFGreenLight} Merging & \hf{Weyaxi/Einstein-v6.1-Llama3-8B} & Positive & 0.550 & \textbf{1.000} & 0.430 & 0.889 & 0.120 & 0.833 & 0.290 & 0.875 \\
\midrule
\rowcolor{TCFBlueLight} Independent & \hf{01-ai/Yi-6B} & Negative & 0.150 & -- & 0.330 & -- & 0.120 & -- & 0.180 & -- \\
\rowcolor{TCFBlueLight} Independent & \hf{EleutherAI/llemma\_7b} & Negative & 0.350 & -- & 0.200 & -- & 0.060 & -- & 0.170 & -- \\
\rowcolor{TCFBlueLight} Independent & \hf{Qwen/Qwen2.5-7B} & Negative & 0.200 & -- & 0.250 & -- & 0.060 & -- & 0.130 & -- \\
\rowcolor{TCFBlueLight} Independent & \hf{allenai/tulu-2-dpo-7b} & Negative & 0.350 & -- & 0.000 & -- & 0.080 & -- & 0.210 & -- \\
\rowcolor{TCFBlueLight} Independent & \hf{lmsys/vicuna-13b-v1.3} & Negative & 0.350 & -- & 0.560 & -- & 0.020 & -- & 0.000 & -- \\
\rowcolor{TCFBlueLight} Independent & \hf{lmsys/vicuna-7b-v1.5} & Negative & 0.300 & -- & 0.000 & -- & 0.160 & -- & 0.270 & -- \\
\rowcolor{TCFBlueLight} Independent & \hf{meta-llama/Llama-2-13b-chat-hf} & Negative & 0.250 & -- & 0.000 & -- & 0.040 & -- & 0.020 & -- \\
\rowcolor{TCFBlueLight} Independent & \hf{mistralai/Mistral-7B-v0.3} & Negative & 0.350 & -- & 0.200 & -- & 0.060 & -- & 0.330 & -- \\
\rowcolor{TCFBlueLight} Independent & \hf{THUDM/chatglm3-6b-base} & Negative & 0.300 & -- & 0.250 & -- & 0.060 & -- & -- & -- \\
\bottomrule
\end{tabular}
\end{adjustbox}
\end{table}
\FloatBarrier

\begin{table}[htp]
\centering
\caption{Per-suspect comparison using \textbf{Mistral-7B-v0.3} as the protected source model.}
\label{tab:mistral-v03-per-suspect}
\fontsize{6.8}{7.9}\selectfont
\setlength{\tabcolsep}{1.15pt}
\renewcommand{\arraystretch}{1.04}
\begin{adjustbox}{max width=\textwidth}
\begin{tabular}{p{0.10\textwidth}p{0.31\textwidth}p{0.065\textwidth} cc cc cc cc}
\toprule
\rowcolor{TCFBlueLight}
\textbf{Category} & \textbf{Suspect Model (Hugging Face ID)} & \textbf{Truth} & \multicolumn{2}{c}{\textbf{TCF}} & \multicolumn{2}{c}{\textbf{TRAP}} & \multicolumn{2}{c}{\textbf{ProFLingo}} & \multicolumn{2}{c}{\textbf{ZeroPrint}} \\
\cmidrule(lr){4-5}\cmidrule(lr){6-7}\cmidrule(lr){8-9}\cmidrule(lr){10-11}
\rowcolor{TCFBlueLight}
& & & \textbf{Score} & \textbf{AUC} & \textbf{Score} & \textbf{AUC} & \textbf{Score} & \textbf{AUC} & \textbf{Score} & \textbf{AUC} \\
\midrule
\rowcolor{TCFGreenLight} Source & \hf{mistralai/Mistral-7B-v0.3} & Positive & 0.600 & \textbf{1.000} & 0.250 & \textbf{1.000} & 0.660 & \textbf{1.000} & 0.570 & \textbf{1.000} \\
\rowcolor{TCFGreenLight} SFT / task FT & \hf{amdevraj/mistral-7b-ift} & Positive & 0.600 & \textbf{1.000} & 0.000 & 0.500 & 0.220 & \textbf{1.000} & 0.310 & \textbf{1.000} \\
\rowcolor{TCFGreenLight} SFT / task FT & \hf{cypienai/cymist-2-v03-SFT} & Positive & 0.400 & \textbf{1.000} & 0.000 & 0.500 & 0.260 & \textbf{1.000} & 0.310 & \textbf{1.000} \\
\rowcolor{TCFGreenLight} SFT / task FT & \hf{migtissera/Tess-3-7B-SFT} & Positive & 0.450 & \textbf{1.000} & 0.000 & 0.500 & 0.200 & \textbf{1.000} & 0.320 & \textbf{1.000} \\
\rowcolor{TCFGreenLight} Preference / RL & \hf{lamm-mit/mistral-7B-v0.3-Base-CPT-SFT-DPO-09022024} & Positive & 0.300 & \textbf{1.000} & 0.000 & 0.500 & 0.160 & \textbf{1.000} & 0.320 & \textbf{1.000} \\
\rowcolor{TCFGreenLight} Preference / RL & \hf{llmat/Mistral-v0.3-7B-ORPO} & Positive & 0.350 & \textbf{1.000} & 0.000 & 0.500 & 0.200 & \textbf{1.000} & 0.470 & \textbf{1.000} \\
\rowcolor{TCFGreenLight} Language / Domain & \hf{BEE-spoke-data/Mistral-7B-v0.3-stepbasin-books-20k} & Positive & 0.600 & \textbf{1.000} & 0.170 & \textbf{1.000} & 0.500 & \textbf{1.000} & 0.580 & \textbf{1.000} \\
\rowcolor{TCFGreenLight} Language / Domain & \hf{entfane/math-virtuoso-7B} & Positive & 0.450 & \textbf{1.000} & 0.000 & 0.500 & 0.180 & \textbf{1.000} & 0.270 & 0.800 \\
\rowcolor{TCFGreenLight} Language / Domain & \hf{klcsp/mistral7b-fft-coding-11-v1} & Positive & 0.350 & \textbf{1.000} & 0.000 & 0.500 & 0.000 & 0.000 & 0.170 & 0.200 \\
\rowcolor{TCFGreenLight} Language / Domain & \hf{openfoodfacts/spellcheck-mistral-7b} & Positive & 0.550 & \textbf{1.000} & 0.000 & 0.500 & 0.380 & \textbf{1.000} & 0.590 & \textbf{1.000} \\
\rowcolor{TCFGreenLight} Language / Domain & \hf{pszemraj/Mistral-7B-v0.3-sarcasm-scrolls-4k} & Positive & 0.450 & \textbf{1.000} & 0.000 & 0.500 & 0.120 & \textbf{1.000} & 0.390 & \textbf{1.000} \\
\rowcolor{TCFGreenLight} Language / Domain & \hf{pucpr-br/Clinical-BR-Mistral-7B-v0.2} & Positive & 0.250 & \textbf{1.000} & 0.000 & 0.500 & 0.060 & 0.250 & 0.350 & \textbf{1.000} \\
\rowcolor{TCFGreenLight} Language / Domain & \hf{silversword/US-Immigration-Law-Mistral-7B-v0.3-CPT} & Positive & 0.550 & \textbf{1.000} & 0.000 & 0.500 & 0.440 & \textbf{1.000} & 0.570 & \textbf{1.000} \\
\rowcolor{TCFGreenLight} Pruning & \hf{ai-and-society/mistral-7B-Instruct-v0.3-wanda-wanda-unstruct-50} & Positive & 0.350 & \textbf{1.000} & 0.000 & 0.500 & 0.400 & \textbf{1.000} & 0.300 & 0.900 \\
\rowcolor{TCFGreenLight} Pruning & \hf{IntelLabs/sqft-mistral-7b-v0.3-50-base} & Positive & 0.400 & \textbf{1.000} & 0.000 & 0.500 & 0.140 & \textbf{1.000} & 0.360 & \textbf{1.000} \\
\rowcolor{TCFGreenLight} Merging & \hf{mlfoundations-dev/hp\_ablations\_grid\_mistral\_base-mistralv0.3} & Positive & 0.350 & \textbf{1.000} & 0.000 & 0.500 & 0.100 & 0.917 & 0.260 & 0.800 \\
\rowcolor{TCFGreenLight} Merging & \hf{mlfoundations-dev/mistral\_7b\_0-3\_oh-dcft-v3.1-llama-3.1-405b} & Positive & 0.450 & \textbf{1.000} & 0.000 & 0.500 & 0.400 & \textbf{1.000} & 0.270 & 0.800 \\
\midrule
\rowcolor{TCFBlueLight} Independent & \hf{01-ai/Yi-6B} & Negative & 0.150 & -- & 0.000 & -- & 0.060 & -- & 0.250 & -- \\
\rowcolor{TCFBlueLight} Independent & \hf{EleutherAI/llemma\_7b} & Negative & 0.200 & -- & 0.000 & -- & 0.080 & -- & 0.230 & -- \\
\rowcolor{TCFBlueLight} Independent & \hf{Qwen/Qwen2.5-7B} & Negative & 0.200 & -- & 0.000 & -- & 0.060 & -- & 0.120 & -- \\
\rowcolor{TCFBlueLight} Independent & \hf{meta-llama/Meta-Llama-3-8B} & Negative & 0.100 & -- & 0.000 & -- & 0.080 & -- & 0.300 & -- \\
\rowcolor{TCFBlueLight} Independent & \hf{microsoft/Phi-3-small-8k-instruct} & Negative & 0.100 & -- & 0.000 & -- & 0.100 & -- & 0.190 & -- \\
\rowcolor{TCFBlueLight} Independent & \hf{THUDM/chatglm3-6b-base} & Negative & 0.200 & -- & 0.000 & -- & 0.060 & -- & -- & -- \\
\bottomrule
\end{tabular}
\end{adjustbox}
\end{table}
\FloatBarrier

\begin{tcfappendixbox}{Compatibility note}
\small
ChatGLM is included as an independent negative for TCF, TRAP, and ProFLingo whenever a valid hard-label artifact is available. Its ZeroPrint entry remains ``--'' because the official/repository-aligned ZeroPrint pipeline encountered a third-party remote-code compatibility issue for ChatGLM in the evaluated software environment; this missing artifact is excluded only from the ZeroPrint AUC calculation.
\end{tcfappendixbox}
\let\hf\tcfPerSuspectOldHf

\subsection{Extended Construction and Verification Record}
\label{app:extended_reproducibility}

\begin{tcfappendixbox}{Artifact repository}
\small
Project available at: 
\par\smallskip
\centering
\href{https://github.com/Underflow-1/TCF-LLM-Fingerprinting}{\nolinkurl{https://github.com/Underflow-1/TCF-LLM-Fingerprinting}}
\end{tcfappendixbox}

The main paper presents the minimal method definition needed to understand TCF. This section records the complete source-only construction path, the retained fingerprint fields, and the online hard-label decision rule so that every reported result can be reconstructed from explicit intermediate artifacts.

\subsubsection{Offline Construction Workflow}
\label{app:offline_construction_workflow}
Tables~\ref{tab:extended_construction_workflow} and~\ref{tab:fingerprint_registry_schema} summarize the complete offline pipeline and the frozen registry fields passed to the online verifier.

\begin{table}[htp]
\centering
\caption{End-to-end source-only fingerprint construction workflow.}
\label{tab:extended_construction_workflow}
\small
\setlength{\tabcolsep}{4pt}
\renewcommand{\arraystretch}{1.12}
\begin{tabularx}{\textwidth}{>{\bfseries}p{0.12\textwidth}p{0.22\textwidth}X}
\toprule
\rowcolor{HeaderGray}
Stage & Input & Operation and retained evidence \\
\midrule
Question preparation & Clean multiple-choice item & Render the fixed answer-only interface and preserve the original question identifier, category, options, and gold answer. \\
Clean source query & Protected model $M_0$ & Record the clean parsed answer and the valid-label distribution used only for offline target selection. \\
Target selection & Clean label distribution & Choose a non-clean target satisfying the clean-side SCM requirement and reject questions with no feasible target. \\
Perturbation search & Source model and target & Optimize the text perturbation with GCG and save the source loss, valid-label probabilities, and SCM trajectory at every checkpoint. \\
Early stopping / checkpoint retention & Source loss and SCM & Retain the first checkpoint satisfying the main-text source-side stopping condition, then recompute parser-side SCM and source stability. \\
Qualification & Retained checkpoint & Recompute parser-side SCM, source validity, response stability, and prompt-format checks; reject candidates that fail any condition. \\
Diversity filtering & Qualified candidates & Balance categories, clean labels, target labels, and question identities before forming the final registry. \\
Registry export & Final fingerprint set & Store the exact prompt, private target, clean answer, perturbation text, SCM metadata, and parser contract required by online verification. \\
\bottomrule
\end{tabularx}
\end{table}

\begin{table}[htp]
\centering
\caption{Recommended fingerprint-registry fields. Private fields are never sent to the suspect service.}
\label{tab:fingerprint_registry_schema}
\small
\setlength{\tabcolsep}{4pt}
\renewcommand{\arraystretch}{1.1}
\begin{tabularx}{\textwidth}{p{0.20\textwidth}p{0.12\textwidth}X}
\toprule
\rowcolor{HeaderGray}
Field & Visibility & Purpose \\
\midrule
Fingerprint ID & Internal & Stable key linking construction, query, parser, and result artifacts. \\
Source model & Internal & Protected checkpoint from which the fingerprint was constructed. \\
Question ID / category & Internal & Supports category-balanced sampling and reproducibility. \\
Clean prompt & Internal & Records the pre-perturbation source behavior and clean answer. \\
Fingerprint prompt & Queried & Exact prompt sent to the suspect API. \\
Clean answer & Internal & Source label before the counterfactual perturbation. \\
Target answer & Private & Ownership target used by the hard-label verifier. \\
Optimized text & Private & Contiguous prefix or distributed NaturalCarrier fragments. \\
Checkpoint / SCM & Internal & Supports construction audits and early-stopping reconstruction. \\
Parser version & Internal & Pins the deterministic response-to-label contract. \\
Raw response & Audit log & Raw suspect response retained before parsing. \\
Parsed label / hit & Audit log & Final legal label and target-hit indicator used for aggregation. \\
\bottomrule
\end{tabularx}
\end{table}
\FloatBarrier

\subsubsection{Online Hard-Label Verification Workflow}
\label{app:online_verification_workflow}

\begin{tcfappendixbox}{Online verification procedure}
\small
\begin{enumerate}[leftmargin=1.8em,itemsep=2pt,topsep=2pt]
    \item Load the frozen fingerprint registry and pre-registered decision threshold.
    \item Send each stored fingerprint prompt to the suspect service using the declared decoding configuration.
    \item Save the raw response before any parsing or normalization.
    \item Apply the deterministic parser and return one label in $\{A,B,C,D\}$ or $\bot$ for invalid or ambiguous output.
    \item Count a hit only when the parsed label equals the private target recorded for that fingerprint.
    \item Average the binary hits to obtain target accuracy and compare it with the pre-registered threshold.
    \item Export the prompt identifier, raw response, parsed label, and decision trace as the verification audit record.
\end{enumerate}
\end{tcfappendixbox}

\begin{algorithm}[H]
\caption{Hard-label ownership verification}
\label{alg:hard_label_verification}
\small
\begin{algorithmic}[1]
\Require Frozen fingerprint registry $\mathcal{F}$; suspect black-box service $M_s$; deterministic parser $\operatorname{Parse}$; pre-registered threshold $\zeta$.
\Ensure Ownership decision and audit trace.
\State $S\gets0$
\For{each fingerprint record $(p_i,t_i)\in\mathcal{F}$}
    \State Query $M_s$ with $p_i$ and save raw response $r_i$.
    \State $\hat a_i\gets\operatorname{Parse}(r_i)\in\mathcal{A}\cup\{\bot\}$.
    \If{$\hat a_i=t_i$}
        \State $S\gets S+1$
    \EndIf
    \State Save $(i,r_i,\hat a_i,\mathbb{I}[\hat a_i=t_i])$ to the audit trace.
\EndFor
\State $\operatorname{TA}(M_s;\mathcal{F})\gets S/|\mathcal{F}|$.
\If{$\operatorname{TA}(M_s;\mathcal{F})\ge\zeta$}
    \State \Return \textsc{Derived}, audit trace
\Else
    \State \Return \textsc{Independent}, audit trace
\EndIf
\end{algorithmic}
\end{algorithm}
\FloatBarrier

The online phase never accesses suspect logits, hidden states, token probabilities, gradients, embeddings, or semantic-similarity models. Invalid outputs remain failures rather than being repaired by a secondary model. This strict contract keeps the reported score auditable at the final-response interface.

\begin{tcfresultbox}{Takeaway}
\small
Online verification reduces ownership evidence to deterministic final-label target hits under a frozen parser and threshold.
\end{tcfresultbox}

\subsubsection{Reproducibility Checklist}
\label{app:reproducibility_checklist}
Table~\ref{tab:reproducibility_checklist} records the minimum configuration and audit artifacts needed to rebuild the reported evaluation without changing the verifier contract.

\begin{table}[htp]
\centering
\caption{Reproducibility checklist for rebuilding a TCF evaluation.}
\label{tab:reproducibility_checklist}
\small
\setlength{\tabcolsep}{4pt}
\renewcommand{\arraystretch}{1.12}
\begin{tabularx}{\textwidth}{p{0.23\textwidth}X}
\toprule
\rowcolor{HeaderGray}
Item & Required record \\
\midrule
Model identity & Exact source, derivative, and independent checkpoint identifiers plus revisions when available. \\
Question pool & Dataset version, split, category list, question identifiers, and decontamination policy. \\
Prompt interface & Exact clean and fingerprint templates, answer-only instruction, insertion positions, and parser version. \\
Construction & Random seed, prefix length, optimization steps, candidate batch size, top-$k$, checkpoint schedule, and SCM acceptance rule. \\
Verification & Decoding parameters, query repetitions, invalid-output policy, target-hit aggregation, and threshold provenance. \\
Baselines & Original protocol, query budget, official score definition, implementation source, and any compatibility patch. \\
Statistics & Positive/negative taxonomy, AUC tie handling, subgroup definitions, and any pooled-score normalization. \\
Artifacts & Raw responses, parsed outputs, per-suspect scores, plot-data tables, and the script used to produce each figure. \\
\bottomrule
\end{tabularx}
\end{table}
\FloatBarrier

\subsection{Expanded Baseline Documentation}
\label{app:expanded_baseline_documentation}

The baselines are evaluated under their own score definitions rather than being forced into TCF's target-hit score. The following examples make the interface difference explicit and complement the compact configuration table in Appendix~\ref{app:baseline_setup}.

\subsubsection{Representative TRAP Artifact}

\begin{tcfpromptbox}{TRAP retrieval artifact}
{\footnotesize\ttfamily\raggedright
Task: produce a short target string under the official TRAP retrieval protocol.\par
Optimized suffix: lista ( print-phaseko@(?/Topic/TabControlighborhood ... [truncated]\par
Target response: 6532\par
Generated response: 6532\par
Official score contribution: successful target retrieval.\par}
\end{tcfpromptbox}

TRAP verifies whether its optimized suffix retrieves the designated target output. Its score is therefore a retrieval statistic and is not converted into TCF's multiple-choice target-hit rate.

\subsubsection{Representative ProFLingo Artifact}

\begin{tcfpromptbox}{ProFLingo adversarial-example record}
{\footnotesize\ttfamily\raggedright
Question: Where does the sun rise?\par
Target keyword: north\par
Generated answer: The sun rises in the east.\par
Official score contribution: unsuccessful target-keyword retrieval.\par}
\end{tcfpromptbox}

ProFLingo evaluates a list of adversarial examples and reports the corresponding attack-success statistic. The original list length and official target-keyword rule are retained in our comparison.

\subsubsection{Representative ZeroPrint Artifact}

\begin{tcfpromptbox}{ZeroPrint source-row similarity artifact}
{\footnotesize\ttfamily\raggedright
Source row: Qwen3-1.7B-Base-ZP\par
Source similarity: 1.000\par
Derivative similarities: stored continuous source-row values\par
Independent similarities: stored continuous source-row values\par
Official score contribution: source-row similarity/correlation.\par}
\end{tcfpromptbox}

ZeroPrint produces a continuous similarity profile rather than a prompt-to-target hard-label event. We therefore preserve its source-row score and compute AUC directly from that profile.

\begin{tcfresultbox}{Takeaway}
\small
The comparison preserves each baseline's original evidence channel and score definition.
\end{tcfresultbox}

\subsection{Expanded Naturalness Documentation}
\label{app:expanded_naturalness_documentation}

\subsubsection{Naturalness Metrics}

No single metric fully characterizes prompt naturalness. We therefore separate local trigger abnormality, whole-prompt fluency, repetition, and embedding-space diagnostics.

\begin{table}[htp]
\centering
\caption{Naturalness and prompt-quality diagnostics used in the extended analysis.}
\label{tab:naturalness_metrics_extended}
\small
\setlength{\tabcolsep}{4pt}
\renewcommand{\arraystretch}{1.12}
\begin{tabularx}{\textwidth}{p{0.20\textwidth}p{0.24\textwidth}X}
\toprule
\rowcolor{HeaderGray}
Metric & Textual object & Interpretation \\
\midrule
Full-input PPL & Complete rendered query & Measures global fluency after inserting all fingerprint text. \\
Window-PPL & Local windows around insertions & Detects short abnormal regions that can be hidden by a long fluent prompt. \\
Fragment PPL & Inserted prefix or fragment only & Isolates the lexical abnormality of the fingerprint text itself. \\
Abnormal-token ratio & Inserted text & Counts tokenizer fragments that violate the readable-token constraints. \\
Repetition score & Inserted text & Detects repeated words, substrings, or token cycles. \\
Semantic-drift proxy & Candidate versus readable seed & Measures how far a naturalized candidate moves from its seed instruction. \\
Prompt embedding & Full prompt or prompt residual & Provides a qualitative view of proximity to benign multiple-choice prompts. \\
Verification AUC & Suspect-model score & Ensures that improved fluency is not obtained by discarding ownership evidence. \\
\bottomrule
\end{tabularx}
\end{table}
\FloatBarrier

\subsubsection{Naturalization Variant Taxonomy}
All variants use source-only construction. Derived and independent models are queried only after the candidate prompt has been frozen, preventing evaluation leakage into fingerprint generation.

\begin{table}[htp]
\centering
\caption{Naturalness-oriented construction variants and their isolated design questions.}
\label{tab:naturalness_variant_taxonomy}
\small
\setlength{\tabcolsep}{3.5pt}
\renewcommand{\arraystretch}{1.12}
\begin{tabularx}{\textwidth}{p{0.20\textwidth}p{0.24\textwidth}X}
\toprule
\rowcolor{HeaderGray}
Variant & Construction rule & Question isolated by the variant \\
\midrule
Raw GCG & Keep the optimized contiguous trigger. & Reference verification strength before naturalization. \\
SemBound-GCG & Restrict candidate semantic drift from a readable seed. & Can explicit semantic constraints preserve target induction? \\
SourceRerank-GCG & Retain source-effective candidates, then rerank by naturalness. & Does source-first selection avoid sacrificing verification strength? \\
PPLRerank-GCG & Add a prefix-fluency proxy to candidate selection. & Can local fluency be optimized directly during GCG search? \\
Position-Fragment-GCG & Split an effective trigger across prompt slots without carriers. & Is distribution alone sufficient to reduce local abnormality? \\
NaturalCarrier & Wrap distributed fragments with short cue-like phrases. & Can carriers improve readability while preserving the trigger information? \\
TCF-S4 / TCF-S8 & Use four or eight distributed insertion groups. & How does split granularity affect naturalness and erasure robustness? \\
SlotPrefix-GF & Search only over natural slot templates. & Can fully natural gradient-free prompts replace token-level triggers? \\
EvoPrefix-GF & Mutate a population of natural instructions. & Is population-based natural-language search sufficient? \\
\bottomrule
\end{tabularx}
\end{table}
\FloatBarrier

\subsubsection{Prompt-Embedding Diagnostics}
\label{app:naturalness_embedding_diagnostics}

The embedding model is used only for visualization and never participates in fingerprint construction or verification. We report both seed stability and distance-to-reference views because a single t-SNE projection can be sensitive to initialization.

\begin{figure}[htp]
\centering
\includegraphics[width=0.98\textwidth]{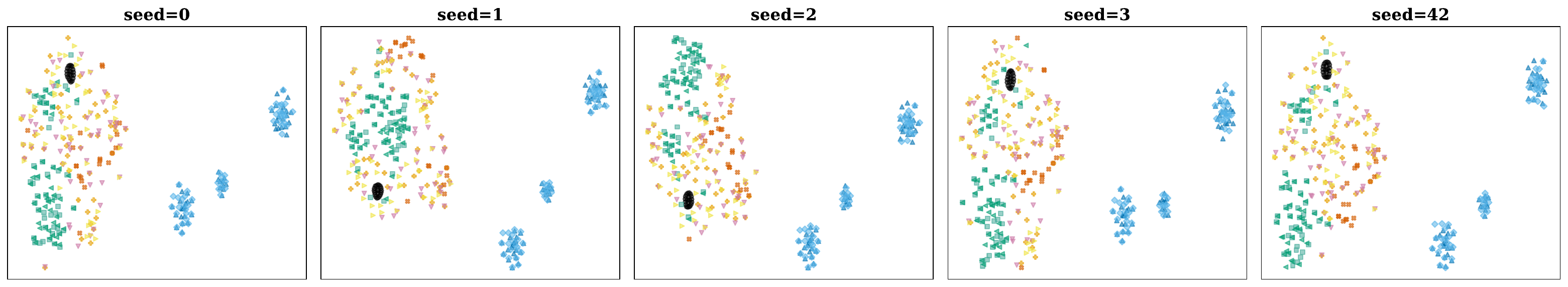}
\caption{Seed-stability diagnostic for the prompt-embedding visualization. Repeated projections test whether the qualitative arrangement is robust to visualization randomness.}
\label{fig:app_tsne_seed_stability}
\end{figure}

\begin{figure}[htp]
\centering
\includegraphics[width=0.64\textwidth]{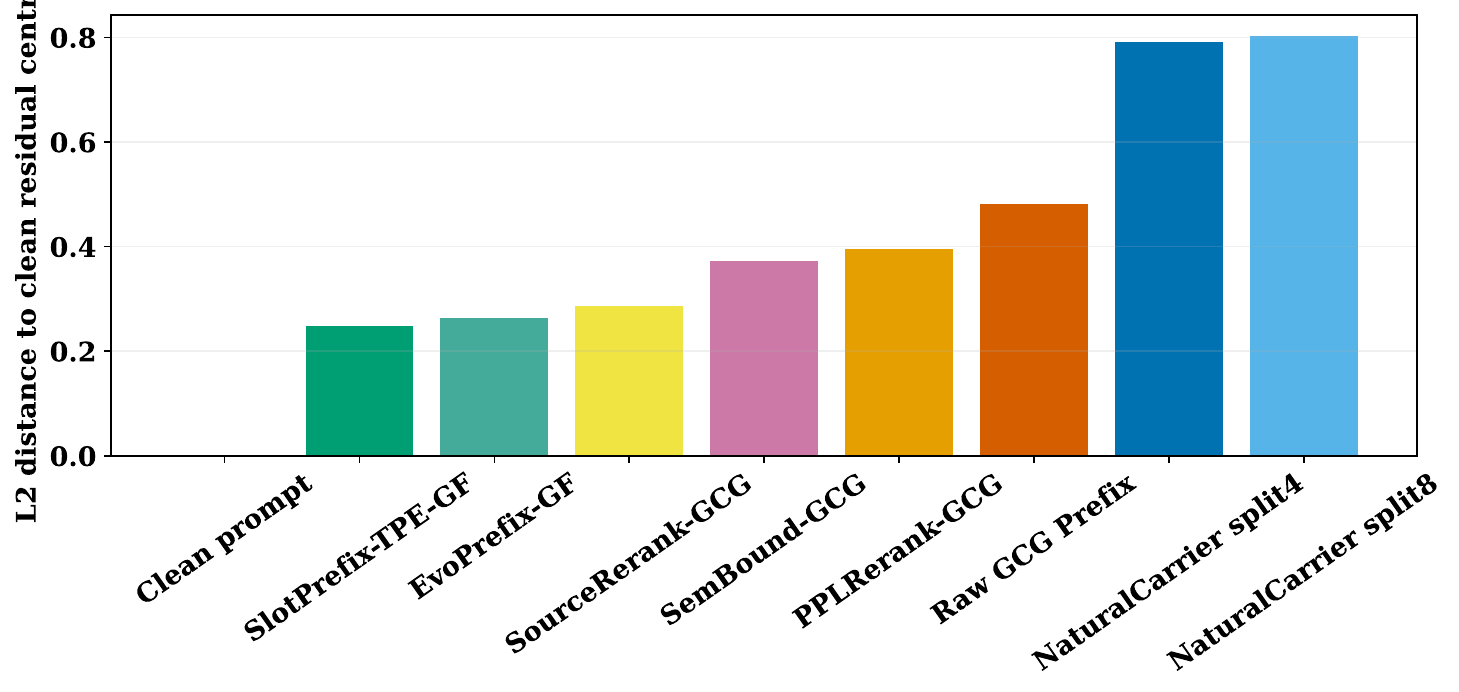}
\caption{Embedding-distance diagnostic for naturalized prompt variants. The external embedding model is used only for offline interpretation.}
\label{fig:app_embedding_distance_bar}
\end{figure}
\FloatBarrier

\begin{tcfresultbox}{Takeaway}
\small
The embedding diagnostics are interpretive rather than monotonic: some natural-language search/reranking variants move closer to the clean-prompt residual centroid, whereas the split NaturalCarrier variants retain a larger structured embedding shift. Their naturalness advantage is therefore supported primarily by the PPL diagnostics in the main text, while verification remains strictly hard-label.
\end{tcfresultbox}
\FloatBarrier

\subsection{Additional Distillation Diagnostics}
\label{app:additional_distillation_diagnostics}

The main text reports the final clean-query extraction result. To show how the ownership signal develops during student training, we additionally expose a checkpoint sweep from a smaller clean-query distillation run. The student never observes the private fingerprint prompts.

\begin{table}[htp]
\centering
\caption{Checkpoint sweep for clean-query distillation from Mistral-7B-v0.3 to Llama-2-7B. Clean100 is clean multiple-choice accuracy, T-Agree is teacher agreement, and the fingerprint rows report student-versus-independent AUC.}
\label{tab:mistral_llama2_clean250_step_sweep}
\scriptsize
\setlength{\tabcolsep}{2.0pt}
\renewcommand{\arraystretch}{1.05}
\begin{adjustbox}{max width=0.92\textwidth}
\begin{tabular}{lccccccccccc}
\toprule
Metric & S0 & S1 & S2 & S3 & S4 & S5 & S6 & S7 & S8 & S9 & S10 \\
\midrule
Clean100 & 0.66 & 0.65 & 0.69 & 0.75 & 0.73 & 0.77 & 0.73 & 0.71 & 0.71 & 0.72 & 0.72 \\
T-Agree  & 0.68 & 0.64 & 0.67 & 0.73 & 0.73 & 0.75 & 0.73 & 0.73 & 0.71 & 0.72 & 0.72 \\
\method{} & 0.31 & 0.56 & 0.75 & 0.56 & 0.75 & 0.88 & 0.81 & 0.88 & 0.88 & 0.88 & 0.88 \\
TRAP     & 0.50 & 0.50 & 0.50 & 0.50 & 0.50 & 0.50 & 0.50 & 0.50 & 0.50 & 0.50 & 0.50 \\
ProFLingo & 0.00 & 0.00 & 0.00 & 0.00 & 0.00 & 0.00 & 0.00 & 0.06 & 0.00 & 0.00 & 0.00 \\
ZeroPrint & 0.61 & 0.22 & 0.22 & 0.11 & 0.11 & 0.00 & 0.11 & 0.11 & 0.11 & 0.11 & 0.11 \\
\bottomrule
\end{tabular}
\end{adjustbox}
\end{table}
\FloatBarrier

The trajectory shows that TCF evidence appears as the student absorbs more of the teacher's clean behavior, even though the extraction data contains no private fingerprint prompts. This distinguishes inherited source behavior from direct memorization of the fingerprint registry.

\begin{tcfresultbox}{Takeaway}
\small
The ownership signal strengthens as the extracted student learns the teacher's clean behavior.
\end{tcfresultbox}

\subsection{Artifact Map and Expected Outputs}
\label{app:artifact_map}

\begin{tcfappendixbox}{Public reproduction versus deployment privacy}
\small
The registry released with this paper is a reproducibility artifact
for reconstructing the reported experiments. It must not be reused as a secret
registry in a real ownership audit. The threat model in Sec.~\ref{sec:threat_model}
requires deployment fingerprints and their targets to remain unknown to the
suspect before verification, so a deployment-grade audit should generate a
fresh private registry and keep its targets secret until the audit is complete.
\par\smallskip
\textbf{Minimum audit bundle.} A third party should be able to reconstruct every reported point from four immutable records: the frozen fingerprint registry, the raw suspect responses, the deterministic parser outputs, and the positive/negative model taxonomy. Plotting scripts should consume these records without querying any model again.
\end{tcfappendixbox}

\begin{table}[htp]
\centering
\caption{Logical artifact map for reproducing the reported experiments. Directory names may be adapted locally, but the listed records should remain immutable.}
\label{tab:artifact_map}
\small
\setlength{\tabcolsep}{4pt}
\renewcommand{\arraystretch}{1.12}
\begin{tabularx}{\textwidth}{p{0.24\textwidth}X}
\toprule
\rowcolor{HeaderGray}
Artifact group & Expected contents \\
\midrule
\texttt{configs/} & Source-model identifiers, generation parameters, decoding settings, thresholds, and random seeds. \\
\texttt{fingerprints/} & Clean prompts, paper-reproduction fingerprint prompts, recorded targets, SCM trajectories, qualification records, and registry manifests; deployment targets must remain private. \\
\texttt{model\_pool/} & Positive/negative taxonomy, checkpoint revisions, derivation categories, and compatibility notes. \\
\texttt{responses/} & Raw suspect outputs before parsing, grouped by source family, method, and robustness condition. \\
\texttt{scores/} & Parsed labels, target-hit vectors, per-suspect scores, AUC inputs, and subgroup summaries. \\
\texttt{baselines/} & Original baseline configurations, generated artifacts, official score fields, and any compatibility patch. \\
\texttt{plots/} & Figure-ready CSV/JSON files and deterministic scripts for every paper figure. \\
\texttt{logs/} & Construction, verification, extraction, robustness, and environment logs required for audit. \\
\texttt{paper/} & LaTeX source, bibliography, figure assets, and the arXiv build script. \\
\bottomrule
\end{tabularx}
\end{table}
\FloatBarrier

\end{document}